\documentclass[%
 reprint,
nofootinbib,
 amsmath,amssymb,
 aps,
 prd,
]{revtex4-2}

\usepackage{multirow}
\usepackage{graphicx}
\usepackage{dcolumn}
\usepackage{bm}
\usepackage[dvipsnames]{xcolor}
\usepackage{hyperref}
\usepackage{enumerate}
\usepackage{amsmath}
\usepackage{CJKutf8} 
\newcommand{\mnras}{MNRAS}
\newcommand{\aap}{Astronomy \& Astrophysics}
\newcommand{\jcap}{Journal of Cosmology and Astroparticle Physics}
\newcommand{\apjl}{Astrophysical Journal Letters}
\newcommand{\physrep}{Physics Reports}
\newcommand{\pasj}{Publications of the Astronomical Society of Japan}

\newcommand{\apjs}{The Astrophysical Journal Supplement Series}
\newcommand{\aapr}{The Astronomy \& Astrophysics Review}

\newcommand{\UM}{Department of Physics, University of Michigan, 450 Church St, Ann Arbor, MI 48109, USA}
\newcommand{\LCTP}{Leinweber Center for Theoretical Physics, 450 Church St, Ann Arbor, MI 48109, USA}

\begin{document}

\preprint{APS/123-QED}

\title{Ten-parameter simulation suite for cosmological emulation beyond $\Lambda$CDM}

\author{\begin{CJK}{UTF8}{gbsn}Yanhui Yang (\CJKfamily{gbsn}杨焱辉)$^1$\end{CJK}}

 \email{yyang440@ucr.edu}
\author{Simeon Bird$^1$}%
 \email{sbird@ucr.edu}
\author{\begin{CJK}{UTF8}{bsmi}Ming-Feng Ho (何銘峰)$^{1,2,3}$\end{CJK}}
\affiliation{%
$^1$Department of Physics \& Astronomy, University of California, Riverside, 900 University Ave., Riverside, CA 92521, USA
}%
\affiliation{$^2$\UM}
\affiliation{$^3$\LCTP}




\date{\today}

\begin{abstract}

We present \texttt{Goku}, a suite of cosmological $N$-body simulations, and the corresponding 10-dimensional emulator, \texttt{GokuEmu}, for the nonlinear matter power spectrum. The simulations span the base parameters of $\Lambda$ Cold Dark Matter ($\Lambda$CDM) cosmology and its extensions, including dynamical dark energy ($w_0$, $w_a$), the sum of the neutrino masses ($\sum m_\nu$), the effective number of neutrinos ($N_\text{eff}$), and the running of the scalar spectral index ($\alpha_\text{s}$), enabling tests of new physics with data from upcoming surveys like the Roman Space Telescope, \textit{Euclid}, and LSST. Designed within the \texttt{MF-Box} framework, which integrates multi-scale and multi-fidelity simulations, the suite includes high-fidelity simulations evolving $3000^3$ particles in $1\,(\text{Gpc}/h)^3$ volumes and low-fidelity simulations with $750^3$ particles across varying box sizes. This approach achieves percent-level accuracy in high-likelihood regions and 5\% accuracy across broader parameter ranges, while reducing computational costs by 94\% compared to single-fidelity methods.

The simulations adopt an accurate treatment of massive neutrinos, enhancing predictions of the matter power spectrum on nonlinear scales. Key innovations include an adaptive sampling strategy and the use of beam search to optimize generalization accuracy. The emulator is valid for redshifts $z \leq 3$ and scales $0.01 \lesssim k / (h \, \text{Mpc}^{-1}) \lesssim 10$. Beyond the matter power spectrum, the simulations also support analyses of other statistical measures, such as the halo mass function. The emulator and its training data are publicly available at \url{https://github.com/astro-YYH/GokuEmu}, providing a valuable resource for cosmological parameter inference and model testing.
\end{abstract}
    
\maketitle


\section{\label{sec:intro}Introduction}

An unprecedented amount of galaxy photometric and spectroscopic data will be made available by ongoing and upcoming cosmological surveys such as DESI~\cite{DESICollaboration2016}, LSST~\cite{LSC2009}, \textit{Euclid}~\cite{Laureijs2011}, the Nancy Grace Roman Space Telescope~\cite{Akeson2019}, the China Space Station Telescope (CSST)~\cite{Gong2019}, and the Prime Focus Spectrograph (PFS) on the Subaru Telescope~\cite{Takada2014}. These surveys will provide complementary datasets across multiple scales, enabling precise measurements of the galaxy power spectrum, as well as the weak lensing shear field.

A common approach to inferring cosmological parameters from these measurements is to evaluate the likelihood of forward models via a summary statistic, such as the matter power spectrum, and use Bayesian techniques. A typical inference run would require $10^6$–$10^7$ matter power spectrum evaluations at different cosmological parameters. Complicating the analysis, mode-counting arguments suggest that much of the cosmological parameter information in these surveys is on scales where the density contrast is too large to be modelled by perturbation theory, necessitating nonlinear models such as $N$-body simulations. However, high-resolution $N$-body simulations are computationally expensive, often requiring thousands of node hours for a single simulation.


Emulation replaces intensive numerical computation for every likelihood evaluation by the evaluation of a cheap pre-trained surrogate model. For instance, emulators have been widely used to replace the Boltzmann codes in cosmological inference~\cite{Auld2007,Auld2008,Arico2021a,Spurio2022,Nygaard2023,Gunther2022,Bonici2024,Bonici2025}. In our case, an emulator builds a model from a number of simulations at a set of cosmologies sampled in the parameter space, which can interpolate to predict summary statistics for arbitrary cosmological parameters. There have been several such cosmological emulators, e.g., {\tt FrankenEmu}~\cite{Heitmann2009,Heitmann2010,Heitmann2013}, the emulators of the Aemulus project~\cite{DeRose2019,McClintock2019,Zhai2019}, {\tt NGenHalofit}~\cite{Smith2019}, the {\tt dark quest} emulator~\cite{Nishimichi2019}, {\tt BE-HaPPY}~\cite{Valcin2019}, the baryonification emulator of the BACCO project~\cite{Arico2021}, the emulators built on the Q{\small UIJOTE} simulations~\cite{Villaescusa2020}, the emulators based on the \textit{Mira-Titan Universe} suite~\cite{Heitmann2016,Lawrence2017,Bocquet2020,Moran2023,Kwan2023}, the {\scriptsize E-MANTIS} emulator~\cite{Casares2024}, {\tt EuclidEmulator}~\cite{Knabenhans2019,Knabenhans2021} and the \texttt{CSST Emulator}~\cite{Chen2025}.
All these emulators can perform as surrogate models, and are able to predict summary statistics within their parameter space at much less computational cost than full simulations.

Despite the success of the standard model of cosmology, several fundamental physics puzzles are still unsolved, e.g., the accelerated expansion of the Universe~\cite{Caldwell2009}, the nature of dark matter (DM)~\cite{Feng2010}, and the sum of the neutrino masses~\cite{Wong2011}. There are several well-motivated $\Lambda$CDM extensions proposed for these questions, which will be tested by upcoming surveys. Some of these extensions have been included in existing emulators.
For instance, {\tt EuclidEmulator2}~\cite{Knabenhans2021}, an emulator for the matter power spectrum, taking into account dynamical dark energy (DE) and massive neutrinos, includes an 8-parameter cosmological model (the 5 base cosmological parameters together with the DE parameters, $w_0$ and $w_a$, and the sum of the neutrino masses, $\sum m_\nu$).
None of the current production emulators cover the full space of currently popular $\Lambda$CDM extensions (e.g., the effective number of neutrinos in the early Universe, $N_\mathrm{eff}$). Current data shows noticeable tensions in $\Lambda$CDM (i.e., the $H_0$ tension~\cite{Riess2021,Riess2022} and the $S_8$ tension~\cite{Asgari2021,Abbott2022}). Resolving these tensions is likely to require larger models, which requires developing new emulators with increased parameter coverage.
 
However, it is computationally expensive to expand the parameter space of an emulator as the number of simulations required to fill the parameter hyperspace grows exponentially~\cite{Ho2023,Ji2021}.
This relation is implied in the empirical formula for the interpolation error within a space-filling design~~\cite{Ji2021,Wendland2004}, $\Phi \propto n^{-\nu/d}$ (simplified here), where $\Phi$ is the estimated error, $n$ the number of simulations,
$d$ the number of dimensions of the parameter space, and $\nu$ a parameter that depends on other conditions of the emulator.

To alleviate this difficulty, Ref.~\cite{Ho2022} developed the multi-fidelity emulation technique for the matter power spectrum, \texttt{MFEmulator}, to reduce the computational cost of building a simulation-based emulator.
\texttt{MFEmulator} reduces the number of high-resolution simulations required by a factor of $\sim 4$, combining simulations with different particle loads in the training set for Gaussian process (GP) regression. This multi-fidelity technique has been used in Ref.~\cite{Bird2023,Fernandez2023} for building emulators for cosmological hydrodynamical simulations, demonstrating its effectiveness.

While \texttt{MFEmulator} has only one fidelity variable, particle load, the extended version of the framework, \texttt{MF-Box}~\cite{Ho2023}, introduces a second fidelity variable, box size, and allows multiple low-fidelity (LF) nodes\footnote{Multiple nodes at the same fidelity contribute to the overall emulation, as long as they form a directed acyclic graph.} to be combined with a high-fidelity (HF) node in training. Specifically, the training set of Ref.~\cite{Ho2023} consists of one HF node and two LF nodes, L1 and L2. The HF node comprises simulations with target box size and resolution (thus particle load), while the LF simulations have a reduced particle load, and L1 (L2) runs have the same box size (resolution) as the HF runs. 
A relatively large number of LF simulations explore the parameter space, while the sparsely sampled HF simulations are used to correct the LF simulations.
\texttt{MF-Box} has been shown to reduce the cost of training an emulator further compared to \texttt{MFEmulator}.

Even with the \texttt{MF-Box} framework, it remains challenging to build an emulator that spans the full range of current cosmological constraints, given the wide parameter ranges that need to be covered. For example, the constraints on dark energy parameters $w_0$ and $w_a$ are still relatively loose (e.g., Ref.~\cite{Adame2024}), and the uncertainty in the effective number of neutrinos, $N_\text{eff}$, remains significant from cosmological surveys (e.g., Ref.~\cite{Aghanim2020}). To balance the trade-off between parameter coverage and emulation accuracy, we employ an adaptive sampling strategy that uses two Latin hypercubes: one covering a large parameter space and another focused on a smaller, more probable region. This inner region encompasses the tighter constraints from Ref.~\cite{Tristram2024} and the combined analyses of Refs.~\cite{Abbott2022,Brout2022}. By concentrating simulation effort in this inner region, we can achieve 1\% accuracy in predicting the matter power spectrum, while maintaining 5\% accuracy in the larger, outer parameter space. This approach allows us to efficiently allocate computational resources, providing high precision where it is most needed while ensuring sufficient coverage across the broader parameter space.\footnote{Similar strategies have been used in emulator construction in some studies. For instance, Ref~\cite{2023MNRAS.518.4818B} iteratively selects training samples in regions of high posterior probability.} 

In this paper, we will present the Suite Of simulatioNs for Gravitational clustering Over $k$-modes in an Ultra-high dimensional parameter space (\texttt{Son Goku}\footnote{A fictional character and the main protagonist of the Dragon Ball manga series created by Akira Toriyama (1955--2024).}, hereafter \texttt{Goku}), that we have designed and performed in the \texttt{MF-Box} framework. The simulation suite is divided into two subsets based on the two Latin hypercubes: \texttt{Goku-W}, which samples a wide parameter range, and \texttt{Goku-N}, which focuses on a narrower, more probable region. And we will introduce the emulator constructed on this simulation suite for the matter power spectrum, \texttt{GokuEmu}.
Our model is the first to include 10 parameters, handling these currently popular $\Lambda$CDM extensions: dynamical DE parameters $w_0$ and $w_a$~\cite{Chevallier2001,Linder2003}, the sum of the neutrino masses $\sum m_\nu$, the effective number of neutrinos $N_\text{eff}$ (for a review of the effects of neutrinos in the context of cosmology, see Refs.~\cite{Lesgourgues2006,Lesgourgues2013}), and the running of the scalar spectral index $\alpha_\text{s}$ ($= \text{d}n_\text{s}/\text{d}\ln k$)~\cite{Aghanim2020,Ade2014,Palanque2015}, which quantifies the scale dependence of the primordial power spectrum slope. Including $\alpha_\text{s}$ provides a crucial means of testing inflation models, as it directly impacts the matter power spectrum on small scales.

Prior to sampling and running our production simulations, we run a suite of small simulations containing both training and test sets, and build exploratory emulators for error estimation. This process enables us to optimize the number of simulations in the LF and HF nodes, and thus minimize the computational budget for achieving our target generalization accuracy. Besides the emulation technique, our simulations improve the theory model by using a more accurate treatment of massive neutrinos than some existing emulators (e.g., \texttt{EuclidEmulator2}), following the technique of Refs.~\cite{Ali2013,Bird2018}, which includes the response of the neutrino component to the nonlinear cold dark matter (CDM) growth.
Our emulator will thus predict the matter power spectrum more accurately, since this effect increases the matter power spectrum by a few percent near $k = 1h\,\text{Mpc}^{-1}$ compared to using a linear theory prediction for the neutrinos (e.g., $\sim 4\%$ at $k=1h\,\text{Mpc}^{-1}$ and $z=0$ for $\sum m_\nu = 0.3\,\text{eV}$)~\cite{Bird2012}.

We organize this paper as follows. Section~\ref{sec:methods} outlines the overall workflow of this study and provides detailed descriptions of the methods employed, including the Sliced Latin Hypercube Design (SLHD) for sampling (Section~\ref{sec:ranges}), the simulation codes MP-Gadget~\cite{Feng2018} and CLASS~\cite{Lesgourgues2011} (Section~\ref{sec:sim_codes}), and the \texttt{MF-Box} emulation framework~\cite{Ho2023} (Section~\ref{sec:MF-Box}). In Section~\ref{sec:results}, we present the results, starting with an overview of the production simulations (Section~\ref{sec:sims}), followed by the matter power spectra (Section~\ref{sec:matter_power}), the application and impact of the pairing-and-fixing (P+F) technique (Section~\ref{sec:pplusf}), performance evaluations of the emulator (Section~\ref{sec:emulator}), and a preliminary exploration of the potential of \texttt{Goku} for emulating the halo mass function (HMF) (Section~\ref{sec:HMF}). Finally, we conclude in Section~\ref{sec:concl} and discuss directions for future improvements in Section~\ref{sec:future}.

\section{\label{sec:methods}Methods}

\begin{figure*}[t]
    \includegraphics[width=\textwidth,trim=75 410 75 95,clip]{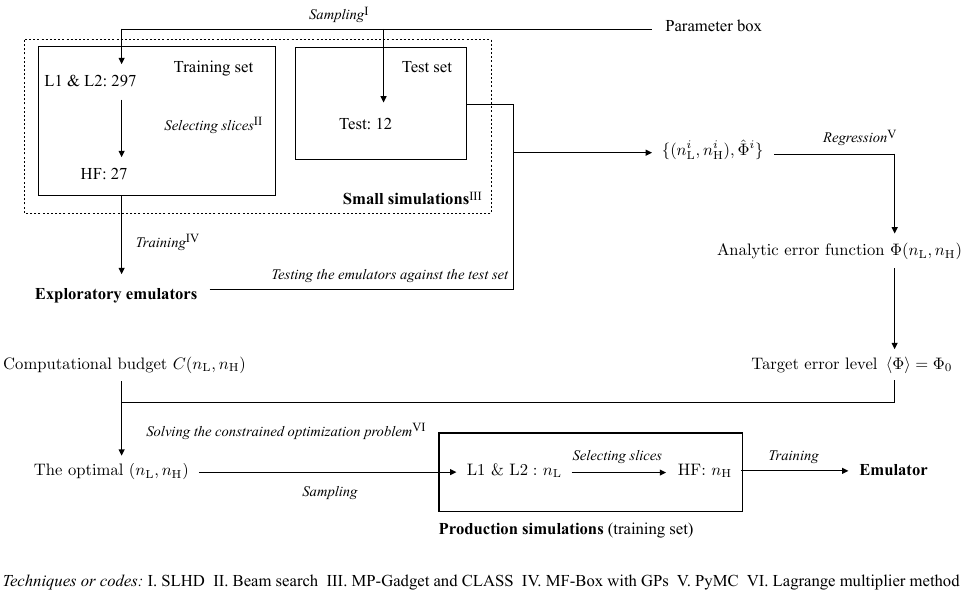}
    \caption{\label{fig:workflow}Illustration of the pipeline of this work. The footnotes are techniques or codes we use in corresponding steps. The set \{$(n_\text{L}^i, n_\text{H}^i)$, $\hat{\Phi}^i$\} represents the various pairings of LF and HF simulation counts used to train the exploratory emulators, along with the associated prediction errors observed in the test set.}
\end{figure*}

Fig.~\ref{fig:workflow} illustrates our pipeline for generating each of \texttt{Goku-W} and \texttt{Goku-N}, presenting an overview of our procedures for obtaining the simulations and emulator. Detailed descriptions of these steps/techniques will be found in dedicated sections.

As shown in Fig.~\ref{fig:workflow}, to develop a preliminary understanding of the emulation error for \texttt{Goku-W} (\texttt{-N}), we first run a series of smaller simulations, referred to as \texttt{Goku-pre-W} (\texttt{-N}), which include both training and test sets. The training set consists of 297 pairs of low-fidelity (LF) simulations (L1 and L2) and 27 high-fidelity (HF) simulations, while the test set includes 12 independently sampled HF simulations. Using the training set, we build a series of exploratory \texttt{MF-Box} emulators with varying allocations of simulations $(n_\text{L}, n_\text{H})$, where $n_\text{L}$ represents the number of LF simulation pairs\footnote{In this work, we assume equal numbers of L1 and L2 simulations, though other configurations could be explored.} and $n_\text{H}$ is the number of HF simulations.
We then derive the error function $\Phi(n_\text{L}, n_\text{H})$, which describes how the emulator’s prediction error depends on the number of LF and HF simulations, using a Markov chain Monte Carlo (MCMC) approach. To find the optimal combination $(n_\text{L}, n_\text{H})$ that minimizes computational cost while achieving 5\% (or 1\%) accuracy, we empirically estimate the computational budget function $C(n_\text{L}, n_\text{H})$, which captures how resource requirements change with $n_\text{L}$ and $n_\text{H}$, and solve the constrained optimization problem using the Lagrange multiplier method.
This optimization assumes that the error function of emulation based on the production simulations can be approximated by that of the exploratory emulators using small simulations.\footnote{This is a conservative assumption, as the exploratory emulators cover the nonlinear $k$ scales that dominate prediction errors, meaning the overall error—averaged over $k$-modes—could be overestimated.} Finally, we perform $n_\text{L}$ LF simulation pairs and $n_\text{H}$ HF simulations, forming the final simulation subset, \texttt{Goku-W} (\texttt{-N}), and construct the corresponding emulators, \texttt{GokuEmu-W} (\texttt{-N}). We also build an emulator based on the entire suite, \texttt{GokuEmu}, and compare the performance of these emulators in Section~\ref{sec:results}.

The ranges of the cosmological parameters of \texttt{Goku-W} are chosen to be sufficiently large that the $5\sigma$ contours of most currently popular constraints are covered, while the definition of the parameter box of \texttt{Goku-N} refers to some tighter constraints. Within the parameter space, the cosmologies (points) for simulations are sampled using SLHD.
We select the optimal slices for the HF training node from the LF cosmologies in each simulation suite using a heuristic search algorithm, beam search.\footnote{Beam search is common in the field of natural language processing (it was first used by Ref.~\cite{Lowerre1976}) but, to our knowledge, rarely used in astronomy or cosmology.}

For each simulation's initial conditions, we use fixed amplitudes for the initial density fields, so that the power spectrum of each mode aligns with the linear theory better. 
To further mitigate cosmic variance on large scales, we perform an additional paired HF simulation at one of the HF cosmologies and compute the paired-and-fixed matter power spectrum. This approach provides a correction factor can be applied to the emulator-predicted matter power spectrum. To evaluate the robustness of this correction, we also conduct several P+F LF simulations to assess potential cosmology-dependent effects. The results of these tests are presented in Section~\ref{sec:pplusf}, and a brief comparison of the P+F-corrected matter power spectrum with that derived from a larger simulation box (double the side length) is provided in Appendix~\ref{app:pplusf_and_largerbox}.

\subsection{\label{sec:specific}Simulation Specifications}

Following the \texttt{MF-Box} framework, \texttt{Goku-W} (or \texttt{-N}) consists of one HF node and two LF nodes, where HF and LF simulations differ in particle loads and/or box sizes. Drawing from \texttt{EuclidEmulator2}~\cite{Knabenhans2021}, our HF simulations evolve $3000^3$ particles in volumes of $1\,{(\text{Gpc}/h)}^3$. For the L1 and L2 simulations, we use a lower particle load of $750^3$. Ref.~\cite{Knabenhans2021} conducted comprehensive convergence tests for the input simulations, considering both volume and resolution effects. They found that a simulation box of $L=1024\,\text{Mpc}\,h^{-1}$ is converged to within $\sim 1\%$. Regarding resolution, they chose a resolution parameter of $\ell^{-1}=3h^{-1}\,\text{Mpc}$, and found that the matter power spectrum is well converged across most $k$-modes, though there are minor systematic deviations (a few percent) at the smallest scales ($k\sim 10h\,\text{Mpc}^{-1}$). This effect was mitigated by applying a $k$- and redshift-dependent, but cosmology-independent, resolution correction factor derived from a single reference cosmology. Users seeking slightly improved accuracy on the smallest scales may apply the correction function from Ref.~\cite{Knabenhans2021}.

We perform a simulation with $L=2\,\text{Gpc}/h$ (denoted as Lgr) that maintains the same resolution parameter as an L1 simulation. This larger box allows us to explore the volume effects on the matter power spectrum and provides a comparison to the pairing-and-fixing technique~\cite{Angulo2016} for mitigating cosmic variance.

The preliminary suite, \texttt{Goku-pre}, serves as a scaled-down version of \texttt{Goku}, with smaller simulation boxes and lower particle loads ($300^3$ for HF and $75^3$ for LF nodes). Table~\ref{tab:sim_specs} summarizes the simulation specifications and estimated computational costs for both \texttt{Goku-pre} and \texttt{Goku}. The rationale for assigning different box sizes to the two LF nodes will be discussed in Section~\ref{sec:MF-Box}.

\begin{table}
    \caption{\label{tab:sim_specs}%
    Specifications and estimated computational costs of simulations in the small suite \texttt{Goku-pre} and the production suite \texttt{Goku}. Computational costs are quoted in Frontera node hours. One Frontera node is $56$ Intel Xeon CPUs.
    }
    \begin{ruledtabular}
    \begin{tabular}{lrrrr}
    \textrm{Simulation}&
    \textrm{Box size}&
    \textrm{Particle}&
    \multicolumn{2}{c}{\textrm{CPU time}}\\
    & ($\text{Mpc}/h$) & \multicolumn{1}{c}{\textrm{load}} & \multicolumn{2}{c}{(node hour)}\\
    \colrule
    \texttt{Goku-pre}: & & & \multicolumn{1}{c}{\texttt{W}} & \multicolumn{1}{c}{\texttt{N}}\\
    HF/Test & 100 & $300^3$& $ 3.19$ & $ 4.35$ \\
    L1 & 100 & $75^3$ & $ 0.0178$ & $ 0.0209$ \\
    L2 & 25 & $75^3$ & $ 0.0329$ & $ 0.0452$ \\
    
    \texttt{Goku}: & & & \multicolumn{1}{c}{\texttt{W}} & \multicolumn{1}{c}{\texttt{N}}\\
    HF & 1000 & $3000^3$& $ 4452$ & $5574$\\  
    L1 & 1000 & $750^3$ & $ 21.4$ & $ 22.0$\\
    L2 & 250 & $750^3$ & $ 44.9$ & $ 61.8$\\
    Lgr & 2000 & $1500^3$ & & $198$\\
    \end{tabular}
    \end{ruledtabular}
\end{table}

\subsection{\label{sec:ranges}Parameter Ranges and Sampling}

The parameter space for our simulations spans the base $\Lambda$CDM model and several widely studied extensions. We parametrize the cosmologies using 10 parameters, such that the emulator inputs are given by $\bm{\theta} = (\Omega_\text{m}, \Omega_\text{b}, A_\text{s}, n_\text{s}, h, \sum m_\nu, w_0, w_a, N_\text{eff}, \alpha_\text{s})$. For each parameter, we have selected a sufficiently broad range to ensure that the 5$\sigma$ contours of most current cosmological constraints are encompassed by \texttt{Goku-W}, while \texttt{Goku-N} focuses on a more probable high-likelihood region. The parameter ranges and their respective bounds are summarized in Table~\ref{tab:ranges}.

\begin{table*}
    \caption{\label{tab:ranges}%
    Parameter boxes for \texttt{Goku-W} and \texttt{Goku-N} defined through their lower  and upper bounds. E.o.S. stands for equation of state.
    }
    \begin{ruledtabular}
    \begin{tabular}{lccccc}
    \textrm{Parameter}&
    \textrm{Definition/description}&
    \textrm{min(\texttt{W})}& \textrm{min(\texttt{N})}& \textrm{max(\texttt{N})}&
    \textrm{max(\texttt{W})}\\
    \colrule
    $\Omega_\text{m}$ & Total matter density parameter (DM and baryons) & $0.22$ &$0.26$ &$0.35$ & $0.40$\\
    $\Omega_\text{b}$ & Total baryon density parameter & $0.040$ &$0.045$&$0.051$ & $0.055$\\
    $h$ & Hubble parameter & $0.60$ &$0.64$& $0.74$ &$0.76$\\
    $A_\text{s}$ & Primordial perturbation amplitude & $1.0\times 10^{-9}$ 
    &$1.7\times 10^{-9}$& $2.5\times 10^{-9}$& $3.0\times 10^{-9}$\\
    $n_\text{s}$ & Primordial spectral index & $0.80$ & $0.95$ & $1.00$ &$1.10$\\
    $w_0$ & Parameter of the time-independent part of the DE E.o.S & $-1.30$ & $-1.30$ & $-0.70$ &$0.25$\\
    $w_a$ & Parameter of the time-dependent part of the DE E.o.S& $-3.0$& $-1.0$ & $0.5$ & $0.5$\\
    $\sum m_\nu$ & Sum of the neutrino masses & $0.00$ & $0.06\,\text{eV}$ & $0.15\,\text{eV}$ &$0.60\,\text{eV}$\\
    $N_\text{eff}$ & Effective number of neutrinos & $2.2$& $2.3$ & $3.7$ & $4.5$\\
    $\alpha_\text{s}$ & Running of the scalar spectral index & $-0.05$ &$-0.03$ &$0.03$ &$0.05$\\
    \end{tabular}
    \end{ruledtabular}
\end{table*}

We have referred to constraints from several recent works to determine the parameter range, e.g., Refs.~\cite{Aghanim2020,Akrami2020,Abbott2018,Ivanov2020,McCarthy2018,Freedman2024}. For \texttt{Goku-W}, we set a relatively large upper bound for the sum of the neutrino masses, $\sum m_\nu$, matching the weakest constraint in the literature, since some of the tightest constraints depend on a restrictive cosmological model. Similarly, the upper bound of the Hubble parameter, $h$, is set larger than the measurement of the Hubble constant by the SH0ES team~\cite{Riess2022} ($\sim 73\,\text{km}\,\text{s}^{-1}\,\text{Mpc}^{-1}$), to allow for the possibility of a larger $h$. For dynamical DE, we cover the recent constraints from baryon acoustic oscillations (BAO) measurements, Ref.~\cite{Adame2024}, while imposing $w_0 + w_a < -0.2$ to avoid highly exotic cosmologies where DE would behave almost like matter shortly after the Big Bang~\cite{Knabenhans2021}. For \texttt{Goku-N}, we choose a narrower range for the parameters, focusing on a more probable region that aligns with the tighter constraints (especially from combined analyses of multiple datasets) from Refs.~\cite{Tristram2024,Abbott2022,Brout2022}.

\begin{figure*}[t]
    \includegraphics[width=\textwidth]{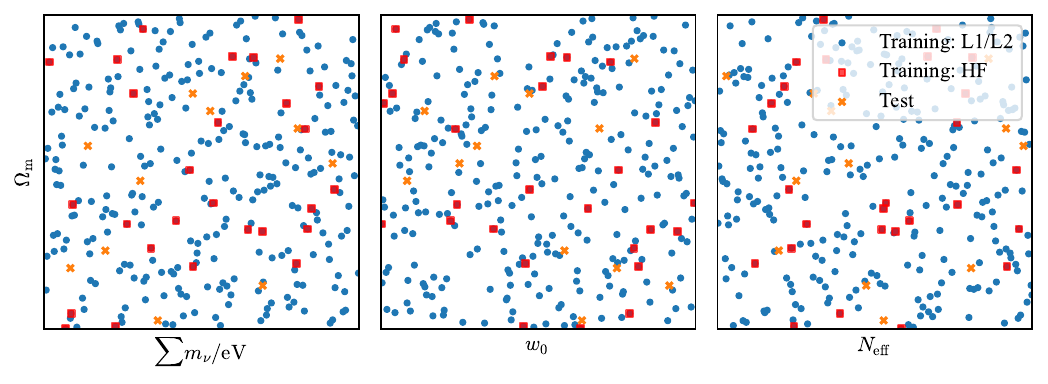}
    \caption{\label{fig:SLHD}Distribution of cosmologies sampled via SLHD for \texttt{Goku-pre-W} emulators in selected parameter planes. 
    Blue dots denote LF training set design points, while red dots represent HF points selected from LF cosmologies (the selection method will be detailed in Section~\ref{sec:optimization_HF}).
    Orange crosses indicate the test set. The parameter planes shown are $\Omega_\text{m}$ versus $\sum m_\nu$, $w_0$, and $N_\text{eff}$. The bounds correspond exactly to those listed in Table~\ref{tab:ranges}. The design points are evenly distributed in the parameter space.}
\end{figure*}

Following the sampling strategy of Ref.~\cite{Ho2023}, we use SLHD to sample the cosmologies for simulations in each parameter box defined in Table~\ref{tab:ranges}.  SLHD is a deterministic sampling technique that is designed to cover the parameter space uniformly and efficiently, which has already been employed in cosmological emulation studies, e.g., Ref.~\cite{Nishimichi2019}. Sliced Latin hypercube is a type of Latin hypercube that can be partitioned by slices or blocks, each of which contains an equal number of design points. Each slice is itself a Latin hypercube. SLHD ensures the space-filling property both in the whole design and in each slice (or any combination of several slices), which means that a few slices selected for the HF node will also be a good Latin hypercube. Therefore, SLHD is an intuitive choice for a multi-fidelity problem. 

For implementation, we use the maximin SLHD package, \texttt{maximinSLHD}\footnote{\url{https://rdrr.io/cran/SLHD/man/maximinSLHD.html}} in \texttt{R}~\cite{Ba2015}. 
For the \texttt{Goku-pre-W} (or \texttt{-N}) suite, we sample 297 design points with 3 points in each slice (thus 99 slices\footnote{In practice, we initially sample more than 99 slices before excluding unrealistic cosmologies with $w_0 + w_a > -0.2$.}) for the LF runs in the training set, while the test set consists of 12 points with 3 in each slice.
The slices for the HF simulations are selected from the LF slices (the selection will be detailed in Section~\ref{sec:optimization_HF}), i.e., the HF sample is a subset of the LF sample.\footnote{A nested design has been shown to be efficient for a multi-fidelity model~\cite{Kennedy2000}.}
We project the SLHD points onto the 2D planes of the \texttt{Goku-pre-W} parameter space in selected dimensions ($\Omega_\text{m}$ against $\sum m_\nu$, $w_0$ and $N_\text{eff}$) in Fig.~\ref{fig:SLHD}. 
The design points are evenly distributed in the whole parameter space. The design of the test set is independent of the training set, and this ensures we will not test on the training simulations.

\subsection{\label{sec:sim_codes}Simulation Codes} 

In this section, we present a brief introduction to the codes for realizing the simulations and how the $\Lambda$CDM extensions are implemented.
We perform DM-only simulations using the open source code MP-Gadget~\cite{Feng2018}, 
an $N$-body and smoothed particle hydrodynamics (SPH) simulation code derived from Gadget-3~\cite{Springel2003}.
It has been used to run the \texttt{Astrid} simulation~\cite{Bird2022,Ni2022,Ni2024}, a large-scale high-resolution cosmological hydrodynamic simulation with a $250\,\text{Mpc}\,h^{-1}$ (periodic) box containing $2\times5500^3$ particles. 
Despite being originally derived from Gadget-3, MP-Gadget has been rewritten to take advantage of shared-memory parallelism and to include the hierarchical timestepping algorithm from Gadget-4~\cite{Springel2021}. Ref.~\cite{Bird2022} describes the gravity solver in more detail.

MP-Gadget includes a fast and accurate model for the effect of massive neutrinos on cosmological structure, 
following Ref.~\cite{Ali2013} (Ref.~\cite{Elbers2021} proposed a similar method). This model includes the dynamic effect of neutrinos in each time step. No explicit neutrino particles are included, but the neutrino perturbation is computed on the particle-mesh (PM) grid, using a linear response formula, from the nonlinear growth of the DM. 
Short-range neutrino forces are neglected, a good approximation as neutrinos are free-streaming on scales smaller than a PM cell. Our treatment of massive neutrinos improves the accuracy of the matter power spectrum prediction on non-linear scales compared to
using a linear theory prediction for the neutrinos~\cite{Bird2012}. For example, the matter power spectrum from a simulation without this improved treatment will be $\sim 2\%$ smaller at $k=1h\,\text{Mpc}^{-1}$ at $z = 0$ for $\sum m_\nu = 0.15\,\text{eV}$. As for the neutrino hierarchy of mass eigenstates, we consider only the normal hierarchy in this work, i.e., $m_1<m_2<m_3$ is assumed when the existence of three different mass eigenstates is allowed by $\sum m_\nu$.

For the time-dependent DE, we adopt the widely used the Chevallier--Polarski--Linder (CPL) parametrization~\cite{Chevallier2001,Linder2003},
\begin{equation}
    w(a) = w_0 + w_a(1-a),
\end{equation}
which describes how the DE equation-of-state parameter $ w $ evolves with the cosmic scale factor $ a $. To account for the effects of DE perturbations on the matter power spectrum (relevant primarily on very large scales), we employ the parametrized post-Friedmann (PPF) framework~\cite{Hu2007} for `phantom-crossing' cosmologies with $w(a)$ that evolves across $w=-1$ over time and the fluid description for `non-phantom-crossing' cosmologies within the context of an effective theory. DE perturbations are included in the Boltzmann solver and thus the simulation initial conditions, but the simulation treats the DE as a homogeneous fluid. For a more comprehensive discussion of dynamical DE, we refer the reader to Refs.~\cite{Dakin2019,Fang2008}.

We enforce that the Universe is spatially flat and determine the density parameter of DE by
\begin{equation}
    \Omega_\text{DE} + \Omega_\text{m} + \Omega_\text{rad} = 1,
\end{equation}
where $\Omega_\text{DE}$, $\Omega_\text{m}$ and $\Omega_\text{rad}$ are the density parameters (present-day energy density fractions) of DE, matter and radiation, respectively. $\Omega_\text{m}$ is the sum of the density parameters of CDM, baryons and massive neutrinos, i.e., $\Omega_\text{m} = \Omega_\text{CDM} + \Omega_\text{b} + \Omega_\nu$.

The total radiation density parameter is the sum of two components:
\begin{equation}
    \Omega_\text{rad} = \Omega_\gamma + \Omega_\text{ur},
\end{equation}
where $\Omega_\gamma$ and $\Omega_\text{ur}$ are the density parameters of photons and (extra) ultra-relativistic neutrinos, respectively. $\Omega_\gamma$ is determined by the temperature of the Cosmic Microwave Background (CMB) which we fix at $T_\text{CMB} = 2.7255\,\text{K}$ (with a given $h$), while $\Omega_\text{ur}$ accounts for the contribution of extra massless neutrinos (beyond the `massive'\footnote{In the early Universe, these neutrinos are also relativistic and `massless'.} neutrinos mentioned before) to radiation. The effective number of neutrinos~\cite{Lesgourgues2006,Shvartsman1969,Steigman1977}, $N_\text{eff}$, is defined by
\begin{equation}
    \rho_\nu = \frac{7}{8}\left(\frac{4}{11}\right)^{4/3}N_\text{eff}\rho_\gamma,
\end{equation}
where $\rho_\nu$ and $\rho_\gamma$ are the energy densities of neutrinos and photons when neutrino decoupling is complete and all neutrinos are relativistic. In practice, we vary $N_\text{eff}$ in the parameter space by adjusting the value of $N_\text{eff}^\text{ur}$ (thus $\Omega_\text{ur}$) given that the total effective number $N_\text{eff} = N_\text{eff}^{\text{`massive'}} + N_\text{eff}^\text{ur}$. For instance, when $N_\text{eff} = 3.044$ with three massive neutrino species ($N_\text{eff}^{\text{`massive'}} = 3.0396$), we are supposed to set $N_\text{eff}^\text{ur} = 0.0044$. Detailed implementation of $N_\text{eff}$ can be found in the CLASS\footnote{http://class-code.net} code.

We set periodic boundary conditions for the simulation boxes, and perform  the simulations from $z = 99$ to $z = 0$. The initial conditions for each simulation are generated using the transfer function and the matter power spectrum produced by CLASS~\cite{Lesgourgues2011}\footnote{In practice, we use the Python binding of CLASS, \texttt{classylss} (\url{https://github.com/nickhand/classylss}), to produce the initial linear power spectrum and transfer function.}. We explicitly confirmed that CLASS and CAMB~\cite{Lewis2000} produce initial matter power spectra for our simulations differing by less than $1\%$. All initial conditions use the same random number seed. The pre-displacement distribution of dark matter particles is a grid, and first-order Lagrangian perturbation theory~\cite{Zel1970,Crocce2006} is used to initialize particle positions and velocities. The effects of the running of the spectral index, $\alpha_s$, on the initial conditions, as well as other primordial power spectrum parameters, are accounted for by CLASS. We use the built-in power spectrum estimator of MP-Gadget to compute the matter power spectrum from the particle distribution at each PM step.

We have made our Python code for running cosmological simulations, \texttt{SimulationRunnerMPG}\footnote{\url{https://github.com/astro-YYH/SimulationRunnerMPG}}, publicly available. \texttt{SimulationRunnerMPG} can generate linear matter power spectra, transfer functions and compute cluster job submission scripts.


\subsection{\label{sec:MF-Box}MF-Box Emulation}
We briefly recap the \texttt{MF-Box} emulation framework here (we refer the reader to Ref.~\cite{Ho2023} for details not covered in this paper). We elaborate on the GP kernel and the procedures that we have improved since Ref.~\cite{Ho2023} in dedicated sections: Section~\ref{sec:gp_kernel} describes the GP kernel we use for emulation, Section~\ref{sec:optimization_HF} introduces the selection of HF cosmologies with the beam search algorithm adopted, and Section~\ref{sec:optimization_assign} details the optimization of the numbers of LF and HF simulations to minimize the computational budget.

\texttt{MF-Box} is a simulation-based emulation technique that utilizes a graphical multi-fidelity Gaussian process (GMGP) model~\cite{Ji2021}.
A GMGP model allows the fidelities to have a directed-in tree structure, rather than assuming the fidelities of simulations form a monotonically increasing sequence in accuracy. 
Thus each HF node can have more than one corresponding LF node. For our application to cosmological emulation, \texttt{MF-Box} uses the simplest case of the tree structure with one HF node and two LF nodes, L1 and L2.
During the training of an \texttt{MF-Box} emulator, information from both LF nodes is passed to the HF node, and then the target quantity can be approximated at the HF level. 
Ref.~\cite{Ho2023} implemented this approach using \texttt{GPy}\footnote{\url{https://sheffieldml.github.io/GPy/}}, a Python package for GP regression.\footnote{Ongoing efforts are underway to port \texttt{MF-Box} to \texttt{GPflow} (\url{https://www.gpflow.org}) to take advantage of GPU acceleration.}

In our case, the emulated quantity is the matter power spectrum, and the HF node comprises the training data from high-resolution simulations with large boxes while the LF nodes are from simulations with a lower particle load.
The L1 simulations match the HF box size, allowing them to capture large-scale structure formation at a significantly reduced computational cost. 
L2 simulations, on the other hand, use a smaller box size than the HF simulations while maintaining the same resolution parameter  $\ell^{-1}$
as the HF simulations.\footnote{This setup improves upon Ref.~\cite{Ho2023}, where L2 simulations have lower resolution parameters than HF simulations, limiting their ability to resolve small-scale structures.} We define the resolution parameter $\ell^{-1}$ as the number of particles per side divided by the side length of the box. For example, a simulation with a box size of $100\,\text{Mpc}/h$ and $200^3$ particles will have a resolution parameter $\ell^{-1} = 2h\,\text{Mpc}^{-1}$.

The high resolution enables L2 simulations to capture nonlinear structure growth on small scales almost as accurately as HF simulations (although cosmic variance is somewhat larger in smaller boxes), while each L2 simulation takes a computation time similar to an L1 simulation.
By combining many LF simulations (for interpolation) with a few HF simulations (for further correction) whose cosmologies are carefully selected from the LF points, we can build an emulator capable of predicting the matter power spectrum at an arbitrary cosmology within our defined parameter volume. The predicted spectrum is expected to closely approximate the spectrum that a corresponding HF simulation would produce.

\subsubsection{\label{sec:gp_kernel}Graphical Gaussian Process Kernel}

In this section, we define the kernel of the GMGP model we use in the \texttt{MF-Box} emulation technique.
The GMGP model is a deep Gaussian process (GP) model~\citep{Damianou:2013} with a graphical structure, where the fidelities of simulations form a tree structure~\citep{Ji2021,Ho2023}. We use a GMGP model with one HF node and two LF nodes, L1 and L2.
Our goal is to build the HF emulator, $f_\mathrm{HF}(\boldsymbol{\theta})$, that predicts the matter power spectrum at a cosmology $\boldsymbol{\theta}$ by combining the outputs from the LF nodes, $f_\mathrm{L1}(\boldsymbol{\theta})$ and $f_\mathrm{L2}(\boldsymbol{\theta})$,
\begin{equation}
    f_\mathrm{HF}(\boldsymbol{\theta}) = \rho(\{ f_\mathrm{t}(\boldsymbol{\theta}) : t \in \mathrm{L1}, \mathrm{L2} \}, \boldsymbol{\theta}) + \delta(\boldsymbol{\theta}),
    \label{eq:gmgp}
\end{equation}
where $\rho$ is a function that combines the outputs from L1 and L2, and $\delta$ is the bias function that corrects the prediction.

In Eq.~(\ref{eq:gmgp}), if we assume both function $\rho$ and function $\delta$ are Gaussian processes, the HF emulator can be represented as a deep GP model,
\begin{equation}
    f_\mathrm{HF}(\boldsymbol{\theta}) \sim \mathcal{GP}(0, K(\boldsymbol{\theta}, \boldsymbol{\theta}')),
    \label{eq:gmgp_gp}
\end{equation}
where $K(\boldsymbol{\theta}, \boldsymbol{\theta}')$ is the kernel function of the GP, and we assume the mean function is zero for simplicity.
The kernel function of Eq.~(\ref{eq:gmgp_gp}) is a composite kernel that combines the LF output power spectra and the HF input parameters, 
\begin{equation}
    \begin{split}
        K(\boldsymbol{\theta}, \boldsymbol{\theta}') = K_\rho &(\boldsymbol{\theta}, \boldsymbol{\theta}')
        \cdot
        K_f(f_\mathrm{*,LF}(\boldsymbol{\theta}), f_\mathrm{*,LF}(\boldsymbol{\theta}')) + \\
        K_\delta & (\boldsymbol{\theta}, \boldsymbol{\theta}'),
    \end{split}
    \label{eq:gmgp_kernel}
\end{equation}
where $f_\mathrm{*,LF}$ is the posterior predictions of the LF nodes.
Here, we make the same approximation as in Ref.~\cite{Perdikaris:2017}, so we can train the deep GP in Eq.~(\ref{eq:gmgp_gp}) recursively as three separate GPs:
We first train the low-fidelity emulators on L1 and L2, $f_\mathrm{L1}$ and $f_\mathrm{L2}$, respectively.
Both L1 and L2 emulators are GPs with squared exponential kernels.
Then, we sample the output posteriors from the L1 and L2 emulators, $f_\mathrm{*,L1}(\boldsymbol{\theta})$ and $f_\mathrm{*,L2}(\boldsymbol{\theta})$,
and use them as the training input for Eq.~(\ref{eq:gmgp_kernel}).

The bias parameter kernel, $K_\delta$, and the scale parameter kernel, $K_\rho$, are squared exponential kernels, also known as radial basis function (RBF) kernels,
\begin{equation}
    K_{\delta,\rho}(\boldsymbol{\theta}, \boldsymbol{\theta}') = \sigma_{\delta,\rho}^2 \exp\left[
        -\frac{||\boldsymbol{\theta} - \boldsymbol{\theta}'||^2}{2\,l_{\delta,\rho}^2}
    \right],
    \label{eq:delta_kernel}
\end{equation}
where $\sigma_{\delta,\rho}$ and $l_{\delta,\rho}$ are the hyperparameters of the kernel.
$\sigma_{\delta,\rho}$ controls the amplitude of the kernel, and $l_{\delta,\rho}$ determines the length scale of the kernel.
A longer $l_{\delta,\rho}$ means the GP prediction is smoother, while a shorter $l_{\delta,\rho}$ means the GP prediction is more wiggly.

The kernel function $K_f$ combines the posterior predictions from the L1 and L2 emulators, i.e.,
\begin{equation}
    \begin{split}
        K_f&(f_\mathrm{*,LF}(\boldsymbol{\theta}), f_\mathrm{*,LF}(\boldsymbol{\theta}')) = K_\mathrm{linear}(f_\mathrm{*,LF}(\boldsymbol{\theta}), f_\mathrm{*,LF}(\boldsymbol{\theta}'))
        +\\
        &K_\mathrm{rbf}(f_\mathrm{*,L1}(\boldsymbol{\theta}), f_\mathrm{*,L1}(\boldsymbol{\theta}')) \cdot
        K_\mathrm{rbf}(f_\mathrm{*,L2}(\boldsymbol{\theta}), f_\mathrm{*,L2}(\boldsymbol{\theta}')),
    \end{split}
    \label{eq:gmgp_kernel_decompose}
\end{equation}
where $K_\mathrm{rbf}$ is a radial basis kernel, and $K_\mathrm{linear}$ is a linear kernel, which can be expressed more explicitly as
\begin{equation}
    \begin{split}
    K_\mathrm{linear}(f_\mathrm{*,LF}(\boldsymbol{\theta}), f_\mathrm{*,LF}(\boldsymbol{\theta}'))
    = \sigma_1^2 & f_\mathrm{*,L1}(\boldsymbol{\theta}) f_\mathrm{*,L1}(\boldsymbol{\theta}')
    +\\ 
    \sigma_2^2 &f_\mathrm{*,L2}(\boldsymbol{\theta}) f_\mathrm{*,L2}(\boldsymbol{\theta}'),
    \end{split}
\end{equation}
where $\sigma_1$ and $\sigma_2$ are the hyperparameters of the linear kernel.
A linear kernel can be interpreted as a Bayesian linear regression.
One way to read the composite kernel is that the multiplication in the kernel means an `AND' operation, showing high covariance only if both kernels have high values.
The addition operator means an `OR' operation, indicating the final covariance is high if either of the kernels gives a high value.
The intuition here is that the linear kernel encodes the linear regression part while the multiplication of RBF kernels encodes the non-linear transformation from the L1 and L2 nodes to the HF node. 

\subsubsection{\label{sec:optimization_HF}Optimization of High-Fidelity Slices}

Recall that the cosmologies for simulations are sampled using SLHD, namely, the points are from a series of slices in a Latin hypercube. Each slice is also a Latin hypercube, ensuring they are space-filling. It is thus natural to select the HF points by choosing some of the slices.

While any slice should produce reasonably good emulation, some slices still perform somewhat better than others, confirmed by Ref.~\cite{Ho2023}.
In this work, we use beam search\footnote{It is not practical to thoroughly iterate over all possible combinations due to the huge number of combinations, e.g., if we had 90 slices and wanted to select 6 of them, the number of combinations would be $\binom{90}{6}\approx 6.23\times 10^{8}$. We thus turn to beam search which compromises between the efficiency of greedy search and the optimality of exhaustive search.}, a heuristic search algorithm, to select the optimal slices for the HF node. We assume the interpolation error of the LF node is correlated with that of the HF node.

We define the interpolation error as follows.
Suppose we have $N$ design points (cosmologies) in a set $Q = \{1,2,\dots, N\}$. For a given subset $S\subset Q$ with cardinality $|S|=N_\text{S}$, the interpolation error of the subset is defined as
\begin{equation}
    \phi_\text{S} = \frac{1}{N_\text{S}^\text{c}}\sum_{i\in S^\text{c}}\phi_{S,i},
\end{equation}  
where $S^\text{c}$ is the complement of $S$ in $Q$ (the remaining points) with $|S^\text{c}|=N_\text{s}^\text{c} = N-N_\text{S}$, and $\phi_{S,i}$ is the error of the emulator trained with $S$ for predicting the output (i.e., the matter power spectrum) of the $i$-th point (only $i$'s in $S^\text{c}$ are considered).
The prediction error tested on point $i$ is averaged over the $z$ bins we consider, i.e., 
\begin{equation}
    \phi_{S,i} = \frac{1}{n_z}\sum_{j=1}^{n_z}L_{S,i}(z_j),
\end{equation}
where $z_j$ is the $j$-th redshift bin, and $L_{S,i}(z_j)$ is the loss function of the emulator trained with $S$ when predicting the matter
power spectrum of the $i$-th point (cosmology) at $z_j$. In practice, we average over 6 distinct redshifts in $\{z_j\,|\,j=1, 2,\dots, n_z\} = \{0.0, 0.
2, 0.5, 2.0, 3.0\}$. We choose mean squared error (MSE) as the loss function,\footnote{The choice of the form of loss function in this context is arbitrary. Other choices instead of MSE should also be fairly good.} i.e., for a given point $i$ at a certain redshift,
\begin{equation}
    L_{S,i} = \frac{1}{n_k}\sum_{l=1}^{n_k}\left[y_{S,i}(k_l) - \hat{y}_{S,i}(k_l)\right]^2,
\end{equation}
where $y_{S,i}$ is the predicted output at cosmology $\bm{\theta}_i$, i.e., $y_{S,i} = f_S(\bm{\theta}_i)$, while $\hat{y}_{S,i}$ is the `true' output (from the corresponding simulation), and $n_k$ is the number of $k$ bins.\footnote{In practice, we emulate the matter power spectrum in log-log space, so $y=\log P$ and $k$ bins are equally spaced in $\log k$.}

We adapt the beam search algorithm to avoid redundant computation. Unlike in natural language processing (NLP), we do not care about the order of the slices (analogous to tokens in NLP) within each combination, and we do not allow repeated slices in a combination (two identical slices do not bring more information than one slice). Therefore, we skip evaluating slice combinations corresponding to these two cases.

\begin{figure*}
    \includegraphics[width=\textwidth,trim=30 265 220 1340,clip]{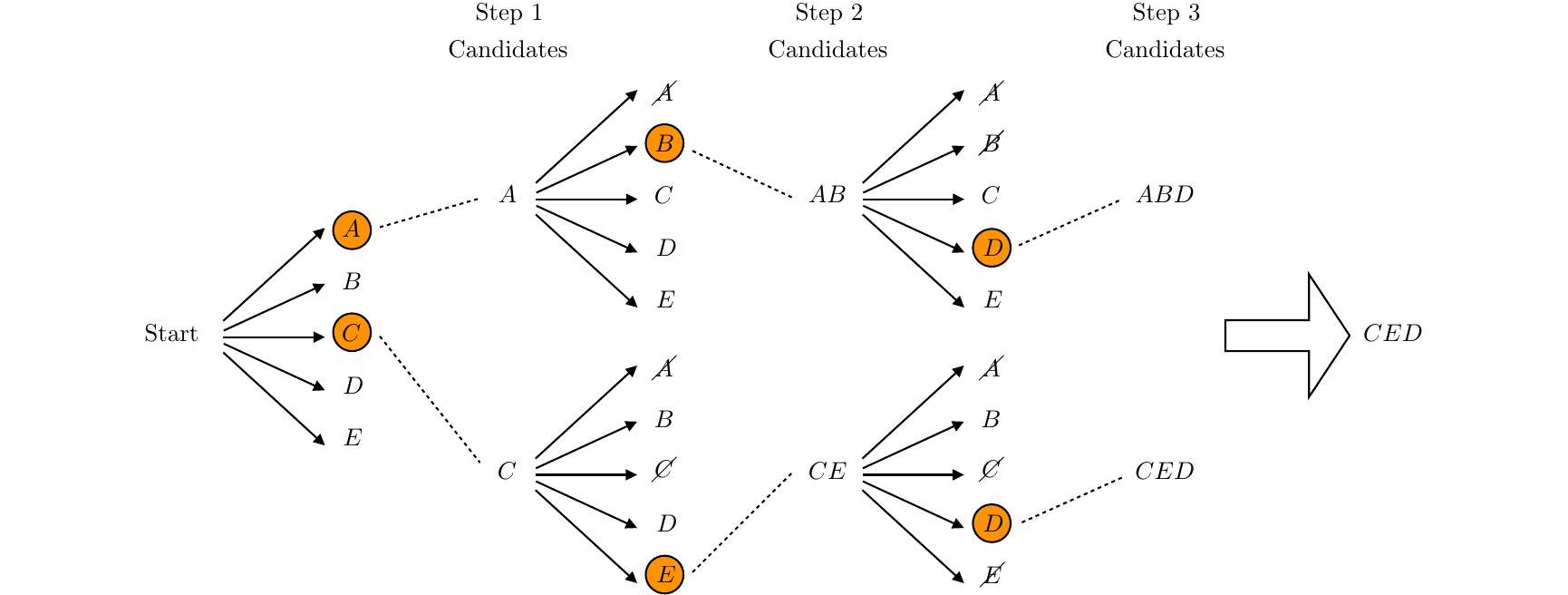}
    \caption{\label{fig:beam}Beam search adapted for selecting the optimal slices for the HF node, with $b=2$, $N_\text{tot}=5$ and $N_\text{HF}=3$. At step 1, we train an emulator based on each of the 5 slices and select the 2 best slices, $A$ and $C$. At step 2, for each candidate from the previous step, we loop over all other slices and combine each of them with the candidate for training (excluding repeated combinations), and then select the best combination. At step 3, we repeat the process and select the best new combination for each candidate. We then compare the 2 best combinations ($ABD$ and $CED$) at the last step and choose the one with the lowest interpolation error, $CED$, as the optimal combination.}
\end{figure*}

Beam search is characterized by a single hyperparameter, the beam size (or beam width), $b$. The beam size is the number of slice combinations (or slices) that are kept at each iteration.
Suppose we have $N_\text{tot}$ slices in total, and we want to choose $N_\text{HF}$ (maximum length of an output combination) slices for the HF node. At the first step, we select the $b$ best slices with the lowest interpolation errors (best tokens with the highest predicted probabilities in NLP). Each of them will be one of the slices in the $b$ candidate combinations. At each subsequent step, based on the $b$ candidate output combinations at the previous step, we continue to select a new set of $b$ candidate combinations with the lowest interpolation errors by adding one slice to each of the candidate combinations. We repeat this process until we have $N_\text{HF}$ slices in each candidate combination. Finally, we select the combination with the lowest interpolation error among all $b$ candidate combinations at the last step as the optimal combination. Within each step, redundant combinations are excluded. Fig.~\ref{fig:beam} shows an example with $b=2$, $N_\text{tot}=5$ and $N_\text{HF}=3$.

In this approach, the larger the beam size is, the more likely we are to find the optimal combination.\footnote{Note that the optimal combination is in fact pseudo-optimal since we do not really iterate over all possible combinations.} We therefore increase the beam size until the selected combination is well converged.

This algorithm is more effective than the previous method in Ref.~\cite{Ho2023}, which simply computes the interpolation error of the emulator for each slice and selects the best $N_\text{HF}$ slices (with the lowest losses). Considering that the combination of the best individual slices is not guaranteed to be the optimal combination (starting with the best slice), the previous method cannot result in a combination better than that from a beam search process with $b=1$. 

\subsubsection{\label{sec:optimization_assign}Optimization of Computational Budget}
Here we explain how we find the optimal numbers of LF and HF simulations for the production suite, \texttt{Goku-W} (\texttt{Goku-N}), that minimize computational cost,
by analyzing a series of exploratory emulators built on the small suite, \texttt{Goku-pre-W} (\texttt{Goku-pre-N}).
Ref.~\cite{Ho2023} includes detailed discussion on the optimization of the computational budget using the Lagrange multiplier method. In this work, we use a similar method to minimize the computational budget required to reach a specific generalization accuracy.

The goal is to find the optimal combination of the number of LF and HF simulations, $(n_\text{L}, n_\text{H})$, that minimizes the computational budget under the constraint that the prediction error of the emulator is $\sim5\%$ ($\sim1\%$). We use mean relative error as the measure of the prediction error, thus the condition is
\begin{equation}
    \left<{\Phi}(n_\text{L}, n_\text{H})\right> = \Phi_0,\label{eq:constraint}
\end{equation}
where the angle brackets denote the expectation and $\Phi_0$ is the target error level. For each exploratory emulator, the mean relative error is computed as
\begin{equation}
    \hat{\Phi}(n_\text{L}, n_\text{H}) = \frac{1}{N_\text{test}}\sum_{i=1}^{N_\text{test}}\left|\frac{y_i - \hat{y}_i}{\hat{y}_i}\right|,\label{eq:relative_error}
\end{equation}
where $\hat{y}_i$ is the true output of the $i$-th point in the test set, and $y_i$ is the predicted counterpart.\footnote{We have averaged over $k$ bins in calculation.} $N_\text{test}$ is the sample size of the test set, which is 12 in our case. We assume the error function takes the form
\begin{eqnarray}
    \Phi(n_\text{L}, n_\text{H}) & = &\eta \left[\frac{\rho}{(n_\text{L}+n_\text{H})(n_\text{L}+\alpha_\text{L})^{\beta_\text{L}-1}(n_\text{H}+\alpha_1)^{\beta_1-1}}\right.\nonumber \\
    & & +\left.\frac{1}{(n_\text{H}+\alpha_2)^{\beta_2}}\right]
    ,\label{eq:error_function}
\end{eqnarray}
which should provide a good approximation to the true error function for emulators when an MCMC approach is used to optimize the parameters $\eta$, $\rho$, $\alpha_\text{L}$, $\alpha_1$, $\alpha_2$, $\beta_\text{L}$, $\beta_1$ and $\beta_2$ with a fairly large sample. This analytic form is inspired by the empirical formulae from Refs.~\cite{Ji2021,Ho2023}, but is more sophisticated and has features that are more sensible for our purpose. In Appendix~\ref{app:error_function}, we explain why this form suits our problem better and provide a comparison between this choice and the form used in Ref.~\cite{Ho2023}.

Based on \texttt{Goku-pre-W} (\texttt{Goku-pre-N}), we vary the number of LF pairs and HF simulations in steps of 3 (the number of points per slice),
and compute the prediction error of the exploratory emulator for each combination $(n_\text{L}, n_\text{H})$. We select simulations at random and build three emulators for each $(n_\text{L}, n_\text{H})$, ultimately producing $99\times 9\times 3 = 2673$ samples\footnote{The three samples for $(n_\text{L},n_\text{H})=(297,27)$ are identical since there is only one unique combination of the simulations.} for the MCMC analysis.
We then fit the error function to the data using the \texttt{PyMC} package~\cite{Salvatier2016}, assuming $\hat{\Phi}$ follows a log-normal\footnote{This assumption is important in practice. If a normal distribution is assumed instead, the derived error function will be biased: the absolute fluctuations of the error at small $n_\text{L}$ and $n_\text{H}$ (thus large error values) are much larger than those at large $n_\text{L}$ and $n_\text{H}$, and the optimization will be biased towards the former and results in an error function that performs poorly in low-error regions (which are of our interest).} distribution, i.e., $\log\hat{\Phi}\sim\mathcal{N}(\mu, \sigma^2)$. Then the expected value of the error function is
\begin{equation}
    \left<\Phi(n_\text{L}, n_\text{H})\right> = \Phi(n_\text{L}, n_\text{H})\exp\left[\frac{1}{2} (\sigma \ln 10)^2\right]. 
    \label{eq:expected_error}
\end{equation}
The objective function to minimize is the computational budget, $C(n_\text{L}, n_\text{H})$, which is the sum of the node hours of the LF and HF simulations, i.e., 
\begin{equation}
    C(n_\text{L}, n_\text{H}) = n_\text{L}C_\text{L} + n_\text{H}C_\text{H},\label{eq:budget}
\end{equation}
where $C_\text{L}$ is the computational cost of a pair of LF simulations (i.e., $C_\text{L1} + C_\text{L2}$) and $C_\text{H}$ is the cost of an HF simulation (refer to Table~\ref{tab:sim_specs} for their values). Then the Lagrangian
\begin{equation}
    \mathcal{L}(n_\text{L}, n_\text{H}, \lambda) = C(n_\text{L}, n_\text{H}) + \lambda(\left<\Phi(n_\text{L}, n_\text{H})\right> - \Phi_0).\label{eq:lagrangian}
\end{equation}
Plugging Eqs.~(\ref{eq:error_function}), (\ref{eq:expected_error}) and~(\ref{eq:budget}) into Eq.~(\ref{eq:lagrangian}) will result in an explicit expression for the Lagrangian. We then numerically solve the problem using the \texttt{SciPy} package~\cite{Virtanen2020} to find the optimal $(n_\text{L}, n_\text{H})$, which essentially solves the following system of equations:
\begin{equation}
    \frac{\partial\mathcal{L}}{\partial n_\text{L}} = 0,\quad \frac{\partial\mathcal{L}}{\partial n_\text{H}} = 0,\quad \frac{\partial\mathcal{L}}{\partial \lambda} = 0.
\end{equation}

We thus sample $n_\text{L}$ points in the 10D parameter space for the LF nodes (L1 \& L2) and select $n_\text{H}$ cosmologies for the HF node of the production suite, and then perform the simulations.
Note that the solution $(n_\text{L}, n_\text{H})$ depends on the ratio of 
$C_\text{L}$ to $C_\text{H}$, rather than their absolute values. So this inference works under the assumption that the ratio $C_\text{L}/C_\text{H}$ is the same in the production suite as in the small suite, and they turn out to be similar indeed (see Table~\ref{tab:sim_specs}).

\section{\label{sec:results}Results}

We summarize the parameters of the error function obtained from our MCMC analysis (Section~\ref{sec:optimization_assign}) in Table~\ref{tab:error_params}, presenting both priors and posteriors. The MCMC analysis was conducted with 6 chains, each running for 6000 iterations, and achieved good convergence. Using these parameters along with the target accuracy and values of $C_\text{L}$ and $C_\text{H}$, we applied a numerical Lagrange multiplier method (Section~\ref{sec:optimization_assign}) to solve the optimization problem, yielding optimal configurations of $(n_\text{L}, n_\text{H}) = (583.93, 7.28)$ for \texttt{Goku-W} and  $(557.80, 17.58)$  for \texttt{Goku-N}.

In practice, we use the integer values  $(n_\text{L}, n_\text{H}) = (564, 21)$  for \texttt{Goku-W} and $(564, 15)$ for \texttt{Goku-N}, rounded to ensure multiples of 3. The number of HF simulations for \texttt{Goku-W} was set higher than the optimal value to enhance validation reliability by providing a larger sample for leave-one-out cross-validation (LOOCV). A detailed leave-one-out validation of the emulator will be provided in Section~\ref{sec:matter_power}.

\begin{table*}
    \caption{\label{tab:error_params}Parameters of the error function (Eq.~(\ref{eq:error_function})) obtained from our MCMC analysis, under the assumption $\log\hat{\Phi}(n_\text{L}, n_\text{H})\sim\mathcal{N}\left(\mu (n_\text{L}, n_\text{H}), \sigma^2\right)$ where $\mu (n_\text{L}, n_\text{H}) = \log \Phi (n_\text{L}, n_\text{H})$.}
    \begin{ruledtabular}
    \begin{tabular}{lcccc}
    \textrm{Parameter}& \multicolumn{2}{c}{\textrm{Prior}\footnote{Given the lack of strong prior knowldge, we conducted MCMC sampling with a sufficiently large number of iterations (Section~\ref{sec:optimization_assign}) and tested different prior distributions. We verified that the results are well converged and robust against variations in prior choices.}} &\multicolumn{2}{c}{\textrm{Median (16th, 84th percentiles)}}\\
    &\texttt{W}&\texttt{N}&\texttt{W}&\texttt{N}\\
    \colrule
    $\eta$ & Normal($\mu = 0.1, \sigma=0.1$) & ($\mu = 1, \sigma= 1$)& $0.0628^{+0.0188}_{-0.0112}$ & $0.0164^{+0.0006}_{-0.0006}$\\
    $\rho$ & LogNormal($\mu=2$, $\sigma=1$) & ($\mu = 1, \sigma= 1$) & $90.84^{+28.80}_{-26.89}$ & $840.52^{+200.67}_{-152.11}$\\
    $\beta_\text{L}$ & Normal($\mu=1$, $\sigma=1$) & ($\mu = 1, \sigma=0.3$)& $1.012^{+0.016}_{-0.014}$ & $1.464^{+0.036}_{-0.034}$ \\
    $\beta_1$ & Normal($\mu=1$, $\sigma=1$) & ($\mu = 3, \sigma= 1$)& $0.763^{+0.029}_{-0.030}$ & $1.134^{+0.039}_{-0.036}$\\
    $\beta_2$ & Normal($\mu=1$, $\sigma=1$) & ($\mu = 3, \sigma= 1$)& $0.378^{+0.109}_{-0.071}$ & $0.200^{+0.014}_{-0.013}$\\
    $\alpha_\text{L}$ & Normal($\mu=3$, $\sigma=1$) & ($\mu = 10, \sigma=3$)& $3.369^{+1.063}_{-1.054}$ & $20.642^{+2.207}_{-2.084}$\\
    $\alpha_1$ & Normal($\mu=3$, $\sigma=1$) & ($\mu = 10, \sigma=3$)& $3.337^{+0.944}_{-0.772}$ & $11.609^{+2.719}_{-2.680}$\\
    $\alpha_2$ & Normal($\mu=3$, $\sigma=1$) & ($\mu = 1, \sigma=3$)& $-1.558^{+0.931}_{-0.562}$ & $-2.641^{+0.095}_{-0.079}$\\
    $\sigma$ & HalfNormal($\sigma=0.1$) & ($\sigma=0.1$)& $0.0857^{+0.0011}_{-0.0011}$ & $0.0772^{+0.0011}_{-0.0011}$\\
    \end{tabular}
    \end{ruledtabular}
\end{table*}

\begin{figure}[t]
    \includegraphics[width=\linewidth]{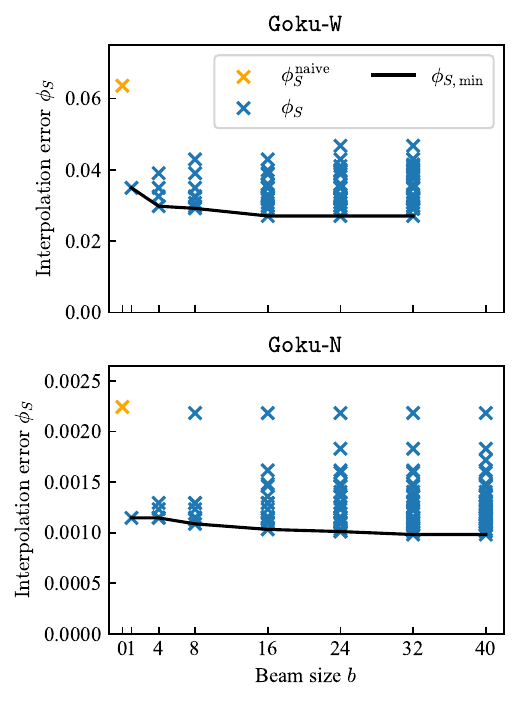}
    \caption{\label{fig:beam_size}\textit{Upper (Lower) panel:} Interpolation errors of emulators trained with 7 (5) slices of LF cosmologies in \texttt{Goku-W} (\texttt{Goku-N}) in the process of beam search. Blue crosses denote the interpolation errors for all candidate combinations at the final step of the beam search. The solid line shows the dependence of the minimum interpolation error on beam size. The minimum interpolation error converges at $b=16\ (32)$, indicating that the optimal combination of slices has been found. The orange cross marks the error of the naive selection approach, which is over $50\%$ higher than the converged error.}
\end{figure}

\begin{figure*}[t]
    \includegraphics[width=\textwidth]{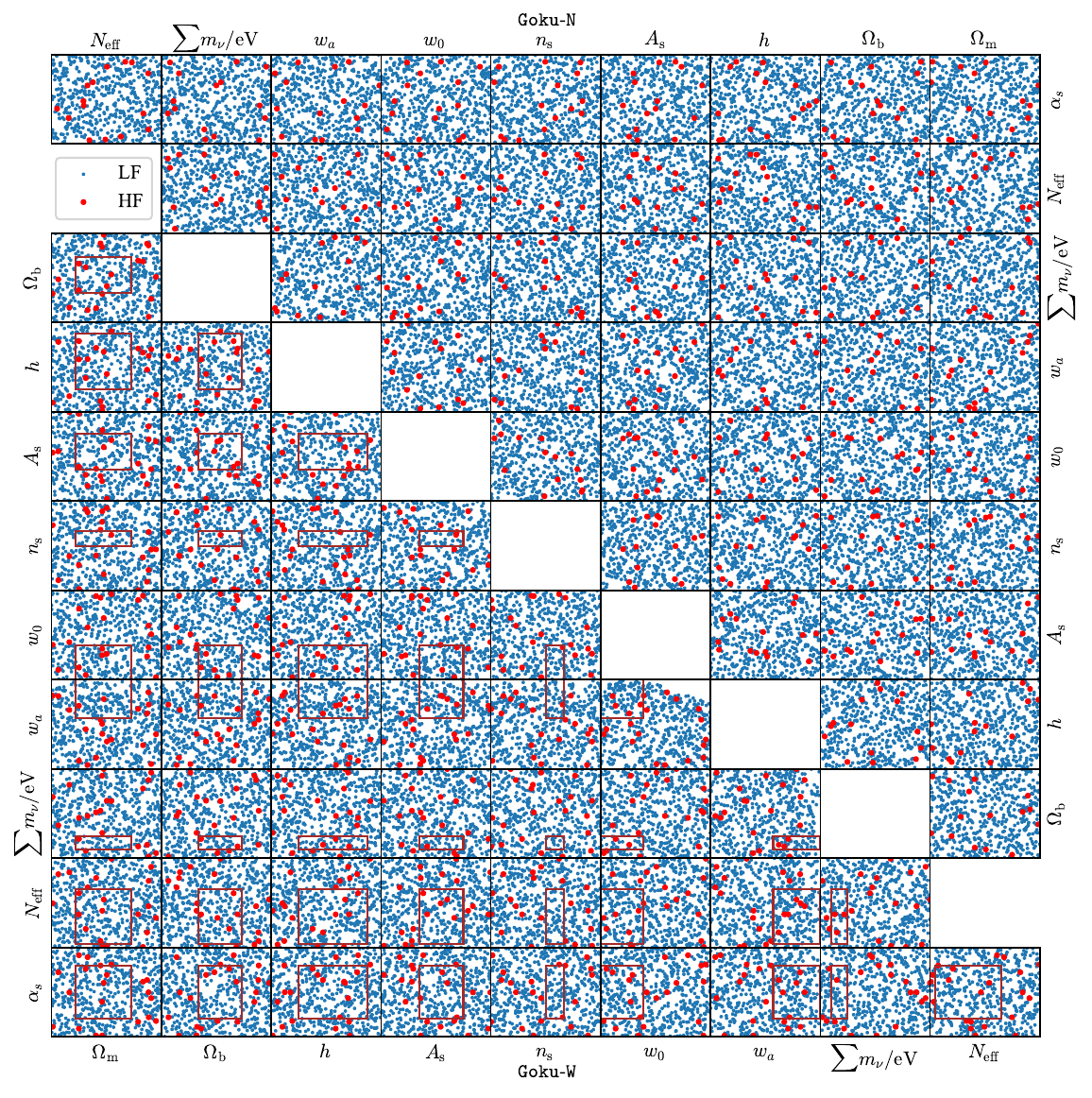}
    \caption{\label{fig:SLHD_pro}Cosmologies sampled via SLHD for \texttt{Goku-W} (lower left) and \texttt{Goku-N} (upper right) in the 10D parameter space. 
    As in Fig.~\ref{fig:SLHD}, blue dots indicate LF points, while larger red dots mark HF points selected from the LF cosmologies.
    Plot bounds for each parameter align with those in Table~\ref{tab:ranges}. The constraint $w_0 + w_a < -0.2$ is visible in the $w_0$--$w_a$ plane (lower). Brown rectangles denote the parameter ranges of \texttt{Goku-N} within the broader \texttt{Goku-W} space.
    }
\end{figure*}

We sample 564 cosmologies using SLHD in our parameter box with 3 points in each slice for the LF nodes (thus 188 slices in total), and then select the HF slices from the LF slices, using beam search (Section~\ref{sec:optimization_HF}).\footnote{Our beam search was performed solely on the L2 node, though both L1 and L2 could be used in principle.} As Fig.~\ref{fig:beam_size} shows, the minimum interpolation error from the beam search performed on \texttt{Goku-W} (\texttt{Goku-N}) is well converged at $b=16\ (32)$, indicating that we have found the optimal combination of slices for the HF node within the ability of the beam search algorithm. The converged error is $57.42\%$ ($56.21\%$) lower than the error obtained in the naive approach, implying that beam search has significantly improved the interpolation accuracy over the algorithm in Ref.~\cite{Ho2023}. Note that the $b=1$ case has already significantly outperformed the naive approach.

The distribution of the sampled cosmologies is plotted in Fig.~\ref{fig:SLHD_pro}, which shows that the cosmologies span the whole parameter volume roughly uniformly. Appendix~\ref{app:HF_cosmologies} shows the cosmologies of the HF runs.

In the following subsections, we present the \texttt{Goku} simulations, including some visualizations and basic information (Section~\ref{sec:sims}), the matter power spectra from the simulations (Section~\ref{sec:matter_power}), the impact of P+F correction (Section~\ref{sec:pplusf}), the performance of \texttt{GokuEmu} (Section~\ref{sec:emulator}) and the HMFs of selected simulations (Section~\ref{sec:HMF}).

\subsection{\label{sec:sims}Simulations}    

\begin{figure*}[t]
    \includegraphics[width=\textwidth,trim=80 1195 60 85,clip]{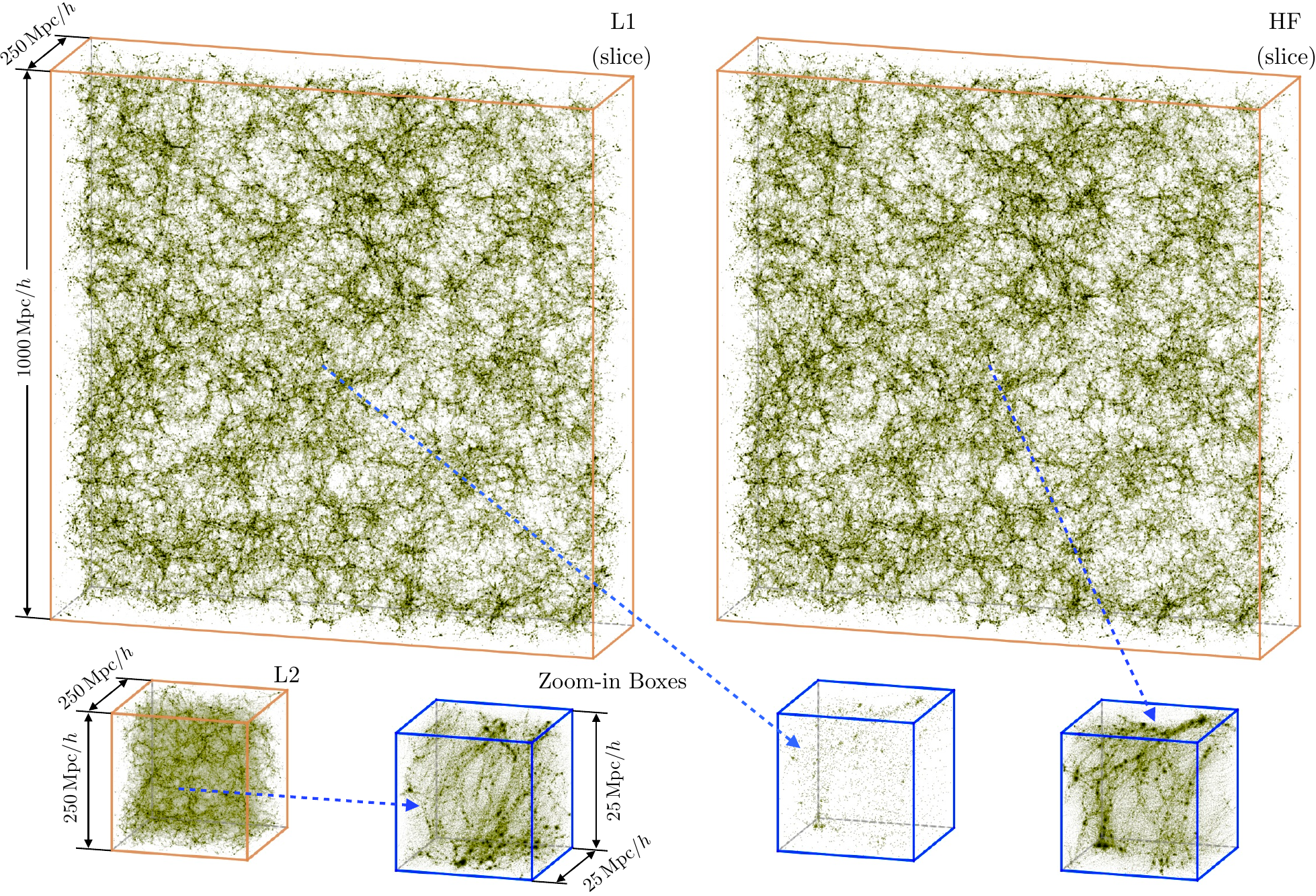}
    \caption{\label{fig:visual}Visualized matter density fields of one of the HF simulations at $z=0$ and its LF counterparts. Boxes with orange frames show the simulations with their original side lengths (except that the L1 and HF boxes are quarterly sliced), while the blue boxes are $10\times$ zoom-in visualizations of the central regions of the orange boxes. All the zoom-in boxes have the same side length of $25\,\text{Mpc}/h$. The L1 simulation successfully replicates the HF simulation on large scales, and the L2 simulation resolves small-scale structures as accurately as the HF simulation.}
\end{figure*}

\begin{figure}[t]
    \includegraphics[width=\linewidth]{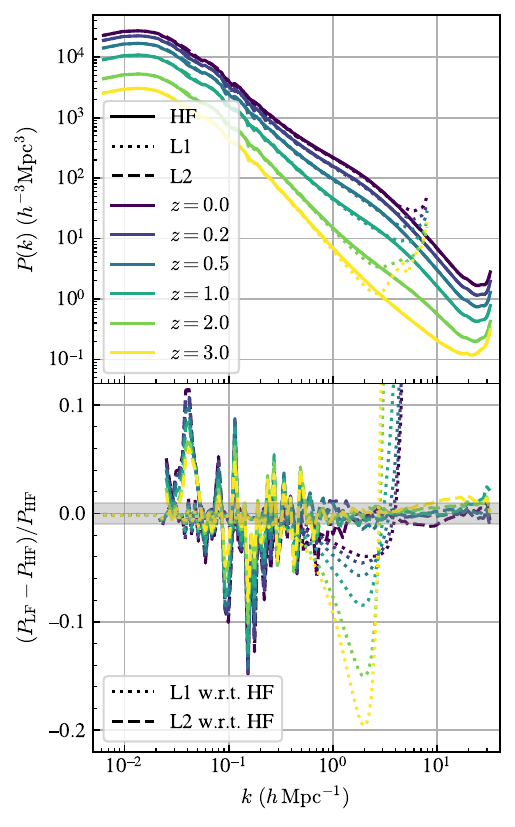}
    \caption{\label{fig:matter_pow_LFHF_eg}\textit{Upper panel:} The matter power spectra measured from the HF, L1 and L2 simulations at \texttt{Goku-N-0195}. Solid curves are spectra of the HF run, and dotted (dashed) curves are from the L1 (L2) run. Different colors represent different redshifts. \textit{Lower panel:} Relative differences of the matter power spectra of the L1 and L2 simulations with respect to the HF simulation. Dotted (dashed) curves are the relative differences of L1 (L2) with respect to HF. The relationship between the colors and redshifts is the same as in the upper panel. The gray-shaded region shows $< 1\%$ relative differences. The differences between the L1 (L2) and HF spectra stay within $1\%$ at large (small) scales, with minor exceptions at a few specific $k$-modes.}
\end{figure}

At the sampled cosmologies, we perform 564 pairs of LF simulations and 21 (15) HF simulations, which compose the \texttt{Goku-W} (\texttt{Goku-N}) suite. We visualize one of the HF simulations (hereafter we call the corresponding cosmology \texttt{Goku-N-0195}\footnote{It is point 195 ($\bm{\theta}_{195}$) from the 564 design points, $\{\bm{\theta}_i\,|\,i=0,1,\dots,563\}$.}) at $z=0$ together with the LF counterparts in Fig.~\ref{fig:visual}.
The density fields shown in the orange boxes demonstrate that the L1 simulation successfully replicates the HF simulation on large scales, while the L2 simulation is not able to do so due to the smaller box size. In contrast, the zoom-in boxes show that the L1 simulation fails to capture the structure formation on relatively small scales, while the L2 simulation resolves the structure as accurately as the HF simulation.

We also plot the matter power spectra of these three \texttt{Goku-N-0195} simulations, as well as the relative differences of L1 and L2 with respect to HF, in Fig.~\ref{fig:matter_pow_LFHF_eg}, which confirms the visual impression. 
The L1 simulation is accurate on scales with $k\lesssim 0.4h\,\text{Mpc}^{-1}$, but fails to capture structure on smaller scales, while the L2 simulation is precise on these smaller scales. The differences are more clearly shown in the lower panel. The difference between the L1 simulation and the HF simulation is $< 1\%$ for $k\lesssim 0.4h\,\text{Mpc}^{-1}$, and becomes significant on smaller scales (larger $k$). The difference increases with redshift on these relatively small scales. In contrast, the L2 simulation is not as accurate as L1 for  $k\lesssim 0.4h\,\text{Mpc}^{-1}$, but performs well on smaller scales, with the relative differences within $1\%$ for $k\gtrsim 0.7h\,\text{Mpc}^{-1}$ (though some are slightly beyond $1\%$ at $k\gtrsim 10h\,\text{Mpc}^{-1}$, which is out of our emulation range). The relative difference between L2 and HF is redshift-dependent on large scales (small $k$), with the difference decreasing with redshift.


Each of the \texttt{Goku} simulations has particle snapshots and halo catalogues\footnote{We use a friends-of-friends (FoF) halo finder~\cite{Davis1985}.} saved at $z = 9, 4, 3, 2.5, 2, 1, 0.5, 0.2$ and $0$, and the matter power spectra are saved at every PM-step. We have also generated gravitational potential planes for weak lensing investigations, which will be detailed in our next paper, though early access can be provided upon reasonable request. The complete simulation suite occupies $\sim 650\,\text{TB}$ of disk space.\footnote{The original data volume exceeded $1\,$PB, but was compressed to reduce storage requirements.}
    

\subsubsection*{Computational Cost}

\begin{figure}[t]
    \includegraphics[width=\linewidth]{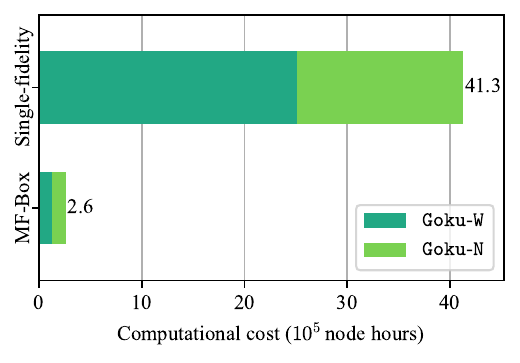}
    \caption{\label{fig:comp_cost}Comparison of the computational cost between the \texttt{Goku} simulations (using \texttt{MF-Box}) and a hypothetical single-fidelity counterpart. Each bar is color-coded to differentiate the contributions of the simulations performed for the \texttt{Goku-W} and \texttt{Goku-N} parameter boxes. The total computational costs for the MF-Box and single-fidelity approaches are annotated on the right side of their respective bars. The \texttt{MF-Box} technique achieves a 93.7\% reduction in computational expense compared to the single-fidelity method.}
\end{figure}

In this section, we demonstrate the effectiveness of the \texttt{MF-Box} framework in reducing the computational cost of building a simulation-based emulator. We compare the computational costs of the \texttt{Goku} simulations to a hypothetical single-fidelity counterpart in Fig.\ref{fig:comp_cost}. For the single-fidelity counterpart to \texttt{Goku-N}, we estimate the number of HF simulations required to achieve the same accuracy by solving Eq.(\ref{eq:constraint}) with $n_\mathrm{L}$ fixed at zero. The corresponding computational cost is then computed for this estimated number of HF simulations.

For \texttt{Goku-W}, applying the same method to estimate $n_\text{H}$ would be less accurate due to the relatively higher target error level (5\%). At this level, interpolation error dominates over the LF-to-HF correction, making the fitted error function less accurate for predicting the dependence on $n_\text{H}$. Instead, we approximate the computational cost of the single-fidelity counterpart by simply assuming $n_\text{H} = 564$, matching the number of LF simulations in \texttt{Goku-W}. This assumption is reasonable given the predominance of the interpolation error in the overall generalization error in this regime.

In Fig.~\ref{fig:comp_cost}, we see that the \texttt{Goku} simulation suite, designed using the \texttt{MF-Box} framework, 
costs only $2.6\times 10^5$ Frontera node hours\footnote{We do not take the paired simulations into account here.}. 
In contrast, constructing a single-fidelity emulator with equivalent accuracy and prior range would consume $4.13\times 10^6$ node hours. This corresponds to a 93.7\% reduction in computational expense. The significant efficiency achieved by the \texttt{MF-Box} framework underscores its utility in enabling the construction of high-dimensional emulators within practical computational budgets. Additionally, the storage space required for the simulations is significantly reduced, aligning with the savings in computational cost.



\subsection{\label{sec:matter_power}Matter Power Spectra}

\begin{figure*}[t]
    \includegraphics[width=\textwidth]{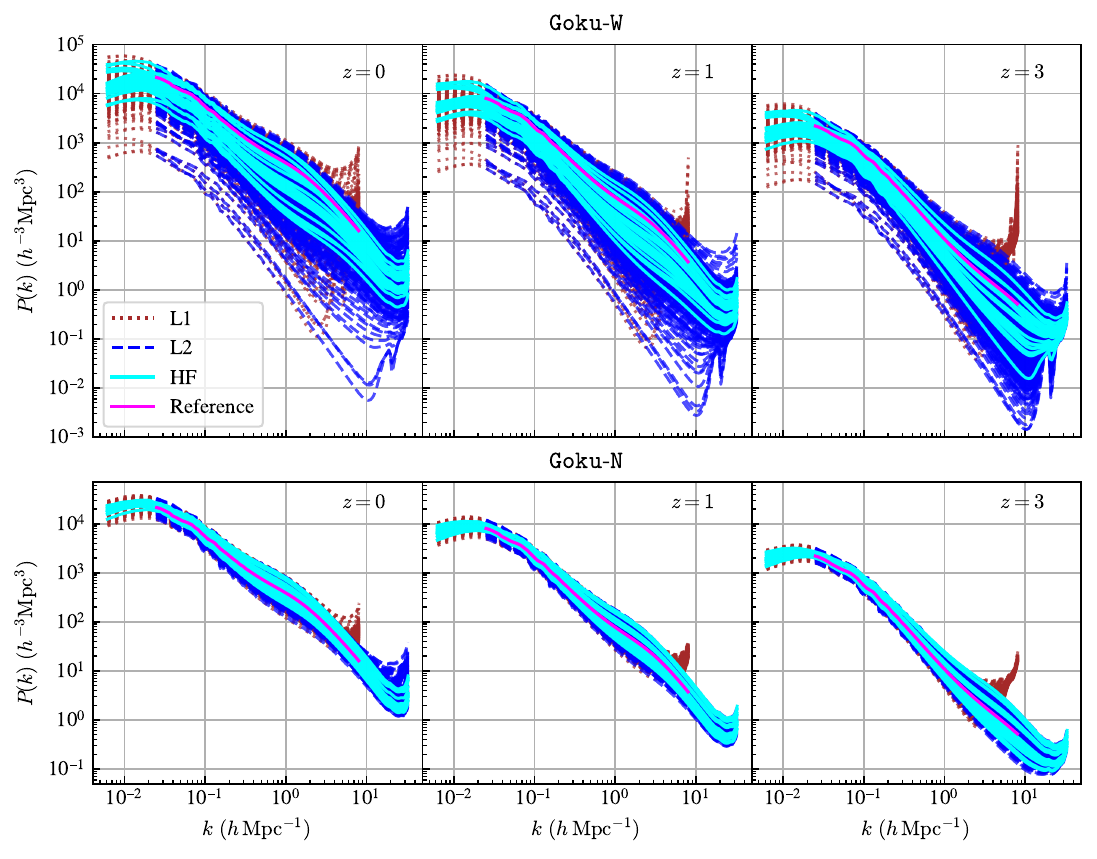}
    \caption{\label{fig:matter_pow_all}The matter power spectra of \texttt{Goku-W} (\textit{upper}) and \texttt{Goku-N} (\textit{lower}) at redshifts $z=0, 1$ and $3$ (different columns). The dotted-brown (dashed-blue) curves are the matter power spectra of the L1 (L2) simulations, the solid-cyan curves are from the HF simulations, and the solid-magenta curves are the matter power spectra predicted by \texttt{GokuEmu} at the reference cosmology. The \texttt{Goku-W} power spectra cover a wide range (2--3 orders of magnitude) due to the broad parameter space, whereas the \texttt{Goku-N} spectra are more concentrated, spanning about one order of magnitude around $k=1h\,$Mpc$^{-1}$.}
\end{figure*}

We plot the matter power spectra of the LF and HF simulations at redshifts 
$z=0, 1$ and $3$ 
in Fig.~\ref{fig:matter_pow_all}, showing the overall variation of the matter power spectra across cosmologies.
We define a reference cosmology, similar to the best-fit $\Lambda$CDM model derived from the combined analysis of TTTEEE (the CMB temperature and polarization power spectra: TT, TE, and EE), lensing, and BAO measurements in Ref.~\cite{Tristram2024}, as $\bm{\theta}_\text{ref} = (\Omega_\text{m}, \Omega_\text{b}, h, A_\text{s}, n_\text{s}, w_0, w_a, \sum m_\nu, N_\text{eff}, \alpha_\text{s})_\text{ref} = (0.31, 0.048, 0.68, 2.1\times10^{-9}, 0.97, -1, 0, 0.1, 3.08, 0)$, and plot the matter power spectra predicted by the emulator at this cosmology. The emulator, \texttt{GokuEmu}, is trained on data of all LF and HF simulations (\texttt{Goku-W} + \texttt{Goku-N}). One observes that the variation of the matter power spectrum due to varying cosmologies is significant -- the power spectra with the highest amplitudes are approximately three (one) order(s) of magnitude larger than the lowest ones over a broad scale range, especially on small scales (large $k$) in \texttt{Goku-W} (\texttt{Goku-N}). The HF spectra distribute over a large range around the middle (in log scales), and the reference cosmology spectra safely reside near the middle of the HF spectra at all redshifts.

\begin{figure}[t]
    \includegraphics[width=\linewidth]{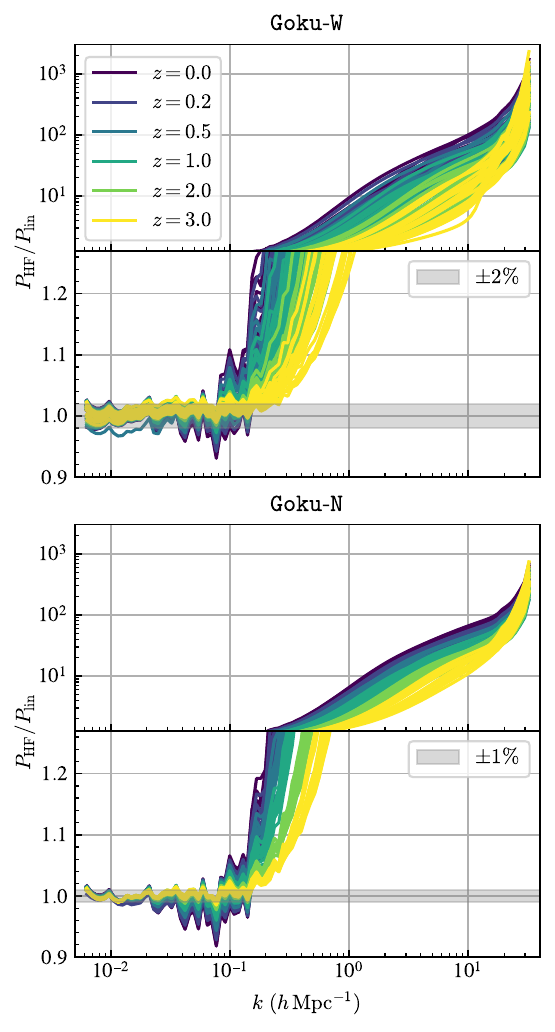}
    \caption{\label{fig:matter_pow_HF_linear}Comparison of the HF matter power spectra of \texttt{Goku-W} (\textit{upper panel}) and \texttt{Goku-N} (\textit{lower panel}) with their linear counterparts at $z=0,0.2,0.5,1,2$ and $3$. Different colors represent different redshifts. The gray-shaded regions denote the range of relative differences within $2\%$ (\textit{upper}) and $1\%$ (\textit{lower}) respectively. The spectra measured from the simulations show good agreement with the linear theory on the largest scales.}
\end{figure}

Fig.~\ref{fig:matter_pow_HF_linear} compares the matter power spectra from the HF simulations with their linear counterparts at $z=0,0.2,0.5,1,2$ and $3$. On large scales ($k\lesssim 0.1h\,\text{Mpc}^{-1}$), the spectra from \texttt{Goku-W} (\texttt{Goku-N}) simulations align closely with their linear counterparts, with the relative differences limited to within $2\%$ ($1\%$) across most of the range. A notable deviation around $k\sim 0.08h\,\text{Mpc}^{-1}$ at lower redshifts is observed in the \texttt{Goku-N-0197} simulation (lower panel), while differences in other simulations remain within $4\%$ at similar scales. These deviations are consistent with expectations due to cosmic variance, given the simulation box size and the range of cosmological parameters, and remain acceptable for the target accuracy. On smaller scales, the nonlinear spectra deviate significantly from the linear spectra, with the differences increasing with time (decreasing redshift), as expected for nonlinear structure growth.

\subsection{\label{sec:pplusf}P+F Correction}
\begin{figure}[t]
    \includegraphics[width=\linewidth]{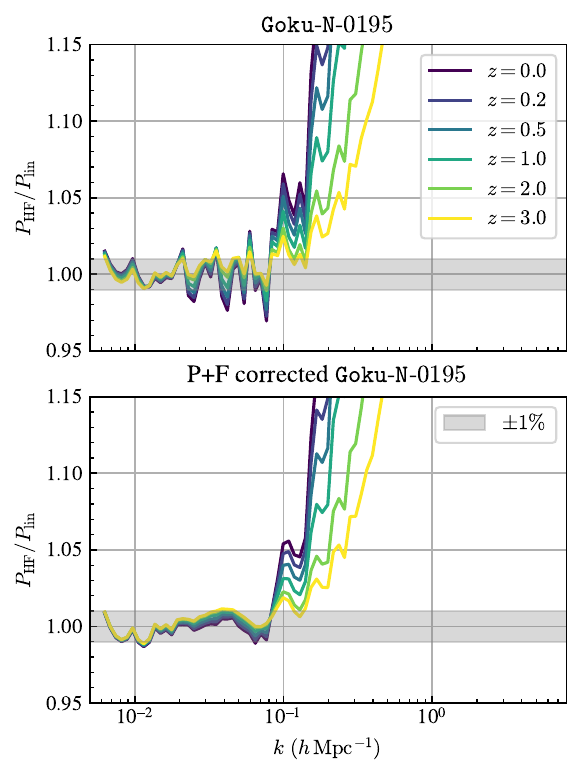}
    \caption{\label{fig:pplusf_correction}\textit{Upper panel:} The ratio of the matter power spectrum from the HF simulation of \texttt{Goku-N-0195} to the corresponding linear theory prediction at various redshifts ($z=0, 0.2, 0.5, 1, 2, 3$). Each color represents a different redshift. The gray-shaded region highlights the range of relative differences within $1\%$. \textit{Lower panel:} The ratio of the matter power spectrum, corrected using the pairing-and-fixing technique, to the linear theory prediction for the same cosmology and redshifts. The P+F corrected spectrum is smoother than the original spectrum and shows better agreement with the linear theory at large scales.}
\end{figure}

We run a paired simulation at the \texttt{Goku-N-0195} cosmology and obtain the P+F matter power spectrum. Fig.~\ref{fig:pplusf_correction} shows the ratio of the original HF matter power spectrum to the linear theory and the P+F-corrected counterpart in the upper and lower panels respectively. Compared to the original spectrum, the P+F-corrected spectrum demonstrates better agreement with linear theory at the largest scales, where the relative differences are rarely greater than $1\%$ for $k\lesssim 0.08h\,$Mpc$^{-1}$ across all redshifts. In addition, the corrected spectrum is smoother than the original over both linear and nonlinear scales, indicating the P+F correction reduces cosmic variance at large scales efficiently.

\begin{figure}[t]
    \includegraphics[width=\linewidth]{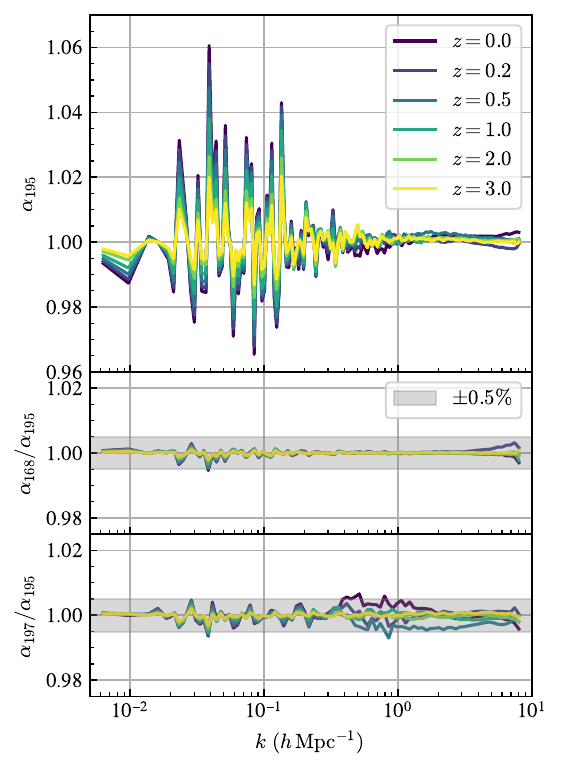}
    \caption{\label{fig:correction_fac_ratio}\textit{Upper panel:} The P+F correction factor, $\alpha_{195} = (P_\text{P+F}/P_\text{orig})_{195}$, for the L1 simulation of \texttt{Goku-N-0195}, at $z=0, 0.2, 0.5, 1, 2$ and $3$. \textit{Middle panel:} The ratio of the P+F correction factor for \texttt{Goku-N-0168} to that of \texttt{Goku-N-0195}. \textit{Lower panel:} The ratio of the P+F correction factor for \texttt{Goku-N-0197} to that of \texttt{Goku-N-0195}. All correction factors presented in this figure are derived from L1 simulations. The gray-shaded regions denote the range of relative differences within $0.5\%$, showcasing the consistency of the correction factors across cosmologies.}
\end{figure}

In Fig.~\ref{fig:matter_pow_HF_linear}, we observe similar patterns of deviation of the HF matter power spectra from linear theory across cosmologies (expected with the random number seed fixed for initial condition generation), suggesting similar P+F correction factors may apply. To verify this, we perform paired realizations for the L1 simulations of \texttt{Goku-N-0195}, \texttt{Goku-N-0168} and \texttt{Goku-N-0197}, and compare their P+F correction factors. Among these, \texttt{Goku-N-0197} is selected as a conservative case due to its large deviation from the linear spectrum around $k=0.08h\,$Mpc$^{-1}$. The correction factor, defined as $\alpha = P_\text{P+F}/P_\text{orig}$, where $P_\text{P+F}$ and $P_\text{orig}$ are the P+F-corrected and original matter power spectra respectively, is compared in Fig.~\ref{fig:correction_fac_ratio}. The correction factors are consistent across different cosmologies, with relative differences well within $0.5\%$ for most $k$-modes, significantly smaller than the variation in the correction factor itself. This consistency suggests that the P+F correction factors are nearly cosmology-independent (at least within the \texttt{Goku-N} parameter box), allowing the factor derived from one cosmology (we choose \texttt{Goku-N-0195} in this study) to be applied to other cosmologies with minimal error.

\subsection{\label{sec:emulator}Emulators for the matter power spectrum}
\subsubsection{\label{sec:loo}Leave-One-Out Validation}

\begin{figure}[t]
    \includegraphics[width=\linewidth]{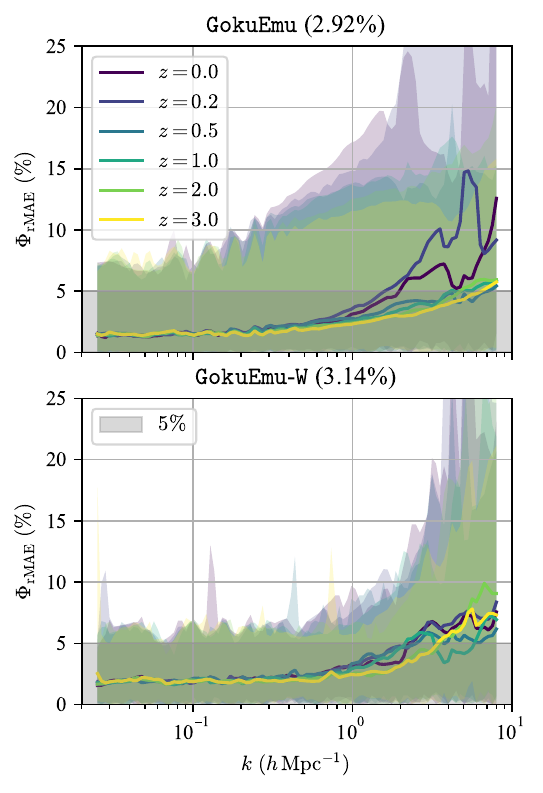}
    \caption{\label{fig:matter_pow_loo_W}\textit{Upper panel:} Leave-one-out rMAE of \texttt{GokuEmu} as a function of $k$ and $z$, averaged over the HF cosmologies in the \texttt{Goku-W} dataset. \textit{Lower panel:} Leave-one-out rMAE of \texttt{GokuEmu-W} (trained exclusively on \texttt{Goku-W}), tested against the same HF simulations. Both panels are titled with the corresponding model names and their overall average leave-one-out errors (shown in parentheses). Shaded regions represent the range of relative absolute errors for individual cosmologies, with the maximum and minimum bounds shown for each redshift. Gray-shaded areas highlight regions where the error remains below $5\%$. In both panels, mean errors are within $5\%$ across most scales and redshifts, with the largest errors occurring at the smallest scales.}
\end{figure}

\begin{figure}[t]
    \includegraphics[width=\linewidth]{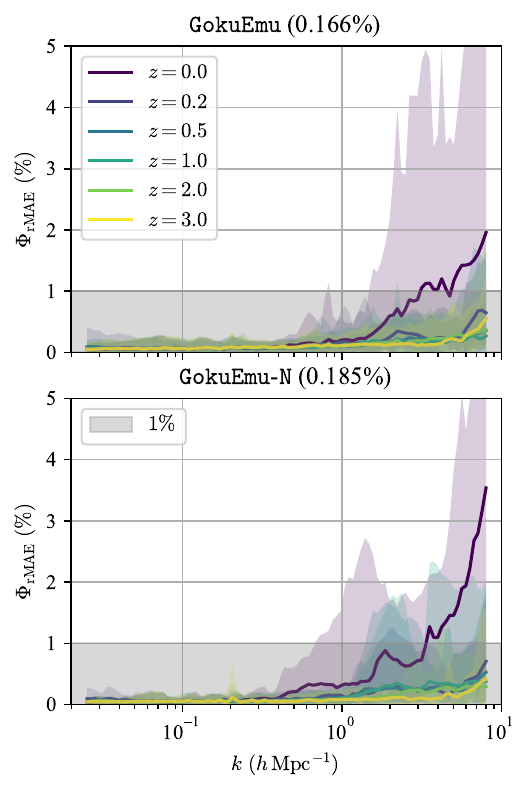}
    \caption{\label{fig:matter_pow_loo_N}\textit{Upper panel:} Leave-one-out rMAE of \texttt{GokuEmu} as a function of $k$ and $z$, averaged over the HF cosmologies in the \texttt{Goku-N} dataset. \textit{Lower panel:} Leave-one-out rMAE of \texttt{GokuEmu-N} (trained exclusively on \texttt{Goku-N}), tested against the same HF simulations. Both panels are titled with the corresponding model names and their overall average leave-one-out errors (shown in parentheses). Shaded regions represent the range of relative absolute errors for individual cosmologies, with the maximum and minimum bounds shown for each redshift. Gray-shaded areas highlight regions where the error remains below $1\%$. In both cases, mean errors remain well below $1\%$ across most scales and redshifts, except at $z=0$ on the smallest scales, where they reach a few percent.}
\end{figure}

We employ LOOCV to estimate the generalization accuracy of the emulator in predicting the matter power spectrum, specifically its ability to approximate the HF matter power spectrum. LOOCV is a method for assessing emulator performance without requiring a separate test set, as previously adopted in similar studies (e.g., Ref.~\cite{Bird2023}). This approach is particularly advantageous in our case, as producing additional HF simulations would be computationally prohibitive. 

When testing against each of the HF simulations, we leave out the corresponding LF and HF spectra and train the emulator on the remaining data. The emulator is then used to predict the omitted HF spectrum. The relative absolute error of these predictions is computed and averaged over the HF cosmologies to derive the relative mean absolute error (rMAE, denoted as $\Phi_\text{rMAE}$). We perform the procedure to estimate the generalization error of \texttt{GokuEmu} for both \texttt{Goku-W} and \texttt{Goku-N} parameter ranges. In addition, we also test the emulators trained exclusively on the \texttt{Goku-W} and \texttt{Goku-N} datasets, denoted as \texttt{GokuEmu-W} and \texttt{GokuEmu-N}, respectively.

In Fig.~\ref{fig:matter_pow_loo_W}, we test \texttt{GokuEmu} and \texttt{GokuEmu-W} against the \texttt{Goku-W} HF simulations and present the resulting leave-one-out (LOO) errors.
The upper panel shows the LOO errors of \texttt{GokuEmu} as functions of $k$ at different redshifts. 
The errors remains below $5\%$ across most scales and redshifts, with relatively high values at the smallest scales ($k\gtrsim 2h\,\text{Mpc}^{-1}$) and the lowest redshifts ($z\lesssim 0.2$). Even in these regions, the errors are still considerably smaller than the typical parameter-driven variations in the matter power spectrum within the parameter space, which span orders of magnitude (see Fig.~\ref{fig:matter_pow_all}). The lower panel shows the LOO errors for \texttt{GokuEmu-W}. While its errors remain below $5\%$ across most scales and redshifts, they display weaker dependence on redshift.
At small scales, \texttt{GokuEmu-W} outperforms \texttt{GokuEmu} at lower redshifts but performs slightly worse at higher redshifts. 
In contrast, \texttt{GokuEmu} tends to be slightly more accurate at large scales. 
Overall, the mean errors are $2.92\%$ for \texttt{GokuEmu} and $3.14\%$ for \texttt{GokuEmu-W}, both well within the predefined target of $5\%$, though the former can perform worse in some extreme regions of the parameter space (indicated by the larger errors at individual cosmologies shown in the upper panel\footnote{The worst two cases are found to be \texttt{Goku-W-0056} and \texttt{Goku-W-0242}.}). We leave the choice of which emulator to use to the user, depending on the specific application.

Fig.~\ref{fig:matter_pow_loo_N} presents similar LOO error analyses for the \texttt{Goku-N} parameter space. Here, both \texttt{GokuEmu} and \texttt{GokuEmu-N} perform exceptionally well, with errors well below 1\% across most scales and redshifts. The only exceptions occur at $z=0$ and the smallest scales ($k > 3$ h/Mpc), where errors can reach a few percent. Overall, \texttt{GokuEmu} demonstrates a lower mean LOO error (0.166\%) compared to \texttt{GokuEmu-N} (0.185\%), and it also delivers better accuracy at $z=0$ on the smallest scales. Therefore, \texttt{GokuEmu} is likely the more suitable choice for most applications within the \texttt{Goku-N} parameter range. We note that the HF cosmologies in both \texttt{Goku-W} and \texttt{Goku-N} are predominantly extreme examples, typically near the boundaries of the parameter space; hence the true generalization error for typical cosmologies is likely even lower than the reported LOOCV estimates.
 
\subsubsection{\label{sec:}Parameter Sensitivity}
\begin{figure*}
    \includegraphics[width=\textwidth]{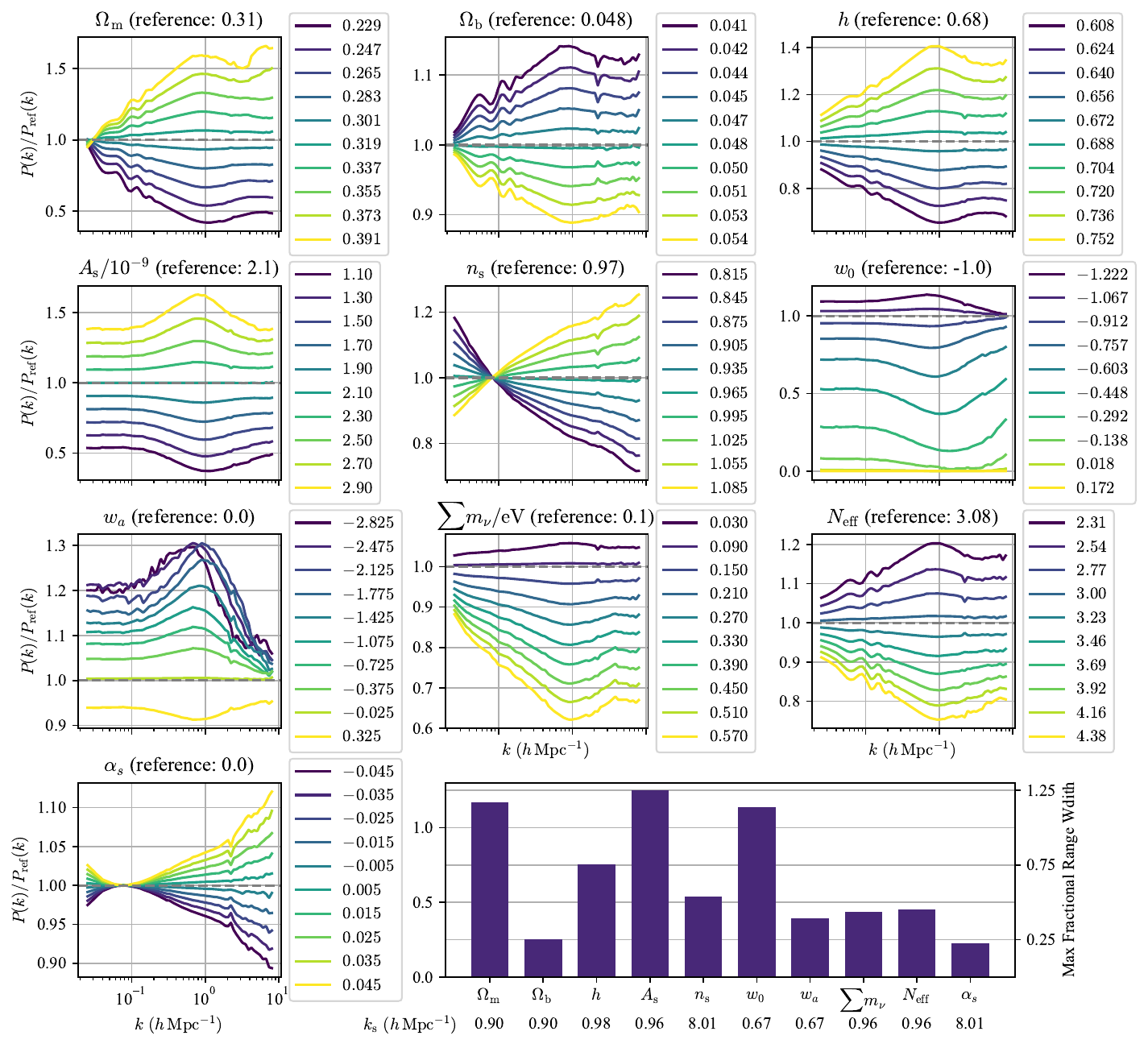}
    \caption{\label{fig:parameter_sensitivity}\textit{Panels 1 to 10:} Variations in the matter power spectrum at $z=0$ induced by each of 10 cosmological parameters, evaluated at the reference cosmology (defined in Section~\ref{sec:matter_power}). In each panel, one parameter is varied independently across 10 values, spanning from $5\%$ to $95\%$ of its prior range, while all other parameters are fixed at their reference values. The power spectra are normalized to the reference spectrum at each scale, with distinct colors indicating different parameter values. When we refer to Panels 1 to 10, they are arranged sequentially from left to right and then top to bottom. The dashed line in each panel corresponds the reference cosmology. \textit{Panel 11:} The size of the fractional range (difference between the largest and smallest power values across $k$-modes) of the matter power spectrum, relative to the reference cosmology, is shown for each parameter. The scale at which this maximum range occurs is indicated below the corresponding parameter name.}
\end{figure*}

To illustrate the impact of individual cosmological parameters on the matter power spectrum within our parameter space, we present a series of plots (panels 1 to 10 of Fig.~\ref{fig:parameter_sensitivity}) where each parameter is varied independently while the others are held fixed at their reference values. The reference cosmology is the same one as defined in Section~\ref{sec:matter_power}. The power spectra are normalized to the reference spectrum at each scale, enabling direct comparison of relative changes. In the last panel, we present the maximum range of the fractional change in the matter power spectrum, relative to the reference cosmology, for each cosmological parameter. The scale at which this maximum range occurs, i.e., the $k$-mode at which the power spectrum is most sensitive to the change of the cosmological parameter ($k_\text{s}$), is also indicated.

We observe that, within the prior ranges we defined (see Table~\ref{tab:ranges}), the matter power spectrum exhibits the highest sensitivity to the parameters $\Omega_\text{m}$, $A_\text{s}$ and $w_0$, with the largest fractional variations exceeding $100\%$ at certain scales. The other parameters have a maximum fractional range wider than or roughly equal to $25\%$. In all cases, the $k$-modes most sensitive to parameter changes are typically located in the nonlinear regime, at relatively small scales ($k\sim 1h\,$Mpc$^{-1}$).

More specifically, for $\Omega_\text{m}$, $\Omega_\text{b}$, $h$, $A_\text{s}$, $\sum m_\nu$ and $N_\text{eff}$, the $k$-modes most responsive to these parameters lie between $0.9h\,\text{Mpc}^{-1}$ and $1h\,\text{Mpc}^{-1}$. When varying dynamical DE parameters, the matter power spectrum changes most at $k\approx 0.67h\,\text{Mpc}^{-1}$.\footnote{Note that $k_\text{s}$ depends on both the chosen reference cosmology and the parameter range. For example, if $w_0$ were varied up to $-0.138$ instead of $0.172$, $k_\text{s}$ would shift closer to $1h\,\text{Mpc}^{-1}$.} As for $n_\text{s}$ and $\alpha_\text{s}$, the most sensitive $k$-modes are at the upper limit of the emulator's $k$-range, with no physical meaning but determined by the resolution of L1 simulation.

\subsubsection{\label{sec:compare_EE2}Comparison with \texttt{EuclidEmulator2}}
\begin{figure}[t]
    \includegraphics[width=\linewidth]{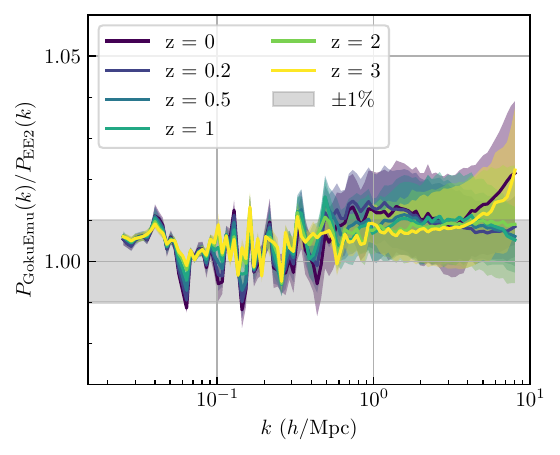}
    \caption{\label{fig:emu_comparison_EE2}Comparison of the matter power spectra predicted by \texttt{GokuEmu} and \texttt{EuclidEmulator2}. The plotted ratios, averaged over 50 cosmologies within the overlapping parameter range, show results at redshifts  $z = 0, 0.2, 0.5, 1, 2$,  and  $3$. Solid lines indicate the mean ratios, while the shaded regions represent the standard deviation at each scale. The gray-shaded band marks a $\pm 1\%$ region around unity, highlighting the level of agreement. Overall good agreement is observed between the two emulators, while deviations at relatively small scales (\(k \gtrsim 0.5h\,\text{Mpc}^{-1}\)) likely arise from the improved treatment of massive neutrinos in MP-Gadget.}
\end{figure}

We compare the matter power spectra predicted by \texttt{GokuEmu} and \texttt{EuclidEmulator2}~\cite{Knabenhans2021} in Fig.~\ref{fig:emu_comparison_EE2}. The comparison spans redshifts  $z = 0, 0.2, 0.5, 1, 2$,  and  $3$, with relative differences averaged over 50 cosmologies sampled using the SLHD method within the overlapping parameter space of \texttt{Goku-N} and \texttt{EuclidEmulator2}.

At large scales (\( k \lesssim 0.5h\,\mathrm{Mpc}^{-1} \)), the two emulators exhibit strong agreement, with relative differences mostly within $1\%$, decreasing with redshift. On smaller scales, \texttt{GokuEmu} generally predicts slightly higher power ($\sim 1\%$) compared to \texttt{EuclidEmulator2}, particularly around $k \sim 1h\,\mathrm{Mpc}^{-1}$, where the differences are more noticeable at lower redshifts. These discrepancies are likely due to the improved treatment of massive neutrinos in MP-Gadget (see Section~\ref{sec:sim_codes}).

Overall, the agreement between the emulators is robust at large scales, while the small-scale deviations align with expectations given the differences in simulation methodologies.

\subsection{\label{sec:HMF}Halo Mass Functions}

\begin{figure}[t]
    \includegraphics[width=\linewidth]{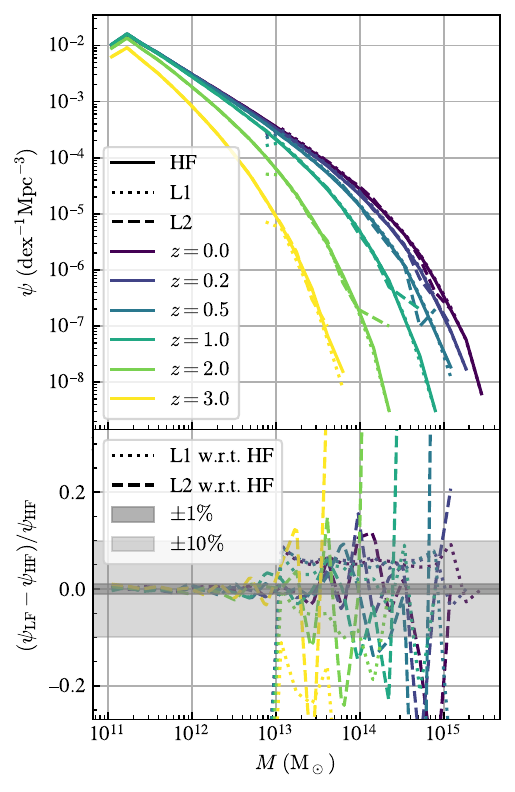}
    \caption{\label{fig:HMF}\textit{Upper panel:} Halo mass functions of the HF and LF simulations at the {\tt Goku-N-0144} cosmology. Solid curves are the HMFs of the HF simulation, and dotted (dashed) curves are of the L1 (L2) simulation. \textit{Lower panel:} The relative differences of the HMFs of the L1 and L2 simulations with respect to the HF simulation. The shaded regions denote the range of relative differences within $1\%$ and $10\%$ respectively. The L2 simulation shows excellent agreement with the HF simulation at the low-mass end, with relative differences mostly within $1\%$. The L1 simulation, limited by its mass resolution, shows a cutoff at $M\sim 10^{13}\,\mathrm{M}_\odot$, but generally agrees with the HF simulation at high masses, with relative differences mostly within $10\%$.}
\end{figure}

We preliminarily assess the potential of \texttt{Goku} for emulating the halo mass function (HMF) by comparing the HMFs of the HF and LF simulations. We define the HMF as
\begin{equation}
    \psi \equiv \frac{\mathrm{d} n(M)}{\mathrm{d}\log (M/\mathrm{M}_\odot)},
\end{equation}
where $n(M)$ is the number density of halos with mass below $M$. Fig.~\ref{fig:HMF} shows the HMFs of the HF and LF simulations for the {\tt Goku-N-0144} cosmology. We see that the L2 simulation provides excellent agreement with the corresponding HF simulation at the low-mass end, with relative differences at the percent level, while the difference becomes more pronounced at the high-mass end. The L1 simulation, limited by its relatively poor mass resolution, shows a cutoff at $M\sim 10^{13}\,\mathrm{M}_\odot$. Nonetheless, its HMF generally agrees with the HF simulation at high masses, with relative differences mostly within $10\%$ (though the agreement weakens at $z=3$), comparable to the generalization errors of some existing emulators (e.g., Ref.~\cite{Bocquet2020}). The consistency between the L1 and HF simulations varies with redshift, with a broader range of agreement at lower redshifts.

Based on these findings, \texttt{Goku} shows promise for emulating the HMF, and the emulator is expected to perform with higher accuracy at lower masses and lower redshifts, which is also observed in single-fidelity HMF emulators (e.g., Ref.~\cite{Bocquet2020}). For $M<10^{13}\,\mathrm{M}_\odot$, the HMF could be emulated using data from the L2 and HF simulations without the need for L1 data, while a standard MF-Box emulator (trained on HF and both LF nodes) could be constructed for the high mass range.

Compared to existing HMF emulators (e.g., Refs.~\cite{McClintock2019,Bocquet2020,Shen2024}), \texttt{Goku} offers the advantage of spanning a higher-dimensional parameter space, enabling more flexible exploration of cosmological models. We leave the exploration of the HMF emulation to future work. We note in passing that baryonic physics, such as feedback processes altering the distribution of matter in and around galaxies (see, e.g., Refs.~\cite{Salucci2019,Marinacci2018,Schaye2015}), could impact the HMF, but these effects are not included in our simulations.

\section{\label{sec:concl}conclusions}

In this work, we have performed a large suite of cosmological $N$-body simulations, \texttt{Goku}, and built an emulator for the matter power spectrum with high accuracy and large parameter space using the output from these simulations, \texttt{GokuEmu}. This work builds on and further develops a pipeline for constructing cosmological emulators using the \texttt{MF-Box} framework. Below, we summarize the contributions of this work.

\subsection{\label{sec:contrib_pipeline}Emulator Construction Pipeline with \texttt{MF-Box}}

The construction pipeline for the matter power spectrum emulator is outlined in Fig.~\ref{fig:workflow}. The process begins with the development of exploratory emulators using a suite of small, computationally inexpensive simulations, \texttt{Goku-pre}, followed by the optimization of computational resources for the production suite, \texttt{Goku}. 

We enhance the \texttt{MF-Box} framework by introducing a beam search algorithm for selecting HF cosmologies. This approach reduces the corresponding interpolation error of the selected cosmologies by more than $50\%$ compared to the naive selection method used in previous work (Fig.~\ref{fig:beam_size}). Additionally, we propose a new analytic formula for multi-fidelity error function fitting. This formula not only provides superior accuracy compared to the previously used method but also offers greater generality, allowing it to predict errors for single-fidelity counterparts.

Our pipeline effectively optimizes the computational budget to achieve a specific generalization accuracy, resulting in a reduction of computational cost by $94\%$ compared to a single-fidelity approach (Fig.~\ref{fig:comp_cost}). 

It is important to note that the \texttt{MF-Box} emulation framework is general and not restricted to Gaussian processes. In principle, other regression techniques, such as polynomial chaos expansion~\cite{Knabenhans2021} and neural networks~\cite{Guo2022,Diao2023}, can also be integrated with \texttt{MF-Box} to build emulators and optimize computational budgets. Furthermore, these alternative regression techniques have the potential to accelerate both the training process and prediction computations, making them particularly advantageous for applications requiring rapid emulation across large parameter spaces.

Overall, the pipeline we developed has the potential to significantly advance cosmological research by enabling the construction of accurate, high-dimensional emulators with drastically reduced computational costs. This advancement will empower the community to explore a wider range of cosmological models, without the prohibitive computational burden traditionally associated with $N$-body simulations. This contribution not only complements existing efforts but also establishes a foundation for future advancements in cosmological emulation frameworks.

\subsection{\label{sec:contrib_sims}10-Parameter Simulation Suite: \texttt{Goku}}

We present the \texttt{Goku} simulation suite, designed for cosmological emulation, with simulation specifications listed in Table~\ref{tab:sim_specs}. The simulations span a 10-dimensional parameter space, enabling investigations of various cosmological models, including the standard $\Lambda$CDM model and its extensions: massive neutrinos, dynamical dark energy, the effective number of neutrino species, and the running of the spectral index of the primordial power spectrum. 

The suite comprises two subsets: \texttt{Goku-W} and \texttt{Goku-N}. The \texttt{Goku-W} subset includes 564 pairs of LF simulations and 21 HF simulations, covering an unprecedentedly large parameter box, while the \texttt{Goku-N} subset contains 564 LF simulations and 15 HF simulations sampled within a narrower parameter range (see Table~\ref{tab:ranges}). The halo catalogues, matter power spectra and other derived products will be made publicly available, with the full particle snapshots available on request. Researchers are encouraged to contact the authors for early access.\footnote{The data are hosted on servers at the Texas Advanced Computing Center (TACC).}

The simulations feature an improved treatment of massive neutrinos, resulting in matter power spectra that are a few percent more accurate (up to $\sim 4\%$, depending on the total neutrino mass and redshift) at $k\sim 1h\,\text{Mpc}^{-1}$, compared to simulations which incorporate neutrinos only at the linear theory level (refer to Ref.~\cite{Bird2012} for detailed comparisons).

We have also explored the potential of this suite of simulations for emulating the HMF. The differences between the HMF from an HF simulation and its LF counterparts can be down to the percent level across a significant fraction of the range (e.g., $10^{11} < M/\mathrm{M}_\odot < 10^{13}$ at $z=3$) and $\lesssim 10\%$ near the high-mass end. An MF-Box emulator for the HMF could be constructed based on the simulations, and the emulator is likely to be more accurate at lower masses and lower redshifts. 

In summary, the \texttt{Goku} simulation suite enables precise exploration of diverse cosmological models and efficient construction of emulators for various summary statistics. It spans a comprehensive 10-dimensional parameter space, including all currently popular $\Lambda$CDM extensions, such as massive neutrinos and dynamical dark energy. The suite's accessibility and versatility will drive advancements in emulator development and cosmological parameter constraints, establishing a foundation for future advances in cosmology.

\subsection{\label{sec:contrib_emu}10-Dimensional Emulator: \texttt{GokuEmu}}

We have constructed a 10D emulator, \texttt{GokuEmu}, for the matter power spectrum, representing the first 10D cosmological emulator. \texttt{GokuEmu} allows significant variations in the spectrum arising from each cosmological parameter (Fig.~\ref{fig:parameter_sensitivity}).

In the broad parameter range of \texttt{Goku-W}, the emulator achieves an average generalization error of approximately $3\%$ across tested cosmologies, redshifts, and $k$-bins. Errors are notably lower at larger scales and higher redshifts (Fig.~\ref{fig:matter_pow_loo_W}). Within the higher-likelihood parameter range, \texttt{Goku-N}, the generalization error decreases to an average of $\sim 0.2\%$, with negligible errors at large scales and higher redshifts. At small scales, errors only exceed $1\%$ for $k > 3$ h/Mpc at $z=0$, where baryonic effects are significant (Fig.~\ref{fig:matter_pow_loo_N}).

To further enhance accuracy, we provide a universal P+F correction factor for the matter power spectrum. This correction effectively reduces cosmic variance at large scales and is nearly cosmology-independent within the \texttt{Goku-N} parameter range, enabling its application to emulator results as a post-processing step.

The emulator is publicly available at Ref.~\cite{GokuEmu2025}, where users can also access the training data (matter power spectra).

\texttt{GokuEmu} marks a major advancement in cosmological modeling by extending emulation into a 10D parameter space. This capability enables precise predictions of the matter power spectrum across diverse cosmological scenarios, enhancing the utility of data from surveys such as \textit{Euclid} and the Roman Space Telescope to constrain cosmological parameters. By providing publicly accessible training data and emulation tools, it supports both emulator development and theoretical model testing.

\section{\label{sec:future}Future Work}

This work has several limitations that present opportunities for future improvements. Below, we state some of these limitations and our plans to address them.

Currently, \texttt{GokuEmu} emulates the matter power spectrum over the overlapping scales of the L1 and L2 simulations within the \texttt{MF-Box} framework, covering the range $0.025 \lesssim k/(h\,\text{Mpc}^{-1}) \lesssim 8$. However, the full spatial range allowed by the simulations is not yet utilized. The lower limit could be extended to $0.006h\,\text{Mpc}^{-1}$ by training the emulator using only L1 and HF simulations over the additional $k$ range. Similarly, the upper limit could be extended to approximately $10h\,\text{Mpc}^{-1}$ using L2 and HF simulations. Future work will focus on enhancing the emulator to cover this broader $k$ range, thereby increasing its applicability.

While the emulation-only (generalization) uncertainty is already minimal for the \texttt{Goku-N} parameter range, the overall prediction uncertainty for the matter power spectrum remains influenced by errors inherent in the simulations themselves. To address this, several approaches will be considered: (i) Cosmology-dependent P+F correction: This technique can further mitigate the effects of cosmic variance, particularly at large scales. (ii) Small-scale improvements: On scales $1 \lesssim k/(h\,\text{Mpc}^{-1}) \lesssim 10$, baryonic physics can introduce significant uncertainties, while resolution effects are generally subdominant but still relevant at the smallest scales. Future work will focus on incorporating cosmology-dependent resolution corrections and modeling baryonic suppression effects~\cite{vanDaalen2011,Mead2021} to enhance the emulator's accuracy on these scales.

Lastly, while the analytic formula for multi-fidelity error function estimation introduced in this work performs well, further refinements are possible to improve its accuracy and generality. Addressing unexpected negative values in the fitting parameters (Table~\ref{tab:error_params}) or exploring more flexible, non-parametric approaches could provide a more robust framework for optimizing the computational budget without relying on predefined error function forms.

\section*{Acknowledgments}
We express our gratitude to José Manuel Zorrilla Matilla for useful suggestions. We thank Julien Lesgourgues and Antony Lewis for valuable discussions on the treatments of massive neutrinos and dynamical dark energy in the Boltzmann codes (CLASS and CAMB). We thank Mahdi (Sum) Qezlou for helpful discussions on the parameter ranges.
We also thank Yueying Ni for hints on visualizing the simulations and Nianyi Chen for assistance with compressing simulation data. YY and SB acknowledge funding from NASA ATP 80NSSC22K1897. MFH is supported by a NASA FINESST grant No. ASTRO20-0022, the Leinweber Foundation and DOE grant DE-SC0019193. Computing resources were provided by Frontera LRAC AST21005.
The authors acknowledge the Frontera computing project at the Texas Advanced Computing Center (TACC) for providing HPC and storage resources that have contributed to the research results reported within this paper.
Frontera is made possible by National Science Foundation award OAC-1818253.
URL: \url{http://www.tacc.utexas.edu}.
Computations were performed using the computer clusters and data storage resources of the HPCC, which were funded by grants from NSF (MRI-2215705, MRI-1429826) and NIH (1S10OD016290-01A1).

\bibliography{Goku}

\begin{thebibliography}{100}%
\makeatletter
\providecommand \@ifxundefined [1]{%
 \@ifx{#1\undefined}
}%
\providecommand \@ifnum [1]{%
 \ifnum #1\expandafter \@firstoftwo
 \else \expandafter \@secondoftwo
 \fi
}%
\providecommand \@ifx [1]{%
 \ifx #1\expandafter \@firstoftwo
 \else \expandafter \@secondoftwo
 \fi
}%
\providecommand \natexlab [1]{#1}%
\providecommand \enquote  [1]{``#1''}%
\providecommand \bibnamefont  [1]{#1}%
\providecommand \bibfnamefont [1]{#1}%
\providecommand \citenamefont [1]{#1}%
\providecommand \href@noop [0]{\@secondoftwo}%
\providecommand \href [0]{\begingroup \@sanitize@url \@href}%
\providecommand \@href[1]{\@@startlink{#1}\@@href}%
\providecommand \@@href[1]{\endgroup#1\@@endlink}%
\providecommand \@sanitize@url [0]{\catcode `\\12\catcode `\$12\catcode
  `\&12\catcode `\#12\catcode `\^12\catcode `\_12\catcode `\%12\relax}%
\providecommand \@@startlink[1]{}%
\providecommand \@@endlink[0]{}%
\providecommand \url  [0]{\begingroup\@sanitize@url \@url }%
\providecommand \@url [1]{\endgroup\@href {#1}{\urlprefix }}%
\providecommand \urlprefix  [0]{URL }%
\providecommand \Eprint [0]{\href }%
\providecommand \doibase [0]{https://doi.org/}%
\providecommand \selectlanguage [0]{\@gobble}%
\providecommand \bibinfo  [0]{\@secondoftwo}%
\providecommand \bibfield  [0]{\@secondoftwo}%
\providecommand \translation [1]{[#1]}%
\providecommand \BibitemOpen [0]{}%
\providecommand \bibitemStop [0]{}%
\providecommand \bibitemNoStop [0]{.\EOS\space}%
\providecommand \EOS [0]{\spacefactor3000\relax}%
\providecommand \BibitemShut  [1]{\csname bibitem#1\endcsname}%
\let\auto@bib@innerbib\@empty
\bibitem [{\citenamefont {{DESI Collaboration}}\ \emph
  {et~al.}(2016)\citenamefont {{DESI Collaboration}}, \citenamefont
  {{Aghamousa}},\ and\ \citenamefont {{Aguilar et
  al.}}}]{DESICollaboration2016}%
  \BibitemOpen
  \bibfield  {author} {\bibinfo {author} {\bibnamefont {{DESI Collaboration}}},
  \bibinfo {author} {\bibfnamefont {A.}~\bibnamefont {{Aghamousa}}},\ and\
  \bibinfo {author} {\bibfnamefont {J.}~\bibnamefont {{Aguilar et al.}}},\
  }\bibfield  {title} {\bibinfo {title} {{The DESI Experiment Part I:
  Science,Targeting, and Survey Design}},\ }\href
  {https://doi.org/10.48550/arXiv.1611.00036} {\bibfield  {journal} {\bibinfo
  {journal} {arXiv e-prints}\ ,\ \bibinfo {eid} {arXiv:1611.00036}} (\bibinfo
  {year} {2016})},\ \Eprint {https://arxiv.org/abs/1611.00036}
  {arXiv:1611.00036 [astro-ph.IM]} \BibitemShut {NoStop}%
\bibitem [{\citenamefont {{Abell et al.}}(2009)}]{LSC2009}%
  \BibitemOpen
  \bibfield  {author} {\bibinfo {author} {\bibfnamefont {P.~A.}\ \bibnamefont
  {{Abell et al.}}},\ }\href {https://doi.org/10.48550/ARXIV.0912.0201} {\emph
  {\bibinfo {title} {LSST Science Book, Version 2.0}}}\ (\bibinfo  {publisher}
  {arXiv},\ \bibinfo {year} {2009})\ \Eprint {https://arxiv.org/abs/0912.0201}
  {arXiv:0912.0201 [astro-ph.IM]} \BibitemShut {NoStop}%
\bibitem [{\citenamefont {{Laureijs et al.}}(2011)}]{Laureijs2011}%
  \BibitemOpen
  \bibfield  {author} {\bibinfo {author} {\bibfnamefont {R.}~\bibnamefont
  {{Laureijs et al.}}},\ }\bibfield  {title} {\bibinfo {title} {{Euclid
  Definition Study Report}},\ }\href {https://doi.org/10.48550/arXiv.1110.3193}
  {\bibfield  {journal} {\bibinfo  {journal} {arXiv e-prints}\ ,\ \bibinfo
  {eid} {arXiv:1110.3193}} (\bibinfo {year} {2011})},\ \Eprint
  {https://arxiv.org/abs/1110.3193} {arXiv:1110.3193 [astro-ph.CO]}
  \BibitemShut {NoStop}%
\bibitem [{\citenamefont {{Akeson et al.}}(2019)}]{Akeson2019}%
  \BibitemOpen
  \bibfield  {author} {\bibinfo {author} {\bibfnamefont {R.}~\bibnamefont
  {{Akeson et al.}}},\ }\bibfield  {title} {\bibinfo {title} {{The Wide Field
  Infrared Survey Telescope: 100 Hubbles for the 2020s}},\ }\href
  {https://doi.org/10.48550/arXiv.1902.05569} {\bibfield  {journal} {\bibinfo
  {journal} {arXiv e-prints}\ ,\ \bibinfo {eid} {arXiv:1902.05569}} (\bibinfo
  {year} {2019})},\ \Eprint {https://arxiv.org/abs/1902.05569}
  {arXiv:1902.05569 [astro-ph.IM]} \BibitemShut {NoStop}%
\bibitem [{\citenamefont {{Gong}}\ \emph {et~al.}(2019)\citenamefont {{Gong}},
  \citenamefont {{Liu}}, \citenamefont {{Cao}}, \citenamefont {{Chen}},
  \citenamefont {{Fan}}, \citenamefont {{Li}}, \citenamefont {{Li}},
  \citenamefont {{Li}}, \citenamefont {{Zhang}},\ and\ \citenamefont
  {{Zhan}}}]{Gong2019}%
  \BibitemOpen
  \bibfield  {author} {\bibinfo {author} {\bibfnamefont {Y.}~\bibnamefont
  {{Gong}}}, \bibinfo {author} {\bibfnamefont {X.}~\bibnamefont {{Liu}}},
  \bibinfo {author} {\bibfnamefont {Y.}~\bibnamefont {{Cao}}}, \bibinfo
  {author} {\bibfnamefont {X.}~\bibnamefont {{Chen}}}, \bibinfo {author}
  {\bibfnamefont {Z.}~\bibnamefont {{Fan}}}, \bibinfo {author} {\bibfnamefont
  {R.}~\bibnamefont {{Li}}}, \bibinfo {author} {\bibfnamefont {X.-D.}\
  \bibnamefont {{Li}}}, \bibinfo {author} {\bibfnamefont {Z.}~\bibnamefont
  {{Li}}}, \bibinfo {author} {\bibfnamefont {X.}~\bibnamefont {{Zhang}}},\ and\
  \bibinfo {author} {\bibfnamefont {H.}~\bibnamefont {{Zhan}}},\ }\bibfield
  {title} {\bibinfo {title} {{Cosmology from the Chinese Space Station Optical
  Survey (CSS-OS)}},\ }\href {https://doi.org/10.3847/1538-4357/ab391e}
  {\bibfield  {journal} {\bibinfo  {journal} {\apj}\ }\textbf {\bibinfo
  {volume} {883}},\ \bibinfo {eid} {203} (\bibinfo {year} {2019})},\ \Eprint
  {https://arxiv.org/abs/1901.04634} {arXiv:1901.04634 [astro-ph.CO]}
  \BibitemShut {NoStop}%
\bibitem [{\citenamefont {{Takada}}\ \emph {et~al.}(2014)\citenamefont
  {{Takada}}, \citenamefont {{Ellis}}, \citenamefont {{Chiba}}, \citenamefont
  {{Greene}}, \citenamefont {{Aihara}}, \citenamefont {{Arimoto}},
  \citenamefont {{Bundy}}, \citenamefont {{Cohen}}, \citenamefont {{Dor{\'e}}},
  \citenamefont {{Graves}}, \citenamefont {{Gunn}}, \citenamefont {{Heckman}},
  \citenamefont {{Hirata}}, \citenamefont {{Ho}}, \citenamefont {{Kneib}},
  \citenamefont {{Le F{\`e}vre}}, \citenamefont {{Lin}}, \citenamefont
  {{More}}, \citenamefont {{Murayama}}, \citenamefont {{Nagao}}, \citenamefont
  {{Ouchi}}, \citenamefont {{Seiffert}}, \citenamefont {{Silverman}},
  \citenamefont {{Sodr{\'e}}}, \citenamefont {{Spergel}}, \citenamefont
  {{Strauss}}, \citenamefont {{Sugai}}, \citenamefont {{Suto}}, \citenamefont
  {{Takami}},\ and\ \citenamefont {{Wyse}}}]{Takada2014}%
  \BibitemOpen
  \bibfield  {author} {\bibinfo {author} {\bibfnamefont {M.}~\bibnamefont
  {{Takada}}}, \bibinfo {author} {\bibfnamefont {R.~S.}\ \bibnamefont
  {{Ellis}}}, \bibinfo {author} {\bibfnamefont {M.}~\bibnamefont {{Chiba}}},
  \bibinfo {author} {\bibfnamefont {J.~E.}\ \bibnamefont {{Greene}}}, \bibinfo
  {author} {\bibfnamefont {H.}~\bibnamefont {{Aihara}}}, \bibinfo {author}
  {\bibfnamefont {N.}~\bibnamefont {{Arimoto}}}, \bibinfo {author}
  {\bibfnamefont {K.}~\bibnamefont {{Bundy}}}, \bibinfo {author} {\bibfnamefont
  {J.}~\bibnamefont {{Cohen}}}, \bibinfo {author} {\bibfnamefont
  {O.}~\bibnamefont {{Dor{\'e}}}}, \bibinfo {author} {\bibfnamefont
  {G.}~\bibnamefont {{Graves}}}, \bibinfo {author} {\bibfnamefont {J.~E.}\
  \bibnamefont {{Gunn}}}, \bibinfo {author} {\bibfnamefont {T.}~\bibnamefont
  {{Heckman}}}, \bibinfo {author} {\bibfnamefont {C.~M.}\ \bibnamefont
  {{Hirata}}}, \bibinfo {author} {\bibfnamefont {P.}~\bibnamefont {{Ho}}},
  \bibinfo {author} {\bibfnamefont {J.-P.}\ \bibnamefont {{Kneib}}}, \bibinfo
  {author} {\bibfnamefont {O.}~\bibnamefont {{Le F{\`e}vre}}}, \bibinfo
  {author} {\bibfnamefont {L.}~\bibnamefont {{Lin}}}, \bibinfo {author}
  {\bibfnamefont {S.}~\bibnamefont {{More}}}, \bibinfo {author} {\bibfnamefont
  {H.}~\bibnamefont {{Murayama}}}, \bibinfo {author} {\bibfnamefont
  {T.}~\bibnamefont {{Nagao}}}, \bibinfo {author} {\bibfnamefont
  {M.}~\bibnamefont {{Ouchi}}}, \bibinfo {author} {\bibfnamefont
  {M.}~\bibnamefont {{Seiffert}}}, \bibinfo {author} {\bibfnamefont {J.~D.}\
  \bibnamefont {{Silverman}}}, \bibinfo {author} {\bibfnamefont
  {L.}~\bibnamefont {{Sodr{\'e}}}}, \bibinfo {author} {\bibfnamefont {D.~N.}\
  \bibnamefont {{Spergel}}}, \bibinfo {author} {\bibfnamefont {M.~A.}\
  \bibnamefont {{Strauss}}}, \bibinfo {author} {\bibfnamefont {H.}~\bibnamefont
  {{Sugai}}}, \bibinfo {author} {\bibfnamefont {Y.}~\bibnamefont {{Suto}}},
  \bibinfo {author} {\bibfnamefont {H.}~\bibnamefont {{Takami}}},\ and\
  \bibinfo {author} {\bibfnamefont {R.}~\bibnamefont {{Wyse}}},\ }\bibfield
  {title} {\bibinfo {title} {{Extragalactic science, cosmology, and Galactic
  archaeology with the Subaru Prime Focus Spectrograph}},\ }\href
  {https://doi.org/10.1093/pasj/pst019} {\bibfield  {journal} {\bibinfo
  {journal} {\pasj}\ }\textbf {\bibinfo {volume} {66}},\ \bibinfo {eid} {R1}
  (\bibinfo {year} {2014})},\ \Eprint {https://arxiv.org/abs/1206.0737}
  {arXiv:1206.0737 [astro-ph.CO]} \BibitemShut {NoStop}%
\bibitem [{\citenamefont {{Auld}}\ \emph {et~al.}(2007)\citenamefont {{Auld}},
  \citenamefont {{Bridges}}, \citenamefont {{Hobson}},\ and\ \citenamefont
  {{Gull}}}]{Auld2007}%
  \BibitemOpen
  \bibfield  {author} {\bibinfo {author} {\bibfnamefont {T.}~\bibnamefont
  {{Auld}}}, \bibinfo {author} {\bibfnamefont {M.}~\bibnamefont {{Bridges}}},
  \bibinfo {author} {\bibfnamefont {M.~P.}\ \bibnamefont {{Hobson}}},\ and\
  \bibinfo {author} {\bibfnamefont {S.~F.}\ \bibnamefont {{Gull}}},\ }\bibfield
   {title} {\bibinfo {title} {{Fast cosmological parameter estimation using
  neural networks}},\ }\href {https://doi.org/10.1111/j.1745-3933.2006.00276.x}
  {\bibfield  {journal} {\bibinfo  {journal} {\mnras}\ }\textbf {\bibinfo
  {volume} {376}},\ \bibinfo {pages} {L11} (\bibinfo {year} {2007})},\ \Eprint
  {https://arxiv.org/abs/astro-ph/0608174} {arXiv:astro-ph/0608174 [astro-ph]}
  \BibitemShut {NoStop}%
\bibitem [{\citenamefont {{Auld}}\ \emph {et~al.}(2008)\citenamefont {{Auld}},
  \citenamefont {{Bridges}},\ and\ \citenamefont {{Hobson}}}]{Auld2008}%
  \BibitemOpen
  \bibfield  {author} {\bibinfo {author} {\bibfnamefont {T.}~\bibnamefont
  {{Auld}}}, \bibinfo {author} {\bibfnamefont {M.}~\bibnamefont {{Bridges}}},\
  and\ \bibinfo {author} {\bibfnamefont {M.~P.}\ \bibnamefont {{Hobson}}},\
  }\bibfield  {title} {\bibinfo {title} {{COSMONET: fast cosmological parameter
  estimation in non-flat models using neural networks}},\ }\href
  {https://doi.org/10.1111/j.1365-2966.2008.13279.x} {\bibfield  {journal}
  {\bibinfo  {journal} {\mnras}\ }\textbf {\bibinfo {volume} {387}},\ \bibinfo
  {pages} {1575} (\bibinfo {year} {2008})},\ \Eprint
  {https://arxiv.org/abs/astro-ph/0703445} {arXiv:astro-ph/0703445 [astro-ph]}
  \BibitemShut {NoStop}%
\bibitem [{\citenamefont {{Aric{\`o}}}\ \emph
  {et~al.}(2021{\natexlab{a}})\citenamefont {{Aric{\`o}}}, \citenamefont
  {{Angulo}},\ and\ \citenamefont {{Zennaro}}}]{Arico2021a}%
  \BibitemOpen
  \bibfield  {author} {\bibinfo {author} {\bibfnamefont {G.}~\bibnamefont
  {{Aric{\`o}}}}, \bibinfo {author} {\bibfnamefont {R.~E.}\ \bibnamefont
  {{Angulo}}},\ and\ \bibinfo {author} {\bibfnamefont {M.}~\bibnamefont
  {{Zennaro}}},\ }\bibfield  {title} {\bibinfo {title} {{Accelerating
  Large-Scale-Structure data analyses by emulating Boltzmann solvers and
  Lagrangian Perturbation Theory}},\ }\href
  {https://doi.org/10.48550/arXiv.2104.14568} {\bibfield  {journal} {\bibinfo
  {journal} {arXiv e-prints}\ ,\ \bibinfo {eid} {arXiv:2104.14568}} (\bibinfo
  {year} {2021}{\natexlab{a}})},\ \Eprint {https://arxiv.org/abs/2104.14568}
  {arXiv:2104.14568 [astro-ph.CO]} \BibitemShut {NoStop}%
\bibitem [{\citenamefont {{Spurio Mancini}}\ \emph {et~al.}(2022)\citenamefont
  {{Spurio Mancini}}, \citenamefont {{Piras}}, \citenamefont {{Alsing}},
  \citenamefont {{Joachimi}},\ and\ \citenamefont {{Hobson}}}]{Spurio2022}%
  \BibitemOpen
  \bibfield  {author} {\bibinfo {author} {\bibfnamefont {A.}~\bibnamefont
  {{Spurio Mancini}}}, \bibinfo {author} {\bibfnamefont {D.}~\bibnamefont
  {{Piras}}}, \bibinfo {author} {\bibfnamefont {J.}~\bibnamefont {{Alsing}}},
  \bibinfo {author} {\bibfnamefont {B.}~\bibnamefont {{Joachimi}}},\ and\
  \bibinfo {author} {\bibfnamefont {M.~P.}\ \bibnamefont {{Hobson}}},\
  }\bibfield  {title} {\bibinfo {title} {{COSMOPOWER: emulating cosmological
  power spectra for accelerated Bayesian inference from next-generation
  surveys}},\ }\href {https://doi.org/10.1093/mnras/stac064} {\bibfield
  {journal} {\bibinfo  {journal} {\mnras}\ }\textbf {\bibinfo {volume} {511}},\
  \bibinfo {pages} {1771} (\bibinfo {year} {2022})},\ \Eprint
  {https://arxiv.org/abs/2106.03846} {arXiv:2106.03846 [astro-ph.CO]}
  \BibitemShut {NoStop}%
\bibitem [{\citenamefont {{Nygaard}}\ \emph {et~al.}(2023)\citenamefont
  {{Nygaard}}, \citenamefont {{Holm}}, \citenamefont {{Hannestad}},\ and\
  \citenamefont {{Tram}}}]{Nygaard2023}%
  \BibitemOpen
  \bibfield  {author} {\bibinfo {author} {\bibfnamefont {A.}~\bibnamefont
  {{Nygaard}}}, \bibinfo {author} {\bibfnamefont {E.~B.}\ \bibnamefont
  {{Holm}}}, \bibinfo {author} {\bibfnamefont {S.}~\bibnamefont
  {{Hannestad}}},\ and\ \bibinfo {author} {\bibfnamefont {T.}~\bibnamefont
  {{Tram}}},\ }\bibfield  {title} {\bibinfo {title} {{CONNECT: a neural network
  based framework for emulating cosmological observables and cosmological
  parameter inference}},\ }\href
  {https://doi.org/10.1088/1475-7516/2023/05/025} {\bibfield  {journal}
  {\bibinfo  {journal} {\jcap}\ }\textbf {\bibinfo {volume} {2023}},\ \bibinfo
  {eid} {025} (\bibinfo {year} {2023})},\ \Eprint
  {https://arxiv.org/abs/2205.15726} {arXiv:2205.15726 [astro-ph.IM]}
  \BibitemShut {NoStop}%
\bibitem [{\citenamefont {{G{\"u}nther}}\ \emph {et~al.}(2022)\citenamefont
  {{G{\"u}nther}}, \citenamefont {{Lesgourgues}}, \citenamefont {{Samaras}},
  \citenamefont {{Sch{\"o}neberg}}, \citenamefont {{Stadtmann}}, \citenamefont
  {{Fidler}},\ and\ \citenamefont {{Torrado}}}]{Gunther2022}%
  \BibitemOpen
  \bibfield  {author} {\bibinfo {author} {\bibfnamefont {S.}~\bibnamefont
  {{G{\"u}nther}}}, \bibinfo {author} {\bibfnamefont {J.}~\bibnamefont
  {{Lesgourgues}}}, \bibinfo {author} {\bibfnamefont {G.}~\bibnamefont
  {{Samaras}}}, \bibinfo {author} {\bibfnamefont {N.}~\bibnamefont
  {{Sch{\"o}neberg}}}, \bibinfo {author} {\bibfnamefont {F.}~\bibnamefont
  {{Stadtmann}}}, \bibinfo {author} {\bibfnamefont {C.}~\bibnamefont
  {{Fidler}}},\ and\ \bibinfo {author} {\bibfnamefont {J.}~\bibnamefont
  {{Torrado}}},\ }\bibfield  {title} {\bibinfo {title} {{CosmicNet II:
  emulating extended cosmologies with efficient and accurate neural
  networks}},\ }\href {https://doi.org/10.1088/1475-7516/2022/11/035}
  {\bibfield  {journal} {\bibinfo  {journal} {\jcap}\ }\textbf {\bibinfo
  {volume} {2022}},\ \bibinfo {eid} {035} (\bibinfo {year} {2022})},\ \Eprint
  {https://arxiv.org/abs/2207.05707} {arXiv:2207.05707 [astro-ph.CO]}
  \BibitemShut {NoStop}%
\bibitem [{\citenamefont {{Bonici}}\ \emph {et~al.}(2024)\citenamefont
  {{Bonici}}, \citenamefont {{Bianchini}},\ and\ \citenamefont
  {{Ruiz-Zapatero}}}]{Bonici2024}%
  \BibitemOpen
  \bibfield  {author} {\bibinfo {author} {\bibfnamefont {M.}~\bibnamefont
  {{Bonici}}}, \bibinfo {author} {\bibfnamefont {F.}~\bibnamefont
  {{Bianchini}}},\ and\ \bibinfo {author} {\bibfnamefont {J.}~\bibnamefont
  {{Ruiz-Zapatero}}},\ }\bibfield  {title} {\bibinfo {title} {{Capse.jl:
  efficient and auto-differentiable CMB power spectra emulation}},\ }\href
  {https://doi.org/10.21105/astro.2307.14339} {\bibfield  {journal} {\bibinfo
  {journal} {The Open Journal of Astrophysics}\ }\textbf {\bibinfo {volume}
  {7}},\ \bibinfo {eid} {10} (\bibinfo {year} {2024})},\ \Eprint
  {https://arxiv.org/abs/2307.14339} {arXiv:2307.14339 [astro-ph.CO]}
  \BibitemShut {NoStop}%
\bibitem [{\citenamefont {{Bonici}}\ \emph {et~al.}(2025)\citenamefont
  {{Bonici}}, \citenamefont {{D'Amico}}, \citenamefont {{Bel}},\ and\
  \citenamefont {{Carbone}}}]{Bonici2025}%
  \BibitemOpen
  \bibfield  {author} {\bibinfo {author} {\bibfnamefont {M.}~\bibnamefont
  {{Bonici}}}, \bibinfo {author} {\bibfnamefont {G.}~\bibnamefont {{D'Amico}}},
  \bibinfo {author} {\bibfnamefont {J.}~\bibnamefont {{Bel}}},\ and\ \bibinfo
  {author} {\bibfnamefont {C.}~\bibnamefont {{Carbone}}},\ }\bibfield  {title}
  {\bibinfo {title} {{Effort: a fast and differentiable emulator for the
  Effective Field Theory of the Large Scale Structure of the Universe}},\
  }\href {https://doi.org/10.48550/arXiv.2501.04639} {\bibfield  {journal}
  {\bibinfo  {journal} {arXiv e-prints}\ ,\ \bibinfo {eid} {arXiv:2501.04639}}
  (\bibinfo {year} {2025})},\ \Eprint {https://arxiv.org/abs/2501.04639}
  {arXiv:2501.04639 [astro-ph.CO]} \BibitemShut {NoStop}%
\bibitem [{\citenamefont {Heitmann}\ \emph {et~al.}(2009)\citenamefont
  {Heitmann}, \citenamefont {Higdon}, \citenamefont {White}, \citenamefont
  {Habib}, \citenamefont {Williams}, \citenamefont {Lawrence},\ and\
  \citenamefont {Wagner}}]{Heitmann2009}%
  \BibitemOpen
  \bibfield  {author} {\bibinfo {author} {\bibfnamefont {K.}~\bibnamefont
  {Heitmann}}, \bibinfo {author} {\bibfnamefont {D.}~\bibnamefont {Higdon}},
  \bibinfo {author} {\bibfnamefont {M.}~\bibnamefont {White}}, \bibinfo
  {author} {\bibfnamefont {S.}~\bibnamefont {Habib}}, \bibinfo {author}
  {\bibfnamefont {B.~J.}\ \bibnamefont {Williams}}, \bibinfo {author}
  {\bibfnamefont {E.}~\bibnamefont {Lawrence}},\ and\ \bibinfo {author}
  {\bibfnamefont {C.}~\bibnamefont {Wagner}},\ }\bibfield  {title} {\bibinfo
  {title} {{THE COYOTE UNIVERSE. II. COSMOLOGICAL MODELS AND PRECISION
  EMULATION OF THE NONLINEAR MATTER POWER SPECTRUM}},\ }\href
  {https://doi.org/10.1088/0004-637x/705/1/156} {\bibfield  {journal} {\bibinfo
   {journal} {The Astrophysical Journal}\ }\textbf {\bibinfo {volume} {705}},\
  \bibinfo {pages} {156} (\bibinfo {year} {2009})}\BibitemShut {NoStop}%
\bibitem [{\citenamefont {Heitmann}\ \emph {et~al.}(2010)\citenamefont
  {Heitmann}, \citenamefont {White}, \citenamefont {Wagner}, \citenamefont
  {Habib},\ and\ \citenamefont {Higdon}}]{Heitmann2010}%
  \BibitemOpen
  \bibfield  {author} {\bibinfo {author} {\bibfnamefont {K.}~\bibnamefont
  {Heitmann}}, \bibinfo {author} {\bibfnamefont {M.}~\bibnamefont {White}},
  \bibinfo {author} {\bibfnamefont {C.}~\bibnamefont {Wagner}}, \bibinfo
  {author} {\bibfnamefont {S.}~\bibnamefont {Habib}},\ and\ \bibinfo {author}
  {\bibfnamefont {D.}~\bibnamefont {Higdon}},\ }\bibfield  {title} {\bibinfo
  {title} {{THE COYOTE UNIVERSE. I. PRECISION DETERMINATION OF THE NONLINEAR
  MATTER POWER SPECTRUM}},\ }\href
  {https://doi.org/10.1088/0004-637x/715/1/104} {\bibfield  {journal} {\bibinfo
   {journal} {The Astrophysical Journal}\ }\textbf {\bibinfo {volume} {715}},\
  \bibinfo {pages} {104} (\bibinfo {year} {2010})}\BibitemShut {NoStop}%
\bibitem [{\citenamefont {Heitmann}\ \emph {et~al.}(2013)\citenamefont
  {Heitmann}, \citenamefont {Lawrence}, \citenamefont {Kwan}, \citenamefont
  {Habib},\ and\ \citenamefont {Higdon}}]{Heitmann2013}%
  \BibitemOpen
  \bibfield  {author} {\bibinfo {author} {\bibfnamefont {K.}~\bibnamefont
  {Heitmann}}, \bibinfo {author} {\bibfnamefont {E.}~\bibnamefont {Lawrence}},
  \bibinfo {author} {\bibfnamefont {J.}~\bibnamefont {Kwan}}, \bibinfo {author}
  {\bibfnamefont {S.}~\bibnamefont {Habib}},\ and\ \bibinfo {author}
  {\bibfnamefont {D.}~\bibnamefont {Higdon}},\ }\bibfield  {title} {\bibinfo
  {title} {{THE COYOTE UNIVERSE EXTENDED: PRECISION EMULATION OF THE MATTER
  POWER SPECTRUM}},\ }\href {https://doi.org/10.1088/0004-637x/780/1/111}
  {\bibfield  {journal} {\bibinfo  {journal} {The Astrophysical Journal}\
  }\textbf {\bibinfo {volume} {780}},\ \bibinfo {pages} {111} (\bibinfo {year}
  {2013})}\BibitemShut {NoStop}%
\bibitem [{\citenamefont {DeRose}\ \emph {et~al.}(2019)\citenamefont {DeRose},
  \citenamefont {Wechsler}, \citenamefont {Tinker}, \citenamefont {Becker},
  \citenamefont {Mao}, \citenamefont {McClintock}, \citenamefont {McLaughlin},
  \citenamefont {Rozo},\ and\ \citenamefont {Zhai}}]{DeRose2019}%
  \BibitemOpen
  \bibfield  {author} {\bibinfo {author} {\bibfnamefont {J.}~\bibnamefont
  {DeRose}}, \bibinfo {author} {\bibfnamefont {R.~H.}\ \bibnamefont
  {Wechsler}}, \bibinfo {author} {\bibfnamefont {J.~L.}\ \bibnamefont
  {Tinker}}, \bibinfo {author} {\bibfnamefont {M.~R.}\ \bibnamefont {Becker}},
  \bibinfo {author} {\bibfnamefont {Y.-Y.}\ \bibnamefont {Mao}}, \bibinfo
  {author} {\bibfnamefont {T.}~\bibnamefont {McClintock}}, \bibinfo {author}
  {\bibfnamefont {S.}~\bibnamefont {McLaughlin}}, \bibinfo {author}
  {\bibfnamefont {E.}~\bibnamefont {Rozo}},\ and\ \bibinfo {author}
  {\bibfnamefont {Z.}~\bibnamefont {Zhai}},\ }\bibfield  {title} {\bibinfo
  {title} {{The Aemulus Project. I. Numerical Simulations for Precision
  Cosmology}},\ }\href {https://doi.org/10.3847/1538-4357/ab1085} {\bibfield
  {journal} {\bibinfo  {journal} {The Astrophysical Journal}\ }\textbf
  {\bibinfo {volume} {875}},\ \bibinfo {pages} {69} (\bibinfo {year}
  {2019})}\BibitemShut {NoStop}%
\bibitem [{\citenamefont {McClintock}\ \emph {et~al.}(2019)\citenamefont
  {McClintock}, \citenamefont {Rozo}, \citenamefont {Becker}, \citenamefont
  {DeRose}, \citenamefont {Mao}, \citenamefont {McLaughlin}, \citenamefont
  {Tinker}, \citenamefont {Wechsler},\ and\ \citenamefont
  {Zhai}}]{McClintock2019}%
  \BibitemOpen
  \bibfield  {author} {\bibinfo {author} {\bibfnamefont {T.}~\bibnamefont
  {McClintock}}, \bibinfo {author} {\bibfnamefont {E.}~\bibnamefont {Rozo}},
  \bibinfo {author} {\bibfnamefont {M.~R.}\ \bibnamefont {Becker}}, \bibinfo
  {author} {\bibfnamefont {J.}~\bibnamefont {DeRose}}, \bibinfo {author}
  {\bibfnamefont {Y.-Y.}\ \bibnamefont {Mao}}, \bibinfo {author} {\bibfnamefont
  {S.}~\bibnamefont {McLaughlin}}, \bibinfo {author} {\bibfnamefont {J.~L.}\
  \bibnamefont {Tinker}}, \bibinfo {author} {\bibfnamefont {R.~H.}\
  \bibnamefont {Wechsler}},\ and\ \bibinfo {author} {\bibfnamefont
  {Z.}~\bibnamefont {Zhai}},\ }\bibfield  {title} {\bibinfo {title} {{The
  Aemulus Project. II. Emulating the Halo Mass Function}},\ }\href
  {https://doi.org/10.3847/1538-4357/aaf568} {\bibfield  {journal} {\bibinfo
  {journal} {The Astrophysical Journal}\ }\textbf {\bibinfo {volume} {872}},\
  \bibinfo {pages} {53} (\bibinfo {year} {2019})}\BibitemShut {NoStop}%
\bibitem [{\citenamefont {Zhai}\ \emph {et~al.}(2019)\citenamefont {Zhai},
  \citenamefont {Tinker}, \citenamefont {Becker}, \citenamefont {DeRose},
  \citenamefont {Mao}, \citenamefont {McClintock}, \citenamefont {McLaughlin},
  \citenamefont {Rozo},\ and\ \citenamefont {Wechsler}}]{Zhai2019}%
  \BibitemOpen
  \bibfield  {author} {\bibinfo {author} {\bibfnamefont {Z.}~\bibnamefont
  {Zhai}}, \bibinfo {author} {\bibfnamefont {J.~L.}\ \bibnamefont {Tinker}},
  \bibinfo {author} {\bibfnamefont {M.~R.}\ \bibnamefont {Becker}}, \bibinfo
  {author} {\bibfnamefont {J.}~\bibnamefont {DeRose}}, \bibinfo {author}
  {\bibfnamefont {Y.-Y.}\ \bibnamefont {Mao}}, \bibinfo {author} {\bibfnamefont
  {T.}~\bibnamefont {McClintock}}, \bibinfo {author} {\bibfnamefont
  {S.}~\bibnamefont {McLaughlin}}, \bibinfo {author} {\bibfnamefont
  {E.}~\bibnamefont {Rozo}},\ and\ \bibinfo {author} {\bibfnamefont {R.~H.}\
  \bibnamefont {Wechsler}},\ }\bibfield  {title} {\bibinfo {title} {{The
  Aemulus Project. III. Emulation of the Galaxy Correlation Function}},\ }\href
  {https://doi.org/10.3847/1538-4357/ab0d7b} {\bibfield  {journal} {\bibinfo
  {journal} {The Astrophysical Journal}\ }\textbf {\bibinfo {volume} {874}},\
  \bibinfo {pages} {95} (\bibinfo {year} {2019})}\BibitemShut {NoStop}%
\bibitem [{\citenamefont {Smith}\ and\ \citenamefont
  {Angulo}(2019)}]{Smith2019}%
  \BibitemOpen
  \bibfield  {author} {\bibinfo {author} {\bibfnamefont {R.~E.}\ \bibnamefont
  {Smith}}\ and\ \bibinfo {author} {\bibfnamefont {R.~E.}\ \bibnamefont
  {Angulo}},\ }\bibfield  {title} {\bibinfo {title} {{Precision modelling of
  the matter power spectrum in a Planck-like Universe}},\ }\href
  {https://doi.org/10.1093/mnras/stz890} {\bibfield  {journal} {\bibinfo
  {journal} {Monthly Notices of the Royal Astronomical Society}\ }\textbf
  {\bibinfo {volume} {486}},\ \bibinfo {pages} {1448} (\bibinfo {year}
  {2019})}\BibitemShut {NoStop}%
\bibitem [{\citenamefont {Nishimichi}\ \emph {et~al.}(2019)\citenamefont
  {Nishimichi}, \citenamefont {Takada}, \citenamefont {Takahashi},
  \citenamefont {Osato}, \citenamefont {Shirasaki}, \citenamefont {Oogi},
  \citenamefont {Miyatake}, \citenamefont {Oguri}, \citenamefont {Murata},
  \citenamefont {Kobayashi},\ and\ \citenamefont {Yoshida}}]{Nishimichi2019}%
  \BibitemOpen
  \bibfield  {author} {\bibinfo {author} {\bibfnamefont {T.}~\bibnamefont
  {Nishimichi}}, \bibinfo {author} {\bibfnamefont {M.}~\bibnamefont {Takada}},
  \bibinfo {author} {\bibfnamefont {R.}~\bibnamefont {Takahashi}}, \bibinfo
  {author} {\bibfnamefont {K.}~\bibnamefont {Osato}}, \bibinfo {author}
  {\bibfnamefont {M.}~\bibnamefont {Shirasaki}}, \bibinfo {author}
  {\bibfnamefont {T.}~\bibnamefont {Oogi}}, \bibinfo {author} {\bibfnamefont
  {H.}~\bibnamefont {Miyatake}}, \bibinfo {author} {\bibfnamefont
  {M.}~\bibnamefont {Oguri}}, \bibinfo {author} {\bibfnamefont
  {R.}~\bibnamefont {Murata}}, \bibinfo {author} {\bibfnamefont
  {Y.}~\bibnamefont {Kobayashi}},\ and\ \bibinfo {author} {\bibfnamefont
  {N.}~\bibnamefont {Yoshida}},\ }\bibfield  {title} {\bibinfo {title} {{Dark
  Quest. I. Fast and Accurate Emulation of Halo Clustering Statistics and Its
  Application to Galaxy Clustering}},\ }\href
  {https://doi.org/10.3847/1538-4357/ab3719} {\bibfield  {journal} {\bibinfo
  {journal} {The Astrophysical Journal}\ }\textbf {\bibinfo {volume} {884}},\
  \bibinfo {pages} {29} (\bibinfo {year} {2019})}\BibitemShut {NoStop}%
\bibitem [{\citenamefont {Valcin}\ \emph {et~al.}(2019)\citenamefont {Valcin},
  \citenamefont {Villaescusa-Navarro}, \citenamefont {Verde},\ and\
  \citenamefont {Raccanelli}}]{Valcin2019}%
  \BibitemOpen
  \bibfield  {author} {\bibinfo {author} {\bibfnamefont {D.}~\bibnamefont
  {Valcin}}, \bibinfo {author} {\bibfnamefont {F.}~\bibnamefont
  {Villaescusa-Navarro}}, \bibinfo {author} {\bibfnamefont {L.}~\bibnamefont
  {Verde}},\ and\ \bibinfo {author} {\bibfnamefont {A.}~\bibnamefont
  {Raccanelli}},\ }\bibfield  {title} {\bibinfo {title} {{BE-HaPPY: bias
  emulator for halo power spectrum including massive neutrinos}},\ }\href
  {https://doi.org/10.1088/1475-7516/2019/12/057} {\bibfield  {journal}
  {\bibinfo  {journal} {Journal of Cosmology and Astroparticle Physics}\
  }\textbf {\bibinfo {volume} {2019}}\bibinfo  {number} { (12)},\ \bibinfo
  {pages} {057}}\BibitemShut {NoStop}%
\bibitem [{\citenamefont {{Aric{\`o}}}\ \emph
  {et~al.}(2021{\natexlab{b}})\citenamefont {{Aric{\`o}}}, \citenamefont
  {{Angulo}}, \citenamefont {{Contreras}}, \citenamefont {{Ondaro-Mallea}},
  \citenamefont {{Pellejero-Iba{\~n}ez}},\ and\ \citenamefont
  {{Zennaro}}}]{Arico2021}%
  \BibitemOpen
\bibfield  {number} {  }\bibfield  {author} {\bibinfo {author} {\bibfnamefont
  {G.}~\bibnamefont {{Aric{\`o}}}}, \bibinfo {author} {\bibfnamefont {R.~E.}\
  \bibnamefont {{Angulo}}}, \bibinfo {author} {\bibfnamefont {S.}~\bibnamefont
  {{Contreras}}}, \bibinfo {author} {\bibfnamefont {L.}~\bibnamefont
  {{Ondaro-Mallea}}}, \bibinfo {author} {\bibfnamefont {M.}~\bibnamefont
  {{Pellejero-Iba{\~n}ez}}},\ and\ \bibinfo {author} {\bibfnamefont
  {M.}~\bibnamefont {{Zennaro}}},\ }\bibfield  {title} {\bibinfo {title} {{The
  BACCO simulation project: a baryonification emulator with neural networks}},\
  }\href {https://doi.org/10.1093/mnras/stab1911} {\bibfield  {journal}
  {\bibinfo  {journal} {\mnras}\ }\textbf {\bibinfo {volume} {506}},\ \bibinfo
  {pages} {4070} (\bibinfo {year} {2021}{\natexlab{b}})},\ \Eprint
  {https://arxiv.org/abs/2011.15018} {arXiv:2011.15018 [astro-ph.CO]}
  \BibitemShut {NoStop}%
\bibitem [{\citenamefont {{Villaescusa-Navarro}}\ \emph
  {et~al.}(2020)\citenamefont {{Villaescusa-Navarro}}, \citenamefont {{Hahn}},
  \citenamefont {{Massara}}, \citenamefont {{Banerjee}}, \citenamefont
  {{Delgado}}, \citenamefont {{Ramanah}}, \citenamefont {{Charnock}},
  \citenamefont {{Giusarma}}, \citenamefont {{Li}}, \citenamefont {{Allys}},
  \citenamefont {{Brochard}}, \citenamefont {{Uhlemann}}, \citenamefont
  {{Chiang}}, \citenamefont {{He}}, \citenamefont {{Pisani}}, \citenamefont
  {{Obuljen}}, \citenamefont {{Feng}}, \citenamefont {{Castorina}},
  \citenamefont {{Contardo}}, \citenamefont {{Kreisch}}, \citenamefont
  {{Nicola}}, \citenamefont {{Alsing}}, \citenamefont {{Scoccimarro}},
  \citenamefont {{Verde}}, \citenamefont {{Viel}}, \citenamefont {{Ho}},
  \citenamefont {{Mallat}}, \citenamefont {{Wandelt}},\ and\ \citenamefont
  {{Spergel}}}]{Villaescusa2020}%
  \BibitemOpen
  \bibfield  {author} {\bibinfo {author} {\bibfnamefont {F.}~\bibnamefont
  {{Villaescusa-Navarro}}}, \bibinfo {author} {\bibfnamefont {C.}~\bibnamefont
  {{Hahn}}}, \bibinfo {author} {\bibfnamefont {E.}~\bibnamefont {{Massara}}},
  \bibinfo {author} {\bibfnamefont {A.}~\bibnamefont {{Banerjee}}}, \bibinfo
  {author} {\bibfnamefont {A.~M.}\ \bibnamefont {{Delgado}}}, \bibinfo {author}
  {\bibfnamefont {D.~K.}\ \bibnamefont {{Ramanah}}}, \bibinfo {author}
  {\bibfnamefont {T.}~\bibnamefont {{Charnock}}}, \bibinfo {author}
  {\bibfnamefont {E.}~\bibnamefont {{Giusarma}}}, \bibinfo {author}
  {\bibfnamefont {Y.}~\bibnamefont {{Li}}}, \bibinfo {author} {\bibfnamefont
  {E.}~\bibnamefont {{Allys}}}, \bibinfo {author} {\bibfnamefont
  {A.}~\bibnamefont {{Brochard}}}, \bibinfo {author} {\bibfnamefont
  {C.}~\bibnamefont {{Uhlemann}}}, \bibinfo {author} {\bibfnamefont {C.-T.}\
  \bibnamefont {{Chiang}}}, \bibinfo {author} {\bibfnamefont {S.}~\bibnamefont
  {{He}}}, \bibinfo {author} {\bibfnamefont {A.}~\bibnamefont {{Pisani}}},
  \bibinfo {author} {\bibfnamefont {A.}~\bibnamefont {{Obuljen}}}, \bibinfo
  {author} {\bibfnamefont {Y.}~\bibnamefont {{Feng}}}, \bibinfo {author}
  {\bibfnamefont {E.}~\bibnamefont {{Castorina}}}, \bibinfo {author}
  {\bibfnamefont {G.}~\bibnamefont {{Contardo}}}, \bibinfo {author}
  {\bibfnamefont {C.~D.}\ \bibnamefont {{Kreisch}}}, \bibinfo {author}
  {\bibfnamefont {A.}~\bibnamefont {{Nicola}}}, \bibinfo {author}
  {\bibfnamefont {J.}~\bibnamefont {{Alsing}}}, \bibinfo {author}
  {\bibfnamefont {R.}~\bibnamefont {{Scoccimarro}}}, \bibinfo {author}
  {\bibfnamefont {L.}~\bibnamefont {{Verde}}}, \bibinfo {author} {\bibfnamefont
  {M.}~\bibnamefont {{Viel}}}, \bibinfo {author} {\bibfnamefont
  {S.}~\bibnamefont {{Ho}}}, \bibinfo {author} {\bibfnamefont {S.}~\bibnamefont
  {{Mallat}}}, \bibinfo {author} {\bibfnamefont {B.}~\bibnamefont
  {{Wandelt}}},\ and\ \bibinfo {author} {\bibfnamefont {D.~N.}\ \bibnamefont
  {{Spergel}}},\ }\bibfield  {title} {\bibinfo {title} {{The Quijote
  Simulations}},\ }\href {https://doi.org/10.3847/1538-4365/ab9d82} {\bibfield
  {journal} {\bibinfo  {journal} {\apjs}\ }\textbf {\bibinfo {volume} {250}},\
  \bibinfo {eid} {2} (\bibinfo {year} {2020})},\ \Eprint
  {https://arxiv.org/abs/1909.05273} {arXiv:1909.05273 [astro-ph.CO]}
  \BibitemShut {NoStop}%
\bibitem [{\citenamefont {Heitmann}\ \emph {et~al.}(2016)\citenamefont
  {Heitmann}, \citenamefont {Bingham}, \citenamefont {Lawrence}, \citenamefont
  {Bergner}, \citenamefont {Habib}, \citenamefont {Higdon}, \citenamefont
  {Pope}, \citenamefont {Biswas}, \citenamefont {Finkel}, \citenamefont
  {Frontiere},\ and\ \citenamefont {Bhattacharya}}]{Heitmann2016}%
  \BibitemOpen
  \bibfield  {author} {\bibinfo {author} {\bibfnamefont {K.}~\bibnamefont
  {Heitmann}}, \bibinfo {author} {\bibfnamefont {D.}~\bibnamefont {Bingham}},
  \bibinfo {author} {\bibfnamefont {E.}~\bibnamefont {Lawrence}}, \bibinfo
  {author} {\bibfnamefont {S.}~\bibnamefont {Bergner}}, \bibinfo {author}
  {\bibfnamefont {S.}~\bibnamefont {Habib}}, \bibinfo {author} {\bibfnamefont
  {D.}~\bibnamefont {Higdon}}, \bibinfo {author} {\bibfnamefont
  {A.}~\bibnamefont {Pope}}, \bibinfo {author} {\bibfnamefont {R.}~\bibnamefont
  {Biswas}}, \bibinfo {author} {\bibfnamefont {H.}~\bibnamefont {Finkel}},
  \bibinfo {author} {\bibfnamefont {N.}~\bibnamefont {Frontiere}},\ and\
  \bibinfo {author} {\bibfnamefont {S.}~\bibnamefont {Bhattacharya}},\
  }\bibfield  {title} {\bibinfo {title} {{THE MIRA–TITAN UNIVERSE: PRECISION
  PREDICTIONS FOR DARK ENERGY SURVEYS}},\ }\href
  {https://doi.org/10.3847/0004-637x/820/2/108} {\bibfield  {journal} {\bibinfo
   {journal} {The Astrophysical Journal}\ }\textbf {\bibinfo {volume} {820}},\
  \bibinfo {pages} {108} (\bibinfo {year} {2016})}\BibitemShut {NoStop}%
\bibitem [{\citenamefont {Lawrence}\ \emph {et~al.}(2017)\citenamefont
  {Lawrence}, \citenamefont {Heitmann}, \citenamefont {Kwan}, \citenamefont
  {Upadhye}, \citenamefont {Bingham}, \citenamefont {Habib}, \citenamefont
  {Higdon}, \citenamefont {Pope}, \citenamefont {Finkel},\ and\ \citenamefont
  {Frontiere}}]{Lawrence2017}%
  \BibitemOpen
  \bibfield  {author} {\bibinfo {author} {\bibfnamefont {E.}~\bibnamefont
  {Lawrence}}, \bibinfo {author} {\bibfnamefont {K.}~\bibnamefont {Heitmann}},
  \bibinfo {author} {\bibfnamefont {J.}~\bibnamefont {Kwan}}, \bibinfo {author}
  {\bibfnamefont {A.}~\bibnamefont {Upadhye}}, \bibinfo {author} {\bibfnamefont
  {D.}~\bibnamefont {Bingham}}, \bibinfo {author} {\bibfnamefont
  {S.}~\bibnamefont {Habib}}, \bibinfo {author} {\bibfnamefont
  {D.}~\bibnamefont {Higdon}}, \bibinfo {author} {\bibfnamefont
  {A.}~\bibnamefont {Pope}}, \bibinfo {author} {\bibfnamefont {H.}~\bibnamefont
  {Finkel}},\ and\ \bibinfo {author} {\bibfnamefont {N.}~\bibnamefont
  {Frontiere}},\ }\bibfield  {title} {\bibinfo {title} {{The Mira-Titan
  Universe. II. Matter Power Spectrum Emulation}},\ }\href
  {https://doi.org/10.3847/1538-4357/aa86a9} {\bibfield  {journal} {\bibinfo
  {journal} {The Astrophysical Journal}\ }\textbf {\bibinfo {volume} {847}},\
  \bibinfo {pages} {50} (\bibinfo {year} {2017})}\BibitemShut {NoStop}%
\bibitem [{\citenamefont {{Bocquet}}\ \emph {et~al.}(2020)\citenamefont
  {{Bocquet}}, \citenamefont {{Heitmann}}, \citenamefont {{Habib}},
  \citenamefont {{Lawrence}}, \citenamefont {{Uram}}, \citenamefont
  {{Frontiere}}, \citenamefont {{Pope}},\ and\ \citenamefont
  {{Finkel}}}]{Bocquet2020}%
  \BibitemOpen
  \bibfield  {author} {\bibinfo {author} {\bibfnamefont {S.}~\bibnamefont
  {{Bocquet}}}, \bibinfo {author} {\bibfnamefont {K.}~\bibnamefont
  {{Heitmann}}}, \bibinfo {author} {\bibfnamefont {S.}~\bibnamefont {{Habib}}},
  \bibinfo {author} {\bibfnamefont {E.}~\bibnamefont {{Lawrence}}}, \bibinfo
  {author} {\bibfnamefont {T.}~\bibnamefont {{Uram}}}, \bibinfo {author}
  {\bibfnamefont {N.}~\bibnamefont {{Frontiere}}}, \bibinfo {author}
  {\bibfnamefont {A.}~\bibnamefont {{Pope}}},\ and\ \bibinfo {author}
  {\bibfnamefont {H.}~\bibnamefont {{Finkel}}},\ }\bibfield  {title} {\bibinfo
  {title} {{The Mira-Titan Universe. III. Emulation of the Halo Mass
  Function}},\ }\href {https://doi.org/10.3847/1538-4357/abac5c} {\bibfield
  {journal} {\bibinfo  {journal} {\apj}\ }\textbf {\bibinfo {volume} {901}},\
  \bibinfo {eid} {5} (\bibinfo {year} {2020})},\ \Eprint
  {https://arxiv.org/abs/2003.12116} {arXiv:2003.12116 [astro-ph.CO]}
  \BibitemShut {NoStop}%
\bibitem [{\citenamefont {{Moran}}\ \emph {et~al.}(2023)\citenamefont
  {{Moran}}, \citenamefont {{Heitmann}}, \citenamefont {{Lawrence}},
  \citenamefont {{Habib}}, \citenamefont {{Bingham}}, \citenamefont
  {{Upadhye}}, \citenamefont {{Kwan}}, \citenamefont {{Higdon}},\ and\
  \citenamefont {{Payne}}}]{Moran2023}%
  \BibitemOpen
  \bibfield  {author} {\bibinfo {author} {\bibfnamefont {K.~R.}\ \bibnamefont
  {{Moran}}}, \bibinfo {author} {\bibfnamefont {K.}~\bibnamefont {{Heitmann}}},
  \bibinfo {author} {\bibfnamefont {E.}~\bibnamefont {{Lawrence}}}, \bibinfo
  {author} {\bibfnamefont {S.}~\bibnamefont {{Habib}}}, \bibinfo {author}
  {\bibfnamefont {D.}~\bibnamefont {{Bingham}}}, \bibinfo {author}
  {\bibfnamefont {A.}~\bibnamefont {{Upadhye}}}, \bibinfo {author}
  {\bibfnamefont {J.}~\bibnamefont {{Kwan}}}, \bibinfo {author} {\bibfnamefont
  {D.}~\bibnamefont {{Higdon}}},\ and\ \bibinfo {author} {\bibfnamefont
  {R.}~\bibnamefont {{Payne}}},\ }\bibfield  {title} {\bibinfo {title} {{The
  Mira-Titan Universe - IV. High-precision power spectrum emulation}},\ }\href
  {https://doi.org/10.1093/mnras/stac3452} {\bibfield  {journal} {\bibinfo
  {journal} {\mnras}\ }\textbf {\bibinfo {volume} {520}},\ \bibinfo {pages}
  {3443} (\bibinfo {year} {2023})},\ \Eprint {https://arxiv.org/abs/2207.12345}
  {arXiv:2207.12345 [astro-ph.CO]} \BibitemShut {NoStop}%
\bibitem [{\citenamefont {{Kwan}}\ \emph {et~al.}(2023)\citenamefont {{Kwan}},
  \citenamefont {{Saito}}, \citenamefont {{Leauthaud}}, \citenamefont
  {{Heitmann}}, \citenamefont {{Habib}}, \citenamefont {{Frontiere}},
  \citenamefont {{Guo}}, \citenamefont {{Huang}}, \citenamefont {{Pope}},\ and\
  \citenamefont {{Rodrigu{\'e}z-Torres}}}]{Kwan2023}%
  \BibitemOpen
  \bibfield  {author} {\bibinfo {author} {\bibfnamefont {J.}~\bibnamefont
  {{Kwan}}}, \bibinfo {author} {\bibfnamefont {S.}~\bibnamefont {{Saito}}},
  \bibinfo {author} {\bibfnamefont {A.}~\bibnamefont {{Leauthaud}}}, \bibinfo
  {author} {\bibfnamefont {K.}~\bibnamefont {{Heitmann}}}, \bibinfo {author}
  {\bibfnamefont {S.}~\bibnamefont {{Habib}}}, \bibinfo {author} {\bibfnamefont
  {N.}~\bibnamefont {{Frontiere}}}, \bibinfo {author} {\bibfnamefont
  {H.}~\bibnamefont {{Guo}}}, \bibinfo {author} {\bibfnamefont
  {S.}~\bibnamefont {{Huang}}}, \bibinfo {author} {\bibfnamefont
  {A.}~\bibnamefont {{Pope}}},\ and\ \bibinfo {author} {\bibfnamefont
  {S.}~\bibnamefont {{Rodrigu{\'e}z-Torres}}},\ }\bibfield  {title} {\bibinfo
  {title} {{Galaxy Clustering in the Mira-Titan Universe. I. Emulators for the
  Redshift Space Galaxy Correlation Function and Galaxy-Galaxy Lensing}},\
  }\href {https://doi.org/10.3847/1538-4357/acd92f} {\bibfield  {journal}
  {\bibinfo  {journal} {\apj}\ }\textbf {\bibinfo {volume} {952}},\ \bibinfo
  {eid} {80} (\bibinfo {year} {2023})},\ \Eprint
  {https://arxiv.org/abs/2302.12379} {arXiv:2302.12379 [astro-ph.CO]}
  \BibitemShut {NoStop}%
\bibitem [{\citenamefont {{S{\'a}ez-Casares}}\ \emph
  {et~al.}(2024)\citenamefont {{S{\'a}ez-Casares}}, \citenamefont {{Rasera}},
  \citenamefont {{Richardson}},\ and\ \citenamefont
  {{Corasaniti}}}]{Casares2024}%
  \BibitemOpen
  \bibfield  {author} {\bibinfo {author} {\bibfnamefont {I.}~\bibnamefont
  {{S{\'a}ez-Casares}}}, \bibinfo {author} {\bibfnamefont {Y.}~\bibnamefont
  {{Rasera}}}, \bibinfo {author} {\bibfnamefont {T.~R.~G.}\ \bibnamefont
  {{Richardson}}},\ and\ \bibinfo {author} {\bibfnamefont {P.~S.}\ \bibnamefont
  {{Corasaniti}}},\ }\bibfield  {title} {\bibinfo {title} {{The e-MANTIS
  emulator: Fast and accurate predictions of the halo mass function in f(R)CDM
  and wCDM cosmologies}},\ }\href {https://doi.org/10.1051/0004-6361/202450193}
  {\bibfield  {journal} {\bibinfo  {journal} {\aap}\ }\textbf {\bibinfo
  {volume} {691}},\ \bibinfo {eid} {A323} (\bibinfo {year} {2024})},\ \Eprint
  {https://arxiv.org/abs/2410.05226} {arXiv:2410.05226 [astro-ph.CO]}
  \BibitemShut {NoStop}%
\bibitem [{\citenamefont {{Euclid Collaboration}}\ \emph
  {et~al.}(2019)\citenamefont {{Euclid Collaboration}}, \citenamefont
  {{Knabenhans}},\ and\ \citenamefont {{Stadel et al.}}}]{Knabenhans2019}%
  \BibitemOpen
  \bibfield  {author} {\bibinfo {author} {\bibnamefont {{Euclid
  Collaboration}}}, \bibinfo {author} {\bibfnamefont {M.}~\bibnamefont
  {{Knabenhans}}},\ and\ \bibinfo {author} {\bibfnamefont {J.}~\bibnamefont
  {{Stadel et al.}}},\ }\bibfield  {title} {\bibinfo {title} {{Euclid
  preparation: II. The EuclidEmulator – a tool to compute the cosmology
  dependence of the nonlinear matter power spectrum}},\ }\href
  {https://doi.org/10.1093/mnras/stz197} {\bibfield  {journal} {\bibinfo
  {journal} {Monthly Notices of the Royal Astronomical Society}\ }\textbf
  {\bibinfo {volume} {484}},\ \bibinfo {pages} {5509} (\bibinfo {year}
  {2019})},\ \Eprint
  {https://arxiv.org/abs/https://academic.oup.com/mnras/article-pdf/484/4/5509/27790453/stz197.pdf}
  {https://academic.oup.com/mnras/article-pdf/484/4/5509/27790453/stz197.pdf}
  \BibitemShut {NoStop}%
\bibitem [{\citenamefont {{Euclid Collaboration}}\ \emph
  {et~al.}(2021)\citenamefont {{Euclid Collaboration}}, \citenamefont
  {{Knabenhans}},\ and\ \citenamefont {{Stadel et al.}}}]{Knabenhans2021}%
  \BibitemOpen
  \bibfield  {author} {\bibinfo {author} {\bibnamefont {{Euclid
  Collaboration}}}, \bibinfo {author} {\bibfnamefont {M.}~\bibnamefont
  {{Knabenhans}}},\ and\ \bibinfo {author} {\bibfnamefont {J.}~\bibnamefont
  {{Stadel et al.}}},\ }\bibfield  {title} {\bibinfo {title} {{Euclid
  preparation: IX. EuclidEmulator2 – power spectrum emulation with massive
  neutrinos and self-consistent dark energy perturbations}},\ }\href
  {https://doi.org/10.1093/mnras/stab1366} {\bibfield  {journal} {\bibinfo
  {journal} {Monthly Notices of the Royal Astronomical Society}\ }\textbf
  {\bibinfo {volume} {505}},\ \bibinfo {pages} {2840} (\bibinfo {year}
  {2021})}\BibitemShut {NoStop}%
\bibitem [{\citenamefont {{Chen}}\ \emph {et~al.}(2025)\citenamefont {{Chen}},
  \citenamefont {{Yu}}, \citenamefont {{Han}},\ and\ \citenamefont
  {{Jing}}}]{Chen2025}%
  \BibitemOpen
  \bibfield  {author} {\bibinfo {author} {\bibfnamefont {Z.}~\bibnamefont
  {{Chen}}}, \bibinfo {author} {\bibfnamefont {Y.}~\bibnamefont {{Yu}}},
  \bibinfo {author} {\bibfnamefont {J.}~\bibnamefont {{Han}}},\ and\ \bibinfo
  {author} {\bibfnamefont {Y.~P.}\ \bibnamefont {{Jing}}},\ }\bibfield  {title}
  {\bibinfo {title} {{CSST Cosmological Emulator I: Matter Power Spectrum
  Emulation with one percent accuracy}},\ }\href
  {https://doi.org/10.48550/arXiv.2502.11160} {\bibfield  {journal} {\bibinfo
  {journal} {arXiv e-prints}\ ,\ \bibinfo {eid} {arXiv:2502.11160}} (\bibinfo
  {year} {2025})},\ \Eprint {https://arxiv.org/abs/2502.11160}
  {arXiv:2502.11160 [astro-ph.CO]} \BibitemShut {NoStop}%
\bibitem [{\citenamefont {{Caldwell}}\ and\ \citenamefont
  {{Kamionkowski}}(2009)}]{Caldwell2009}%
  \BibitemOpen
  \bibfield  {author} {\bibinfo {author} {\bibfnamefont {R.~R.}\ \bibnamefont
  {{Caldwell}}}\ and\ \bibinfo {author} {\bibfnamefont {M.}~\bibnamefont
  {{Kamionkowski}}},\ }\bibfield  {title} {\bibinfo {title} {{The Physics of
  Cosmic Acceleration}},\ }\href
  {https://doi.org/10.1146/annurev-nucl-010709-151330} {\bibfield  {journal}
  {\bibinfo  {journal} {Annual Review of Nuclear and Particle Science}\
  }\textbf {\bibinfo {volume} {59}},\ \bibinfo {pages} {397} (\bibinfo {year}
  {2009})},\ \Eprint {https://arxiv.org/abs/0903.0866} {arXiv:0903.0866
  [astro-ph.CO]} \BibitemShut {NoStop}%
\bibitem [{\citenamefont {{Feng}}(2010)}]{Feng2010}%
  \BibitemOpen
  \bibfield  {author} {\bibinfo {author} {\bibfnamefont {J.~L.}\ \bibnamefont
  {{Feng}}},\ }\bibfield  {title} {\bibinfo {title} {{Dark Matter Candidates
  from Particle Physics and Methods of Detection}},\ }\href
  {https://doi.org/10.1146/annurev-astro-082708-101659} {\bibfield  {journal}
  {\bibinfo  {journal} {Annual Review of Astronomy and Astrophysics}\ }\textbf
  {\bibinfo {volume} {48}},\ \bibinfo {pages} {495} (\bibinfo {year} {2010})},\
  \Eprint {https://arxiv.org/abs/1003.0904} {arXiv:1003.0904 [astro-ph.CO]}
  \BibitemShut {NoStop}%
\bibitem [{\citenamefont {{Wong}}(2011)}]{Wong2011}%
  \BibitemOpen
  \bibfield  {author} {\bibinfo {author} {\bibfnamefont {Y.~Y.~Y.}\
  \bibnamefont {{Wong}}},\ }\bibfield  {title} {\bibinfo {title} {{Neutrino
  Mass in Cosmology: Status and Prospects}},\ }\href
  {https://doi.org/10.1146/annurev-nucl-102010-130252} {\bibfield  {journal}
  {\bibinfo  {journal} {Annual Review of Nuclear and Particle Science}\
  }\textbf {\bibinfo {volume} {61}},\ \bibinfo {pages} {69} (\bibinfo {year}
  {2011})},\ \Eprint {https://arxiv.org/abs/1111.1436} {arXiv:1111.1436
  [astro-ph.CO]} \BibitemShut {NoStop}%
\bibitem [{\citenamefont {{Riess}}\ \emph {et~al.}(2021)\citenamefont
  {{Riess}}, \citenamefont {{Casertano}}, \citenamefont {{Yuan}}, \citenamefont
  {{Bowers}}, \citenamefont {{Macri}}, \citenamefont {{Zinn}},\ and\
  \citenamefont {{Scolnic}}}]{Riess2021}%
  \BibitemOpen
  \bibfield  {author} {\bibinfo {author} {\bibfnamefont {A.~G.}\ \bibnamefont
  {{Riess}}}, \bibinfo {author} {\bibfnamefont {S.}~\bibnamefont
  {{Casertano}}}, \bibinfo {author} {\bibfnamefont {W.}~\bibnamefont {{Yuan}}},
  \bibinfo {author} {\bibfnamefont {J.~B.}\ \bibnamefont {{Bowers}}}, \bibinfo
  {author} {\bibfnamefont {L.}~\bibnamefont {{Macri}}}, \bibinfo {author}
  {\bibfnamefont {J.~C.}\ \bibnamefont {{Zinn}}},\ and\ \bibinfo {author}
  {\bibfnamefont {D.}~\bibnamefont {{Scolnic}}},\ }\bibfield  {title} {\bibinfo
  {title} {{Cosmic Distances Calibrated to 1\% Precision with Gaia EDR3
  Parallaxes and Hubble Space Telescope Photometry of 75 Milky Way Cepheids
  Confirm Tension with {\ensuremath{\Lambda}}CDM}},\ }\href
  {https://doi.org/10.3847/2041-8213/abdbaf} {\bibfield  {journal} {\bibinfo
  {journal} {\apjl}\ }\textbf {\bibinfo {volume} {908}},\ \bibinfo {eid} {L6}
  (\bibinfo {year} {2021})},\ \Eprint {https://arxiv.org/abs/2012.08534}
  {arXiv:2012.08534 [astro-ph.CO]} \BibitemShut {NoStop}%
\bibitem [{\citenamefont {{Riess et al.}}(2022)}]{Riess2022}%
  \BibitemOpen
  \bibfield  {author} {\bibinfo {author} {\bibfnamefont {A.~G.}\ \bibnamefont
  {{Riess et al.}}},\ }\bibfield  {title} {\bibinfo {title} {{A Comprehensive
  Measurement of the Local Value of the Hubble Constant with 1 km s$^{-1}$
  Mpc$^{-1}$ Uncertainty from the Hubble Space Telescope and the SH0ES Team}},\
  }\href {https://doi.org/10.3847/2041-8213/ac5c5b} {\bibfield  {journal}
  {\bibinfo  {journal} {\apjl}\ }\textbf {\bibinfo {volume} {934}},\ \bibinfo
  {eid} {L7} (\bibinfo {year} {2022})},\ \Eprint
  {https://arxiv.org/abs/2112.04510} {arXiv:2112.04510 [astro-ph.CO]}
  \BibitemShut {NoStop}%
\bibitem [{\citenamefont {{Asgari}}\ \emph {et~al.}(2021)\citenamefont
  {{Asgari}}, \citenamefont {{Lin}}, \citenamefont {{Joachimi}}, \citenamefont
  {{Giblin}}, \citenamefont {{Heymans}}, \citenamefont {{Hildebrandt}},
  \citenamefont {{Kannawadi}}, \citenamefont {{St{\"o}lzner}}, \citenamefont
  {{Tr{\"o}ster}}, \citenamefont {{van den Busch}}, \citenamefont {{Wright}},
  \citenamefont {{Bilicki}}, \citenamefont {{Blake}}, \citenamefont {{de
  Jong}}, \citenamefont {{Dvornik}}, \citenamefont {{Erben}}, \citenamefont
  {{Getman}}, \citenamefont {{Hoekstra}}, \citenamefont {{K{\"o}hlinger}},
  \citenamefont {{Kuijken}}, \citenamefont {{Miller}}, \citenamefont
  {{Radovich}}, \citenamefont {{Schneider}}, \citenamefont {{Shan}},\ and\
  \citenamefont {{Valentijn}}}]{Asgari2021}%
  \BibitemOpen
  \bibfield  {author} {\bibinfo {author} {\bibfnamefont {M.}~\bibnamefont
  {{Asgari}}}, \bibinfo {author} {\bibfnamefont {C.-A.}\ \bibnamefont {{Lin}}},
  \bibinfo {author} {\bibfnamefont {B.}~\bibnamefont {{Joachimi}}}, \bibinfo
  {author} {\bibfnamefont {B.}~\bibnamefont {{Giblin}}}, \bibinfo {author}
  {\bibfnamefont {C.}~\bibnamefont {{Heymans}}}, \bibinfo {author}
  {\bibfnamefont {H.}~\bibnamefont {{Hildebrandt}}}, \bibinfo {author}
  {\bibfnamefont {A.}~\bibnamefont {{Kannawadi}}}, \bibinfo {author}
  {\bibfnamefont {B.}~\bibnamefont {{St{\"o}lzner}}}, \bibinfo {author}
  {\bibfnamefont {T.}~\bibnamefont {{Tr{\"o}ster}}}, \bibinfo {author}
  {\bibfnamefont {J.~L.}\ \bibnamefont {{van den Busch}}}, \bibinfo {author}
  {\bibfnamefont {A.~H.}\ \bibnamefont {{Wright}}}, \bibinfo {author}
  {\bibfnamefont {M.}~\bibnamefont {{Bilicki}}}, \bibinfo {author}
  {\bibfnamefont {C.}~\bibnamefont {{Blake}}}, \bibinfo {author} {\bibfnamefont
  {J.}~\bibnamefont {{de Jong}}}, \bibinfo {author} {\bibfnamefont
  {A.}~\bibnamefont {{Dvornik}}}, \bibinfo {author} {\bibfnamefont
  {T.}~\bibnamefont {{Erben}}}, \bibinfo {author} {\bibfnamefont
  {F.}~\bibnamefont {{Getman}}}, \bibinfo {author} {\bibfnamefont
  {H.}~\bibnamefont {{Hoekstra}}}, \bibinfo {author} {\bibfnamefont
  {F.}~\bibnamefont {{K{\"o}hlinger}}}, \bibinfo {author} {\bibfnamefont
  {K.}~\bibnamefont {{Kuijken}}}, \bibinfo {author} {\bibfnamefont
  {L.}~\bibnamefont {{Miller}}}, \bibinfo {author} {\bibfnamefont
  {M.}~\bibnamefont {{Radovich}}}, \bibinfo {author} {\bibfnamefont
  {P.}~\bibnamefont {{Schneider}}}, \bibinfo {author} {\bibfnamefont
  {H.}~\bibnamefont {{Shan}}},\ and\ \bibinfo {author} {\bibfnamefont
  {E.}~\bibnamefont {{Valentijn}}},\ }\bibfield  {title} {\bibinfo {title}
  {{KiDS-1000 cosmology: Cosmic shear constraints and comparison between two
  point statistics}},\ }\href {https://doi.org/10.1051/0004-6361/202039070}
  {\bibfield  {journal} {\bibinfo  {journal} {\aap}\ }\textbf {\bibinfo
  {volume} {645}},\ \bibinfo {eid} {A104} (\bibinfo {year} {2021})},\ \Eprint
  {https://arxiv.org/abs/2007.15633} {arXiv:2007.15633 [astro-ph.CO]}
  \BibitemShut {NoStop}%
\bibitem [{\citenamefont {{DES Collaboration}}\ \emph
  {et~al.}(2022)\citenamefont {{DES Collaboration}}, \citenamefont {{Abbott}},\
  and\ \citenamefont {{Aguena et al.}}}]{Abbott2022}%
  \BibitemOpen
  \bibfield  {author} {\bibinfo {author} {\bibnamefont {{DES Collaboration}}},
  \bibinfo {author} {\bibfnamefont {T.~M.~C.}\ \bibnamefont {{Abbott}}},\ and\
  \bibinfo {author} {\bibfnamefont {M.}~\bibnamefont {{Aguena et al.}}},\
  }\bibfield  {title} {\bibinfo {title} {{Dark Energy Survey Year 3 results:
  Cosmological constraints from galaxy clustering and weak lensing}},\ }\href
  {https://doi.org/10.1103/PhysRevD.105.023520} {\bibfield  {journal} {\bibinfo
   {journal} {\prd}\ }\textbf {\bibinfo {volume} {105}},\ \bibinfo {eid}
  {023520} (\bibinfo {year} {2022})},\ \Eprint
  {https://arxiv.org/abs/2105.13549} {arXiv:2105.13549 [astro-ph.CO]}
  \BibitemShut {NoStop}%
\bibitem [{\citenamefont {{Ho}}\ \emph {et~al.}(2023)\citenamefont {{Ho}},
  \citenamefont {{Bird}}, \citenamefont {{Fernandez}},\ and\ \citenamefont
  {{Shelton}}}]{Ho2023}%
  \BibitemOpen
  \bibfield  {author} {\bibinfo {author} {\bibfnamefont {M.-F.}\ \bibnamefont
  {{Ho}}}, \bibinfo {author} {\bibfnamefont {S.}~\bibnamefont {{Bird}}},
  \bibinfo {author} {\bibfnamefont {M.~A.}\ \bibnamefont {{Fernandez}}},\ and\
  \bibinfo {author} {\bibfnamefont {C.~R.}\ \bibnamefont {{Shelton}}},\
  }\bibfield  {title} {\bibinfo {title} {{MF-Box: multifidelity and multiscale
  emulation for the matter power spectrum}},\ }\href
  {https://doi.org/10.1093/mnras/stad2901} {\bibfield  {journal} {\bibinfo
  {journal} {MNRAS}\ }\textbf {\bibinfo {volume} {526}},\ \bibinfo {pages}
  {2903} (\bibinfo {year} {2023})},\ \Eprint {https://arxiv.org/abs/2306.03144}
  {arXiv:2306.03144 [astro-ph.CO]} \BibitemShut {NoStop}%
\bibitem [{\citenamefont {{Ji}}\ \emph {et~al.}(2021)\citenamefont {{Ji}},
  \citenamefont {{Mak}}, \citenamefont {{Soeder}}, \citenamefont {{Paquet}},\
  and\ \citenamefont {{Bass}}}]{Ji2021}%
  \BibitemOpen
  \bibfield  {author} {\bibinfo {author} {\bibfnamefont {Y.}~\bibnamefont
  {{Ji}}}, \bibinfo {author} {\bibfnamefont {S.}~\bibnamefont {{Mak}}},
  \bibinfo {author} {\bibfnamefont {D.}~\bibnamefont {{Soeder}}}, \bibinfo
  {author} {\bibfnamefont {J.-F.}\ \bibnamefont {{Paquet}}},\ and\ \bibinfo
  {author} {\bibfnamefont {S.~A.}\ \bibnamefont {{Bass}}},\ }\bibfield  {title}
  {\bibinfo {title} {{A graphical multi-fidelity Gaussian process model, with
  application to emulation of heavy-ion collisions}},\ }\href
  {https://doi.org/10.48550/arXiv.2108.00306} {\bibfield  {journal} {\bibinfo
  {journal} {arXiv e-prints}\ ,\ \bibinfo {eid} {arXiv:2108.00306}} (\bibinfo
  {year} {2021})},\ \Eprint {https://arxiv.org/abs/2108.00306}
  {arXiv:2108.00306 [stat.ME]} \BibitemShut {NoStop}%
\bibitem [{\citenamefont {Wendland}(2004)}]{Wendland2004}%
  \BibitemOpen
  \bibfield  {author} {\bibinfo {author} {\bibfnamefont {H.}~\bibnamefont
  {Wendland}},\ }\href@noop {} {\emph {\bibinfo {title} {Scattered Data
  Approximation}}},\ Cambridge Monographs on Applied and Computational
  Mathematics\ (\bibinfo  {publisher} {Cambridge University Press},\ \bibinfo
  {year} {2004})\BibitemShut {NoStop}%
\bibitem [{\citenamefont {{Ho}}\ \emph {et~al.}(2022)\citenamefont {{Ho}},
  \citenamefont {{Bird}},\ and\ \citenamefont {{Shelton}}}]{Ho2022}%
  \BibitemOpen
  \bibfield  {author} {\bibinfo {author} {\bibfnamefont {M.-F.}\ \bibnamefont
  {{Ho}}}, \bibinfo {author} {\bibfnamefont {S.}~\bibnamefont {{Bird}}},\ and\
  \bibinfo {author} {\bibfnamefont {C.~R.}\ \bibnamefont {{Shelton}}},\
  }\bibfield  {title} {\bibinfo {title} {{Multifidelity emulation for the
  matter power spectrum using Gaussian processes}},\ }\href
  {https://doi.org/10.1093/mnras/stab3114} {\bibfield  {journal} {\bibinfo
  {journal} {MNRAS}\ }\textbf {\bibinfo {volume} {509}},\ \bibinfo {pages}
  {2551} (\bibinfo {year} {2022})},\ \Eprint {https://arxiv.org/abs/2105.01081}
  {arXiv:2105.01081 [astro-ph.CO]} \BibitemShut {NoStop}%
\bibitem [{\citenamefont {{Bird}}\ \emph {et~al.}(2023)\citenamefont {{Bird}},
  \citenamefont {{Fernandez}}, \citenamefont {{Ho}}, \citenamefont {{Qezlou}},
  \citenamefont {{Monadi}}, \citenamefont {{Ni}}, \citenamefont {{Chen}},
  \citenamefont {{Croft}},\ and\ \citenamefont {{Di Matteo}}}]{Bird2023}%
  \BibitemOpen
  \bibfield  {author} {\bibinfo {author} {\bibfnamefont {S.}~\bibnamefont
  {{Bird}}}, \bibinfo {author} {\bibfnamefont {M.}~\bibnamefont {{Fernandez}}},
  \bibinfo {author} {\bibfnamefont {M.-F.}\ \bibnamefont {{Ho}}}, \bibinfo
  {author} {\bibfnamefont {M.}~\bibnamefont {{Qezlou}}}, \bibinfo {author}
  {\bibfnamefont {R.}~\bibnamefont {{Monadi}}}, \bibinfo {author}
  {\bibfnamefont {Y.}~\bibnamefont {{Ni}}}, \bibinfo {author} {\bibfnamefont
  {N.}~\bibnamefont {{Chen}}}, \bibinfo {author} {\bibfnamefont
  {R.}~\bibnamefont {{Croft}}},\ and\ \bibinfo {author} {\bibfnamefont
  {T.}~\bibnamefont {{Di Matteo}}},\ }\bibfield  {title} {\bibinfo {title}
  {{PRIYA: a new suite of Lyman-{\ensuremath{\alpha}} forest simulations for
  cosmology}},\ }\href {https://doi.org/10.1088/1475-7516/2023/10/037}
  {\bibfield  {journal} {\bibinfo  {journal} {\jcap}\ }\textbf {\bibinfo
  {volume} {2023}},\ \bibinfo {eid} {037} (\bibinfo {year} {2023})},\ \Eprint
  {https://arxiv.org/abs/2306.05471} {arXiv:2306.05471 [astro-ph.CO]}
  \BibitemShut {NoStop}%
\bibitem [{\citenamefont {{Fernandez}}\ \emph {et~al.}(2023)\citenamefont
  {{Fernandez}}, \citenamefont {{Bird}},\ and\ \citenamefont
  {{Ho}}}]{Fernandez2023}%
  \BibitemOpen
  \bibfield  {author} {\bibinfo {author} {\bibfnamefont {M.~A.}\ \bibnamefont
  {{Fernandez}}}, \bibinfo {author} {\bibfnamefont {S.}~\bibnamefont
  {{Bird}}},\ and\ \bibinfo {author} {\bibfnamefont {M.-F.}\ \bibnamefont
  {{Ho}}},\ }\bibfield  {title} {\bibinfo {title} {{Cosmological Constraints
  from the eBOSS Lyman-$\alpha$ Forest using the PRIYA Simulations}},\ }\href
  {https://doi.org/10.48550/arXiv.2309.03943} {\bibfield  {journal} {\bibinfo
  {journal} {arXiv e-prints}\ ,\ \bibinfo {eid} {arXiv:2309.03943}} (\bibinfo
  {year} {2023})},\ \Eprint {https://arxiv.org/abs/2309.03943}
  {arXiv:2309.03943 [astro-ph.CO]} \BibitemShut {NoStop}%
\bibitem [{\citenamefont {{DESI Collaboration}}\ \emph
  {et~al.}(2024)\citenamefont {{DESI Collaboration}}, \citenamefont {{Adame}},\
  and\ \citenamefont {{Aguilar et al.}}}]{Adame2024}%
  \BibitemOpen
  \bibfield  {author} {\bibinfo {author} {\bibnamefont {{DESI Collaboration}}},
  \bibinfo {author} {\bibfnamefont {A.~G.}\ \bibnamefont {{Adame}}},\ and\
  \bibinfo {author} {\bibfnamefont {J.}~\bibnamefont {{Aguilar et al.}}},\
  }\bibfield  {title} {\bibinfo {title} {{DESI 2024 VI: Cosmological
  Constraints from the Measurements of Baryon Acoustic Oscillations}},\ }\href
  {https://doi.org/10.48550/arXiv.2404.03002} {\bibfield  {journal} {\bibinfo
  {journal} {arXiv e-prints}\ ,\ \bibinfo {eid} {arXiv:2404.03002}} (\bibinfo
  {year} {2024})},\ \Eprint {https://arxiv.org/abs/2404.03002}
  {arXiv:2404.03002 [astro-ph.CO]} \BibitemShut {NoStop}%
\bibitem [{\citenamefont {{Planck Collaboration}}\ \emph
  {et~al.}(2020{\natexlab{a}})\citenamefont {{Planck Collaboration}},
  \citenamefont {{Aghanim}},\ and\ \citenamefont {{Akrami et
  al.}}}]{Aghanim2020}%
  \BibitemOpen
  \bibfield  {author} {\bibinfo {author} {\bibnamefont {{Planck
  Collaboration}}}, \bibinfo {author} {\bibfnamefont {N.}~\bibnamefont
  {{Aghanim}}},\ and\ \bibinfo {author} {\bibfnamefont {Y.}~\bibnamefont
  {{Akrami et al.}}},\ }\bibfield  {title} {\bibinfo {title} {{Planck 2018
  results. VI. Cosmological parameters}},\ }\href
  {https://doi.org/10.1051/0004-6361/201833910} {\bibfield  {journal} {\bibinfo
   {journal} {\aap}\ }\textbf {\bibinfo {volume} {641}},\ \bibinfo {eid} {A6}
  (\bibinfo {year} {2020}{\natexlab{a}})},\ \Eprint
  {https://arxiv.org/abs/1807.06209} {arXiv:1807.06209 [astro-ph.CO]}
  \BibitemShut {NoStop}%
\bibitem [{\citenamefont {{Tristram}}\ \emph {et~al.}(2024)\citenamefont
  {{Tristram}}, \citenamefont {{Banday}}, \citenamefont {{Douspis}},
  \citenamefont {{Garrido}}, \citenamefont {{G{\'o}rski}}, \citenamefont
  {{Henrot-Versill{\'e}}}, \citenamefont {{Hergt}}, \citenamefont {{Ili{\'c}}},
  \citenamefont {{Keskitalo}}, \citenamefont {{Lagache}}, \citenamefont
  {{Lawrence}}, \citenamefont {{Partridge}},\ and\ \citenamefont
  {{Scott}}}]{Tristram2024}%
  \BibitemOpen
  \bibfield  {author} {\bibinfo {author} {\bibfnamefont {M.}~\bibnamefont
  {{Tristram}}}, \bibinfo {author} {\bibfnamefont {A.~J.}\ \bibnamefont
  {{Banday}}}, \bibinfo {author} {\bibfnamefont {M.}~\bibnamefont {{Douspis}}},
  \bibinfo {author} {\bibfnamefont {X.}~\bibnamefont {{Garrido}}}, \bibinfo
  {author} {\bibfnamefont {K.~M.}\ \bibnamefont {{G{\'o}rski}}}, \bibinfo
  {author} {\bibfnamefont {S.}~\bibnamefont {{Henrot-Versill{\'e}}}}, \bibinfo
  {author} {\bibfnamefont {L.~T.}\ \bibnamefont {{Hergt}}}, \bibinfo {author}
  {\bibfnamefont {S.}~\bibnamefont {{Ili{\'c}}}}, \bibinfo {author}
  {\bibfnamefont {R.}~\bibnamefont {{Keskitalo}}}, \bibinfo {author}
  {\bibfnamefont {G.}~\bibnamefont {{Lagache}}}, \bibinfo {author}
  {\bibfnamefont {C.~R.}\ \bibnamefont {{Lawrence}}}, \bibinfo {author}
  {\bibfnamefont {B.}~\bibnamefont {{Partridge}}},\ and\ \bibinfo {author}
  {\bibfnamefont {D.}~\bibnamefont {{Scott}}},\ }\bibfield  {title} {\bibinfo
  {title} {{Cosmological parameters derived from the final Planck data release
  (PR4)}},\ }\href {https://doi.org/10.1051/0004-6361/202348015} {\bibfield
  {journal} {\bibinfo  {journal} {\aap}\ }\textbf {\bibinfo {volume} {682}},\
  \bibinfo {eid} {A37} (\bibinfo {year} {2024})},\ \Eprint
  {https://arxiv.org/abs/2309.10034} {arXiv:2309.10034 [astro-ph.CO]}
  \BibitemShut {NoStop}%
\bibitem [{\citenamefont {{Brout}}\ \emph {et~al.}(2022)\citenamefont
  {{Brout}}, \citenamefont {{Scolnic}}, \citenamefont {{Popovic}},
  \citenamefont {{Riess}}, \citenamefont {{Carr}}, \citenamefont {{Zuntz}},
  \citenamefont {{Kessler}}, \citenamefont {{Davis}}, \citenamefont {{Hinton}},
  \citenamefont {{Jones}}, \citenamefont {{Kenworthy}}, \citenamefont
  {{Peterson}}, \citenamefont {{Said}}, \citenamefont {{Taylor}}, \citenamefont
  {{Ali}}, \citenamefont {{Armstrong}}, \citenamefont {{Charvu}}, \citenamefont
  {{Dwomoh}}, \citenamefont {{Meldorf}}, \citenamefont {{Palmese}},
  \citenamefont {{Qu}}, \citenamefont {{Rose}}, \citenamefont {{Sanchez}},
  \citenamefont {{Stubbs}}, \citenamefont {{Vincenzi}}, \citenamefont {{Wood}},
  \citenamefont {{Brown}}, \citenamefont {{Chen}}, \citenamefont {{Chambers}},
  \citenamefont {{Coulter}}, \citenamefont {{Dai}}, \citenamefont
  {{Dimitriadis}}, \citenamefont {{Filippenko}}, \citenamefont {{Foley}},
  \citenamefont {{Jha}}, \citenamefont {{Kelsey}}, \citenamefont {{Kirshner}},
  \citenamefont {{M{\"o}ller}}, \citenamefont {{Muir}}, \citenamefont
  {{Nadathur}}, \citenamefont {{Pan}}, \citenamefont {{Rest}}, \citenamefont
  {{Rojas-Bravo}}, \citenamefont {{Sako}}, \citenamefont {{Siebert}},
  \citenamefont {{Smith}}, \citenamefont {{Stahl}},\ and\ \citenamefont
  {{Wiseman}}}]{Brout2022}%
  \BibitemOpen
  \bibfield  {author} {\bibinfo {author} {\bibfnamefont {D.}~\bibnamefont
  {{Brout}}}, \bibinfo {author} {\bibfnamefont {D.}~\bibnamefont {{Scolnic}}},
  \bibinfo {author} {\bibfnamefont {B.}~\bibnamefont {{Popovic}}}, \bibinfo
  {author} {\bibfnamefont {A.~G.}\ \bibnamefont {{Riess}}}, \bibinfo {author}
  {\bibfnamefont {A.}~\bibnamefont {{Carr}}}, \bibinfo {author} {\bibfnamefont
  {J.}~\bibnamefont {{Zuntz}}}, \bibinfo {author} {\bibfnamefont
  {R.}~\bibnamefont {{Kessler}}}, \bibinfo {author} {\bibfnamefont {T.~M.}\
  \bibnamefont {{Davis}}}, \bibinfo {author} {\bibfnamefont {S.}~\bibnamefont
  {{Hinton}}}, \bibinfo {author} {\bibfnamefont {D.}~\bibnamefont {{Jones}}},
  \bibinfo {author} {\bibfnamefont {W.~D.}\ \bibnamefont {{Kenworthy}}},
  \bibinfo {author} {\bibfnamefont {E.~R.}\ \bibnamefont {{Peterson}}},
  \bibinfo {author} {\bibfnamefont {K.}~\bibnamefont {{Said}}}, \bibinfo
  {author} {\bibfnamefont {G.}~\bibnamefont {{Taylor}}}, \bibinfo {author}
  {\bibfnamefont {N.}~\bibnamefont {{Ali}}}, \bibinfo {author} {\bibfnamefont
  {P.}~\bibnamefont {{Armstrong}}}, \bibinfo {author} {\bibfnamefont
  {P.}~\bibnamefont {{Charvu}}}, \bibinfo {author} {\bibfnamefont
  {A.}~\bibnamefont {{Dwomoh}}}, \bibinfo {author} {\bibfnamefont
  {C.}~\bibnamefont {{Meldorf}}}, \bibinfo {author} {\bibfnamefont
  {A.}~\bibnamefont {{Palmese}}}, \bibinfo {author} {\bibfnamefont
  {H.}~\bibnamefont {{Qu}}}, \bibinfo {author} {\bibfnamefont {B.~M.}\
  \bibnamefont {{Rose}}}, \bibinfo {author} {\bibfnamefont {B.}~\bibnamefont
  {{Sanchez}}}, \bibinfo {author} {\bibfnamefont {C.~W.}\ \bibnamefont
  {{Stubbs}}}, \bibinfo {author} {\bibfnamefont {M.}~\bibnamefont
  {{Vincenzi}}}, \bibinfo {author} {\bibfnamefont {C.~M.}\ \bibnamefont
  {{Wood}}}, \bibinfo {author} {\bibfnamefont {P.~J.}\ \bibnamefont {{Brown}}},
  \bibinfo {author} {\bibfnamefont {R.}~\bibnamefont {{Chen}}}, \bibinfo
  {author} {\bibfnamefont {K.}~\bibnamefont {{Chambers}}}, \bibinfo {author}
  {\bibfnamefont {D.~A.}\ \bibnamefont {{Coulter}}}, \bibinfo {author}
  {\bibfnamefont {M.}~\bibnamefont {{Dai}}}, \bibinfo {author} {\bibfnamefont
  {G.}~\bibnamefont {{Dimitriadis}}}, \bibinfo {author} {\bibfnamefont {A.~V.}\
  \bibnamefont {{Filippenko}}}, \bibinfo {author} {\bibfnamefont {R.~J.}\
  \bibnamefont {{Foley}}}, \bibinfo {author} {\bibfnamefont {S.~W.}\
  \bibnamefont {{Jha}}}, \bibinfo {author} {\bibfnamefont {L.}~\bibnamefont
  {{Kelsey}}}, \bibinfo {author} {\bibfnamefont {R.~P.}\ \bibnamefont
  {{Kirshner}}}, \bibinfo {author} {\bibfnamefont {A.}~\bibnamefont
  {{M{\"o}ller}}}, \bibinfo {author} {\bibfnamefont {J.}~\bibnamefont
  {{Muir}}}, \bibinfo {author} {\bibfnamefont {S.}~\bibnamefont {{Nadathur}}},
  \bibinfo {author} {\bibfnamefont {Y.-C.}\ \bibnamefont {{Pan}}}, \bibinfo
  {author} {\bibfnamefont {A.}~\bibnamefont {{Rest}}}, \bibinfo {author}
  {\bibfnamefont {C.}~\bibnamefont {{Rojas-Bravo}}}, \bibinfo {author}
  {\bibfnamefont {M.}~\bibnamefont {{Sako}}}, \bibinfo {author} {\bibfnamefont
  {M.~R.}\ \bibnamefont {{Siebert}}}, \bibinfo {author} {\bibfnamefont
  {M.}~\bibnamefont {{Smith}}}, \bibinfo {author} {\bibfnamefont {B.~E.}\
  \bibnamefont {{Stahl}}},\ and\ \bibinfo {author} {\bibfnamefont
  {P.}~\bibnamefont {{Wiseman}}},\ }\bibfield  {title} {\bibinfo {title} {{The
  Pantheon+ Analysis: Cosmological Constraints}},\ }\href
  {https://doi.org/10.3847/1538-4357/ac8e04} {\bibfield  {journal} {\bibinfo
  {journal} {\apj}\ }\textbf {\bibinfo {volume} {938}},\ \bibinfo {eid} {110}
  (\bibinfo {year} {2022})},\ \Eprint {https://arxiv.org/abs/2202.04077}
  {arXiv:2202.04077 [astro-ph.CO]} \BibitemShut {NoStop}%
\bibitem [{\citenamefont {{Boruah}}\ \emph {et~al.}(2023)\citenamefont
  {{Boruah}}, \citenamefont {{Eifler}}, \citenamefont {{Miranda}},\ and\
  \citenamefont {{Krishanth}}}]{2023MNRAS.518.4818B}%
  \BibitemOpen
  \bibfield  {author} {\bibinfo {author} {\bibfnamefont {S.~S.}\ \bibnamefont
  {{Boruah}}}, \bibinfo {author} {\bibfnamefont {T.}~\bibnamefont {{Eifler}}},
  \bibinfo {author} {\bibfnamefont {V.}~\bibnamefont {{Miranda}}},\ and\
  \bibinfo {author} {\bibfnamefont {P.~M.~S.}\ \bibnamefont {{Krishanth}}},\
  }\bibfield  {title} {\bibinfo {title} {{Accelerating cosmological inference
  with Gaussian processes and neural networks - an application to LSST Y1 weak
  lensing and galaxy clustering}},\ }\href
  {https://doi.org/10.1093/mnras/stac3417} {\bibfield  {journal} {\bibinfo
  {journal} {\mnras}\ }\textbf {\bibinfo {volume} {518}},\ \bibinfo {pages}
  {4818} (\bibinfo {year} {2023})},\ \Eprint {https://arxiv.org/abs/2203.06124}
  {arXiv:2203.06124 [astro-ph.CO]} \BibitemShut {NoStop}%
\bibitem [{\citenamefont {{Chevallier}}\ and\ \citenamefont
  {{Polarski}}(2001)}]{Chevallier2001}%
  \BibitemOpen
  \bibfield  {author} {\bibinfo {author} {\bibfnamefont {M.}~\bibnamefont
  {{Chevallier}}}\ and\ \bibinfo {author} {\bibfnamefont {D.}~\bibnamefont
  {{Polarski}}},\ }\bibfield  {title} {\bibinfo {title} {{Accelerating
  Universes with Scaling Dark Matter}},\ }\href
  {https://doi.org/10.1142/S0218271801000822} {\bibfield  {journal} {\bibinfo
  {journal} {International Journal of Modern Physics D}\ }\textbf {\bibinfo
  {volume} {10}},\ \bibinfo {pages} {213} (\bibinfo {year} {2001})},\ \Eprint
  {https://arxiv.org/abs/gr-qc/0009008} {arXiv:gr-qc/0009008 [gr-qc]}
  \BibitemShut {NoStop}%
\bibitem [{\citenamefont {{Linder}}(2003)}]{Linder2003}%
  \BibitemOpen
  \bibfield  {author} {\bibinfo {author} {\bibfnamefont {E.~V.}\ \bibnamefont
  {{Linder}}},\ }\bibfield  {title} {\bibinfo {title} {{Exploring the Expansion
  History of the Universe}},\ }\href
  {https://doi.org/10.1103/PhysRevLett.90.091301} {\bibfield  {journal}
  {\bibinfo  {journal} {\prl}\ }\textbf {\bibinfo {volume} {90}},\ \bibinfo
  {eid} {091301} (\bibinfo {year} {2003})},\ \Eprint
  {https://arxiv.org/abs/astro-ph/0208512} {arXiv:astro-ph/0208512 [astro-ph]}
  \BibitemShut {NoStop}%
\bibitem [{\citenamefont {{Lesgourgues}}\ and\ \citenamefont
  {{Pastor}}(2006)}]{Lesgourgues2006}%
  \BibitemOpen
  \bibfield  {author} {\bibinfo {author} {\bibfnamefont {J.}~\bibnamefont
  {{Lesgourgues}}}\ and\ \bibinfo {author} {\bibfnamefont {S.}~\bibnamefont
  {{Pastor}}},\ }\bibfield  {title} {\bibinfo {title} {{Massive neutrinos and
  cosmology}},\ }\href {https://doi.org/10.1016/j.physrep.2006.04.001}
  {\bibfield  {journal} {\bibinfo  {journal} {\physrep}\ }\textbf {\bibinfo
  {volume} {429}},\ \bibinfo {pages} {307} (\bibinfo {year} {2006})},\ \Eprint
  {https://arxiv.org/abs/astro-ph/0603494} {arXiv:astro-ph/0603494 [astro-ph]}
  \BibitemShut {NoStop}%
\bibitem [{\citenamefont {Lesgourgues}\ \emph {et~al.}(2013)\citenamefont
  {Lesgourgues}, \citenamefont {Mangano}, \citenamefont {Miele},\ and\
  \citenamefont {Pastor}}]{Lesgourgues2013}%
  \BibitemOpen
  \bibfield  {author} {\bibinfo {author} {\bibfnamefont {J.}~\bibnamefont
  {Lesgourgues}}, \bibinfo {author} {\bibfnamefont {G.}~\bibnamefont
  {Mangano}}, \bibinfo {author} {\bibfnamefont {G.}~\bibnamefont {Miele}},\
  and\ \bibinfo {author} {\bibfnamefont {S.}~\bibnamefont {Pastor}},\
  }\href@noop {} {\emph {\bibinfo {title} {Neutrino Cosmology}}}\ (\bibinfo
  {publisher} {Cambridge University Press},\ \bibinfo {year}
  {2013})\BibitemShut {NoStop}%
\bibitem [{\citenamefont {{Planck Collaboration}}\ \emph
  {et~al.}(2014)\citenamefont {{Planck Collaboration}}, \citenamefont {{Ade}},\
  and\ \citenamefont {{Aghanim et al.}}}]{Ade2014}%
  \BibitemOpen
  \bibfield  {author} {\bibinfo {author} {\bibnamefont {{Planck
  Collaboration}}}, \bibinfo {author} {\bibfnamefont {P.~A.~R.}\ \bibnamefont
  {{Ade}}},\ and\ \bibinfo {author} {\bibfnamefont {N.}~\bibnamefont {{Aghanim
  et al.}}},\ }\bibfield  {title} {\bibinfo {title} {{Planck 2013 results. XVI.
  Cosmological parameters}},\ }\href
  {https://doi.org/10.1051/0004-6361/201321591} {\bibfield  {journal} {\bibinfo
   {journal} {\aap}\ }\textbf {\bibinfo {volume} {571}},\ \bibinfo {eid} {A16}
  (\bibinfo {year} {2014})},\ \Eprint {https://arxiv.org/abs/1303.5076}
  {arXiv:1303.5076 [astro-ph.CO]} \BibitemShut {NoStop}%
\bibitem [{\citenamefont {{Palanque-Delabrouille}}\ \emph
  {et~al.}(2015)\citenamefont {{Palanque-Delabrouille}}, \citenamefont
  {{Y{\`e}che}}, \citenamefont {{Baur}}, \citenamefont {{Magneville}},
  \citenamefont {{Rossi}}, \citenamefont {{Lesgourgues}}, \citenamefont
  {{Borde}}, \citenamefont {{Burtin}}, \citenamefont {{LeGoff}}, \citenamefont
  {{Rich}}, \citenamefont {{Viel}},\ and\ \citenamefont
  {{Weinberg}}}]{Palanque2015}%
  \BibitemOpen
  \bibfield  {author} {\bibinfo {author} {\bibfnamefont {N.}~\bibnamefont
  {{Palanque-Delabrouille}}}, \bibinfo {author} {\bibfnamefont
  {C.}~\bibnamefont {{Y{\`e}che}}}, \bibinfo {author} {\bibfnamefont
  {J.}~\bibnamefont {{Baur}}}, \bibinfo {author} {\bibfnamefont
  {C.}~\bibnamefont {{Magneville}}}, \bibinfo {author} {\bibfnamefont
  {G.}~\bibnamefont {{Rossi}}}, \bibinfo {author} {\bibfnamefont
  {J.}~\bibnamefont {{Lesgourgues}}}, \bibinfo {author} {\bibfnamefont
  {A.}~\bibnamefont {{Borde}}}, \bibinfo {author} {\bibfnamefont
  {E.}~\bibnamefont {{Burtin}}}, \bibinfo {author} {\bibfnamefont {J.-M.}\
  \bibnamefont {{LeGoff}}}, \bibinfo {author} {\bibfnamefont {J.}~\bibnamefont
  {{Rich}}}, \bibinfo {author} {\bibfnamefont {M.}~\bibnamefont {{Viel}}},\
  and\ \bibinfo {author} {\bibfnamefont {D.}~\bibnamefont {{Weinberg}}},\
  }\bibfield  {title} {\bibinfo {title} {{Neutrino masses and cosmology with
  Lyman-alpha forest power spectrum}},\ }\href
  {https://doi.org/10.1088/1475-7516/2015/11/011} {\bibfield  {journal}
  {\bibinfo  {journal} {\jcap}\ }\textbf {\bibinfo {volume} {2015}},\ \bibinfo
  {pages} {011} (\bibinfo {year} {2015})},\ \Eprint
  {https://arxiv.org/abs/1506.05976} {arXiv:1506.05976 [astro-ph.CO]}
  \BibitemShut {NoStop}%
\bibitem [{\citenamefont {{Ali-Ha{\"\i}moud}}\ and\ \citenamefont
  {{Bird}}(2013)}]{Ali2013}%
  \BibitemOpen
  \bibfield  {author} {\bibinfo {author} {\bibfnamefont {Y.}~\bibnamefont
  {{Ali-Ha{\"\i}moud}}}\ and\ \bibinfo {author} {\bibfnamefont
  {S.}~\bibnamefont {{Bird}}},\ }\bibfield  {title} {\bibinfo {title} {{An
  efficient implementation of massive neutrinos in non-linear structure
  formation simulations}},\ }\href {https://doi.org/10.1093/mnras/sts286}
  {\bibfield  {journal} {\bibinfo  {journal} {MNRAS}\ }\textbf {\bibinfo
  {volume} {428}},\ \bibinfo {pages} {3375} (\bibinfo {year} {2013})},\ \Eprint
  {https://arxiv.org/abs/1209.0461} {arXiv:1209.0461 [astro-ph.CO]}
  \BibitemShut {NoStop}%
\bibitem [{\citenamefont {{Bird}}\ \emph {et~al.}(2018)\citenamefont {{Bird}},
  \citenamefont {{Ali-Ha{\"\i}moud}}, \citenamefont {{Feng}},\ and\
  \citenamefont {{Liu}}}]{Bird2018}%
  \BibitemOpen
  \bibfield  {author} {\bibinfo {author} {\bibfnamefont {S.}~\bibnamefont
  {{Bird}}}, \bibinfo {author} {\bibfnamefont {Y.}~\bibnamefont
  {{Ali-Ha{\"\i}moud}}}, \bibinfo {author} {\bibfnamefont {Y.}~\bibnamefont
  {{Feng}}},\ and\ \bibinfo {author} {\bibfnamefont {J.}~\bibnamefont
  {{Liu}}},\ }\bibfield  {title} {\bibinfo {title} {{An efficient and accurate
  hybrid method for simulating non-linear neutrino structure}},\ }\href
  {https://doi.org/10.1093/mnras/sty2376} {\bibfield  {journal} {\bibinfo
  {journal} {MNRAS}\ }\textbf {\bibinfo {volume} {481}},\ \bibinfo {pages}
  {1486} (\bibinfo {year} {2018})},\ \Eprint {https://arxiv.org/abs/1803.09854}
  {arXiv:1803.09854 [astro-ph.CO]} \BibitemShut {NoStop}%
\bibitem [{\citenamefont {{Bird}}\ \emph {et~al.}(2012)\citenamefont {{Bird}},
  \citenamefont {{Viel}},\ and\ \citenamefont {{Haehnelt}}}]{Bird2012}%
  \BibitemOpen
  \bibfield  {author} {\bibinfo {author} {\bibfnamefont {S.}~\bibnamefont
  {{Bird}}}, \bibinfo {author} {\bibfnamefont {M.}~\bibnamefont {{Viel}}},\
  and\ \bibinfo {author} {\bibfnamefont {M.~G.}\ \bibnamefont {{Haehnelt}}},\
  }\bibfield  {title} {\bibinfo {title} {{Massive neutrinos and the non-linear
  matter power spectrum}},\ }\href
  {https://doi.org/10.1111/j.1365-2966.2011.20222.x} {\bibfield  {journal}
  {\bibinfo  {journal} {\mnras}\ }\textbf {\bibinfo {volume} {420}},\ \bibinfo
  {pages} {2551} (\bibinfo {year} {2012})},\ \Eprint
  {https://arxiv.org/abs/1109.4416} {arXiv:1109.4416 [astro-ph.CO]}
  \BibitemShut {NoStop}%
\bibitem [{\citenamefont {Feng}\ \emph {et~al.}(2018)\citenamefont {Feng},
  \citenamefont {Bird}, \citenamefont {Anderson}, \citenamefont {Font-Ribera},\
  and\ \citenamefont {Pedersen}}]{Feng2018}%
  \BibitemOpen
  \bibfield  {author} {\bibinfo {author} {\bibfnamefont {Y.}~\bibnamefont
  {Feng}}, \bibinfo {author} {\bibfnamefont {S.}~\bibnamefont {Bird}}, \bibinfo
  {author} {\bibfnamefont {L.}~\bibnamefont {Anderson}}, \bibinfo {author}
  {\bibfnamefont {A.}~\bibnamefont {Font-Ribera}},\ and\ \bibinfo {author}
  {\bibfnamefont {C.}~\bibnamefont {Pedersen}},\ }\href
  {https://doi.org/10.5281/zenodo.1451799} {\bibinfo {title}
  {{MP-Gadget/MP-Gadget: A tag for getting a DOI}}} (\bibinfo {year}
  {2018})\BibitemShut {NoStop}%
\bibitem [{\citenamefont {{Lesgourgues}}(2011)}]{Lesgourgues2011}%
  \BibitemOpen
  \bibfield  {author} {\bibinfo {author} {\bibfnamefont {J.}~\bibnamefont
  {{Lesgourgues}}},\ }\bibfield  {title} {\bibinfo {title} {{The Cosmic Linear
  Anisotropy Solving System (CLASS) I: Overview}},\ }\href
  {https://doi.org/10.48550/arXiv.1104.2932} {\bibfield  {journal} {\bibinfo
  {journal} {arXiv e-prints}\ ,\ \bibinfo {eid} {arXiv:1104.2932}} (\bibinfo
  {year} {2011})},\ \Eprint {https://arxiv.org/abs/1104.2932} {arXiv:1104.2932
  [astro-ph.IM]} \BibitemShut {NoStop}%
\bibitem [{\citenamefont {{Lowerre}}(1976)}]{Lowerre1976}%
  \BibitemOpen
  \bibfield  {author} {\bibinfo {author} {\bibfnamefont {B.}~\bibnamefont
  {{Lowerre}}},\ }\bibfield  {title} {\bibinfo {title} {{The Harpy Speech
  Recognition System: performance with large vocabularies}},\ }\href
  {https://doi.org/10.1121/1.2003089} {\bibfield  {journal} {\bibinfo
  {journal} {Acoustical Society of America Journal}\ }\textbf {\bibinfo
  {volume} {60}},\ \bibinfo {pages} {S10} (\bibinfo {year} {1976})}\BibitemShut
  {NoStop}%
\bibitem [{\citenamefont {{Angulo}}\ and\ \citenamefont
  {{Pontzen}}(2016)}]{Angulo2016}%
  \BibitemOpen
  \bibfield  {author} {\bibinfo {author} {\bibfnamefont {R.~E.}\ \bibnamefont
  {{Angulo}}}\ and\ \bibinfo {author} {\bibfnamefont {A.}~\bibnamefont
  {{Pontzen}}},\ }\bibfield  {title} {\bibinfo {title} {{Cosmological N-body
  simulations with suppressed variance}},\ }\href
  {https://doi.org/10.1093/mnrasl/slw098} {\bibfield  {journal} {\bibinfo
  {journal} {\mnras}\ }\textbf {\bibinfo {volume} {462}},\ \bibinfo {pages}
  {L1} (\bibinfo {year} {2016})},\ \Eprint {https://arxiv.org/abs/1603.05253}
  {arXiv:1603.05253 [astro-ph.CO]} \BibitemShut {NoStop}%
\bibitem [{\citenamefont {{Planck Collaboration}}\ \emph
  {et~al.}(2020{\natexlab{b}})\citenamefont {{Planck Collaboration}},
  \citenamefont {{Akrami}},\ and\ \citenamefont {{Arroja et
  al.}}}]{Akrami2020}%
  \BibitemOpen
  \bibfield  {author} {\bibinfo {author} {\bibnamefont {{Planck
  Collaboration}}}, \bibinfo {author} {\bibfnamefont {Y.}~\bibnamefont
  {{Akrami}}},\ and\ \bibinfo {author} {\bibfnamefont {F.}~\bibnamefont
  {{Arroja et al.}}},\ }\bibfield  {title} {\bibinfo {title} {{Planck 2018
  results. X. Constraints on inflation}},\ }\href
  {https://doi.org/10.1051/0004-6361/201833887} {\bibfield  {journal} {\bibinfo
   {journal} {\aap}\ }\textbf {\bibinfo {volume} {641}},\ \bibinfo {eid} {A10}
  (\bibinfo {year} {2020}{\natexlab{b}})},\ \Eprint
  {https://arxiv.org/abs/1807.06211} {arXiv:1807.06211 [astro-ph.CO]}
  \BibitemShut {NoStop}%
\bibitem [{\citenamefont {{Abbott et al.}}(2018)}]{Abbott2018}%
  \BibitemOpen
  \bibfield  {author} {\bibinfo {author} {\bibfnamefont {T.~M.~C.}\
  \bibnamefont {{Abbott et al.}}},\ }\bibfield  {title} {\bibinfo {title}
  {{Dark Energy Survey year 1 results: Cosmological constraints from galaxy
  clustering and weak lensing}},\ }\href
  {https://doi.org/10.1103/PhysRevD.98.043526} {\bibfield  {journal} {\bibinfo
  {journal} {\prd}\ }\textbf {\bibinfo {volume} {98}},\ \bibinfo {eid} {043526}
  (\bibinfo {year} {2018})},\ \Eprint {https://arxiv.org/abs/1708.01530}
  {arXiv:1708.01530 [astro-ph.CO]} \BibitemShut {NoStop}%
\bibitem [{\citenamefont {{Ivanov}}\ \emph {et~al.}(2020)\citenamefont
  {{Ivanov}}, \citenamefont {{Simonovi{\'c}}},\ and\ \citenamefont
  {{Zaldarriaga}}}]{Ivanov2020}%
  \BibitemOpen
  \bibfield  {author} {\bibinfo {author} {\bibfnamefont {M.~M.}\ \bibnamefont
  {{Ivanov}}}, \bibinfo {author} {\bibfnamefont {M.}~\bibnamefont
  {{Simonovi{\'c}}}},\ and\ \bibinfo {author} {\bibfnamefont {M.}~\bibnamefont
  {{Zaldarriaga}}},\ }\bibfield  {title} {\bibinfo {title} {{Cosmological
  parameters from the BOSS galaxy power spectrum}},\ }\href
  {https://doi.org/10.1088/1475-7516/2020/05/042} {\bibfield  {journal}
  {\bibinfo  {journal} {\jcap}\ }\textbf {\bibinfo {volume} {2020}},\ \bibinfo
  {eid} {042} (\bibinfo {year} {2020})},\ \Eprint
  {https://arxiv.org/abs/1909.05277} {arXiv:1909.05277 [astro-ph.CO]}
  \BibitemShut {NoStop}%
\bibitem [{\citenamefont {McCarthy}\ \emph {et~al.}(2018)\citenamefont
  {McCarthy}, \citenamefont {Bird}, \citenamefont {Schaye}, \citenamefont
  {Harnois-Deraps}, \citenamefont {Font},\ and\ \citenamefont {van
  Waerbeke}}]{McCarthy2018}%
  \BibitemOpen
  \bibfield  {author} {\bibinfo {author} {\bibfnamefont {I.~G.}\ \bibnamefont
  {McCarthy}}, \bibinfo {author} {\bibfnamefont {S.}~\bibnamefont {Bird}},
  \bibinfo {author} {\bibfnamefont {J.}~\bibnamefont {Schaye}}, \bibinfo
  {author} {\bibfnamefont {J.}~\bibnamefont {Harnois-Deraps}}, \bibinfo
  {author} {\bibfnamefont {A.~S.}\ \bibnamefont {Font}},\ and\ \bibinfo
  {author} {\bibfnamefont {L.}~\bibnamefont {van Waerbeke}},\ }\bibfield
  {title} {\bibinfo {title} {{The BAHAMAS project: the CMB–large-scale
  structure tension and the roles of massive neutrinos and galaxy formation}},\
  }\href {https://doi.org/10.1093/mnras/sty377} {\bibfield  {journal} {\bibinfo
   {journal} {Monthly Notices of the Royal Astronomical Society}\ }\textbf
  {\bibinfo {volume} {476}},\ \bibinfo {pages} {2999} (\bibinfo {year}
  {2018})},\ \Eprint
  {https://arxiv.org/abs/https://academic.oup.com/mnras/article-pdf/476/3/2999/24446897/sty377.pdf}
  {https://academic.oup.com/mnras/article-pdf/476/3/2999/24446897/sty377.pdf}
  \BibitemShut {NoStop}%
\bibitem [{\citenamefont {{Freedman}}\ \emph {et~al.}(2024)\citenamefont
  {{Freedman}}, \citenamefont {{Madore}}, \citenamefont {{Jang}}, \citenamefont
  {{Hoyt}}, \citenamefont {{Lee}},\ and\ \citenamefont
  {{Owens}}}]{Freedman2024}%
  \BibitemOpen
  \bibfield  {author} {\bibinfo {author} {\bibfnamefont {W.~L.}\ \bibnamefont
  {{Freedman}}}, \bibinfo {author} {\bibfnamefont {B.~F.}\ \bibnamefont
  {{Madore}}}, \bibinfo {author} {\bibfnamefont {I.~S.}\ \bibnamefont
  {{Jang}}}, \bibinfo {author} {\bibfnamefont {T.~J.}\ \bibnamefont {{Hoyt}}},
  \bibinfo {author} {\bibfnamefont {A.~J.}\ \bibnamefont {{Lee}}},\ and\
  \bibinfo {author} {\bibfnamefont {K.~A.}\ \bibnamefont {{Owens}}},\
  }\bibfield  {title} {\bibinfo {title} {{Status Report on the Chicago-Carnegie
  Hubble Program (CCHP): Three Independent Astrophysical Determinations of the
  Hubble Constant Using the James Webb Space Telescope}},\ }\href
  {https://doi.org/10.48550/arXiv.2408.06153} {\bibfield  {journal} {\bibinfo
  {journal} {arXiv e-prints}\ ,\ \bibinfo {eid} {arXiv:2408.06153}} (\bibinfo
  {year} {2024})},\ \Eprint {https://arxiv.org/abs/2408.06153}
  {arXiv:2408.06153 [astro-ph.CO]} \BibitemShut {NoStop}%
\bibitem [{\citenamefont {Shan~Ba}\ and\ \citenamefont
  {Brenneman}(2015)}]{Ba2015}%
  \BibitemOpen
  \bibfield  {author} {\bibinfo {author} {\bibfnamefont {W.~R.~M.}\
  \bibnamefont {Shan~Ba}}\ and\ \bibinfo {author} {\bibfnamefont {W.~A.}\
  \bibnamefont {Brenneman}},\ }\bibfield  {title} {\bibinfo {title} {{Optimal
  Sliced Latin Hypercube Designs}},\ }\href
  {https://doi.org/10.1080/00401706.2014.957867} {\bibfield  {journal}
  {\bibinfo  {journal} {Technometrics}\ }\textbf {\bibinfo {volume} {57}},\
  \bibinfo {pages} {479} (\bibinfo {year} {2015})},\ \Eprint
  {https://arxiv.org/abs/https://doi.org/10.1080/00401706.2014.957867}
  {https://doi.org/10.1080/00401706.2014.957867} \BibitemShut {NoStop}%
\bibitem [{\citenamefont {Kennedy}\ and\ \citenamefont
  {O'Hagan}(2000)}]{Kennedy2000}%
  \BibitemOpen
  \bibfield  {author} {\bibinfo {author} {\bibfnamefont {M.}~\bibnamefont
  {Kennedy}}\ and\ \bibinfo {author} {\bibfnamefont {A.}~\bibnamefont
  {O'Hagan}},\ }\bibfield  {title} {\bibinfo {title} {{Predicting the output
  from a complex computer code when fast approximations are available}},\
  }\href {https://doi.org/10.1093/biomet/87.1.1} {\bibfield  {journal}
  {\bibinfo  {journal} {Biometrika}\ }\textbf {\bibinfo {volume} {87}},\
  \bibinfo {pages} {1} (\bibinfo {year} {2000})},\ \Eprint
  {https://arxiv.org/abs/https://academic.oup.com/biomet/article-pdf/87/1/1/590577/870001.pdf}
  {https://academic.oup.com/biomet/article-pdf/87/1/1/590577/870001.pdf}
  \BibitemShut {NoStop}%
\bibitem [{\citenamefont {{Springel}}\ and\ \citenamefont
  {{Hernquist}}(2003)}]{Springel2003}%
  \BibitemOpen
  \bibfield  {author} {\bibinfo {author} {\bibfnamefont {V.}~\bibnamefont
  {{Springel}}}\ and\ \bibinfo {author} {\bibfnamefont {L.}~\bibnamefont
  {{Hernquist}}},\ }\bibfield  {title} {\bibinfo {title} {{Cosmological
  smoothed particle hydrodynamics simulations: a hybrid multiphase model for
  star formation}},\ }\href {https://doi.org/10.1046/j.1365-8711.2003.06206.x}
  {\bibfield  {journal} {\bibinfo  {journal} {\mnras}\ }\textbf {\bibinfo
  {volume} {339}},\ \bibinfo {pages} {289} (\bibinfo {year} {2003})},\ \Eprint
  {https://arxiv.org/abs/astro-ph/0206393} {arXiv:astro-ph/0206393 [astro-ph]}
  \BibitemShut {NoStop}%
\bibitem [{\citenamefont {{Bird}}\ \emph {et~al.}(2022)\citenamefont {{Bird}},
  \citenamefont {{Ni}}, \citenamefont {{Di Matteo}}, \citenamefont {{Croft}},
  \citenamefont {{Feng}},\ and\ \citenamefont {{Chen}}}]{Bird2022}%
  \BibitemOpen
  \bibfield  {author} {\bibinfo {author} {\bibfnamefont {S.}~\bibnamefont
  {{Bird}}}, \bibinfo {author} {\bibfnamefont {Y.}~\bibnamefont {{Ni}}},
  \bibinfo {author} {\bibfnamefont {T.}~\bibnamefont {{Di Matteo}}}, \bibinfo
  {author} {\bibfnamefont {R.}~\bibnamefont {{Croft}}}, \bibinfo {author}
  {\bibfnamefont {Y.}~\bibnamefont {{Feng}}},\ and\ \bibinfo {author}
  {\bibfnamefont {N.}~\bibnamefont {{Chen}}},\ }\bibfield  {title} {\bibinfo
  {title} {{The ASTRID simulation: galaxy formation and reionization}},\ }\href
  {https://doi.org/10.1093/mnras/stac648} {\bibfield  {journal} {\bibinfo
  {journal} {\mnras}\ }\textbf {\bibinfo {volume} {512}},\ \bibinfo {pages}
  {3703} (\bibinfo {year} {2022})},\ \Eprint {https://arxiv.org/abs/2111.01160}
  {arXiv:2111.01160 [astro-ph.GA]} \BibitemShut {NoStop}%
\bibitem [{\citenamefont {Ni}\ \emph {et~al.}(2022)\citenamefont {Ni},
  \citenamefont {Di~Matteo}, \citenamefont {Bird}, \citenamefont {Croft},
  \citenamefont {Feng}, \citenamefont {Chen}, \citenamefont {Tremmel},
  \citenamefont {DeGraf},\ and\ \citenamefont {Li}}]{Ni2022}%
  \BibitemOpen
  \bibfield  {author} {\bibinfo {author} {\bibfnamefont {Y.}~\bibnamefont
  {Ni}}, \bibinfo {author} {\bibfnamefont {T.}~\bibnamefont {Di~Matteo}},
  \bibinfo {author} {\bibfnamefont {S.}~\bibnamefont {Bird}}, \bibinfo {author}
  {\bibfnamefont {R.}~\bibnamefont {Croft}}, \bibinfo {author} {\bibfnamefont
  {Y.}~\bibnamefont {Feng}}, \bibinfo {author} {\bibfnamefont {N.}~\bibnamefont
  {Chen}}, \bibinfo {author} {\bibfnamefont {M.}~\bibnamefont {Tremmel}},
  \bibinfo {author} {\bibfnamefont {C.}~\bibnamefont {DeGraf}},\ and\ \bibinfo
  {author} {\bibfnamefont {Y.}~\bibnamefont {Li}},\ }\bibfield  {title}
  {\bibinfo {title} {{The ASTRID simulation: the evolution of supermassive
  black holes}},\ }\href {https://doi.org/10.1093/mnras/stac351} {\bibfield
  {journal} {\bibinfo  {journal} {Monthly Notices of the Royal Astronomical
  Society}\ }\textbf {\bibinfo {volume} {513}},\ \bibinfo {pages} {670}
  (\bibinfo {year} {2022})},\ \Eprint
  {https://arxiv.org/abs/https://academic.oup.com/mnras/article-pdf/513/1/670/43462607/stac351.pdf}
  {https://academic.oup.com/mnras/article-pdf/513/1/670/43462607/stac351.pdf}
  \BibitemShut {NoStop}%
\bibitem [{\citenamefont {{Ni}}\ \emph {et~al.}(2024)\citenamefont {{Ni}},
  \citenamefont {{Chen}}, \citenamefont {{Zhou}}, \citenamefont {{Park}},
  \citenamefont {{Yang}}, \citenamefont {{DiMatteo}}, \citenamefont {{Bird}},\
  and\ \citenamefont {{Croft}}}]{Ni2024}%
  \BibitemOpen
  \bibfield  {author} {\bibinfo {author} {\bibfnamefont {Y.}~\bibnamefont
  {{Ni}}}, \bibinfo {author} {\bibfnamefont {N.}~\bibnamefont {{Chen}}},
  \bibinfo {author} {\bibfnamefont {Y.}~\bibnamefont {{Zhou}}}, \bibinfo
  {author} {\bibfnamefont {M.}~\bibnamefont {{Park}}}, \bibinfo {author}
  {\bibfnamefont {Y.}~\bibnamefont {{Yang}}}, \bibinfo {author} {\bibfnamefont
  {T.}~\bibnamefont {{DiMatteo}}}, \bibinfo {author} {\bibfnamefont
  {S.}~\bibnamefont {{Bird}}},\ and\ \bibinfo {author} {\bibfnamefont
  {R.}~\bibnamefont {{Croft}}},\ }\bibfield  {title} {\bibinfo {title} {{The
  Astrid Simulation: Evolution of black holes and galaxies to z=0.5 and
  different evolution pathways for galaxy quenching}},\ }\href
  {https://doi.org/10.48550/arXiv.2409.10666} {\bibfield  {journal} {\bibinfo
  {journal} {arXiv e-prints}\ ,\ \bibinfo {eid} {arXiv:2409.10666}} (\bibinfo
  {year} {2024})},\ \Eprint {https://arxiv.org/abs/2409.10666}
  {arXiv:2409.10666 [astro-ph.GA]} \BibitemShut {NoStop}%
\bibitem [{\citenamefont {{Springel}}\ \emph {et~al.}(2021)\citenamefont
  {{Springel}}, \citenamefont {{Pakmor}}, \citenamefont {{Zier}},\ and\
  \citenamefont {{Reinecke}}}]{Springel2021}%
  \BibitemOpen
  \bibfield  {author} {\bibinfo {author} {\bibfnamefont {V.}~\bibnamefont
  {{Springel}}}, \bibinfo {author} {\bibfnamefont {R.}~\bibnamefont
  {{Pakmor}}}, \bibinfo {author} {\bibfnamefont {O.}~\bibnamefont {{Zier}}},\
  and\ \bibinfo {author} {\bibfnamefont {M.}~\bibnamefont {{Reinecke}}},\
  }\bibfield  {title} {\bibinfo {title} {{Simulating cosmic structure formation
  with the GADGET-4 code}},\ }\href {https://doi.org/10.1093/mnras/stab1855}
  {\bibfield  {journal} {\bibinfo  {journal} {\mnras}\ }\textbf {\bibinfo
  {volume} {506}},\ \bibinfo {pages} {2871} (\bibinfo {year} {2021})},\ \Eprint
  {https://arxiv.org/abs/2010.03567} {arXiv:2010.03567 [astro-ph.IM]}
  \BibitemShut {NoStop}%
\bibitem [{\citenamefont {{Elbers}}\ \emph {et~al.}(2021)\citenamefont
  {{Elbers}}, \citenamefont {{Frenk}}, \citenamefont {{Jenkins}}, \citenamefont
  {{Li}},\ and\ \citenamefont {{Pascoli}}}]{Elbers2021}%
  \BibitemOpen
  \bibfield  {author} {\bibinfo {author} {\bibfnamefont {W.}~\bibnamefont
  {{Elbers}}}, \bibinfo {author} {\bibfnamefont {C.~S.}\ \bibnamefont
  {{Frenk}}}, \bibinfo {author} {\bibfnamefont {A.}~\bibnamefont {{Jenkins}}},
  \bibinfo {author} {\bibfnamefont {B.}~\bibnamefont {{Li}}},\ and\ \bibinfo
  {author} {\bibfnamefont {S.}~\bibnamefont {{Pascoli}}},\ }\bibfield  {title}
  {\bibinfo {title} {{An optimal non-linear method for simulating relic
  neutrinos}},\ }\href {https://doi.org/10.1093/mnras/stab2260} {\bibfield
  {journal} {\bibinfo  {journal} {\mnras}\ }\textbf {\bibinfo {volume} {507}},\
  \bibinfo {pages} {2614} (\bibinfo {year} {2021})},\ \Eprint
  {https://arxiv.org/abs/2010.07321} {arXiv:2010.07321 [astro-ph.CO]}
  \BibitemShut {NoStop}%
\bibitem [{\citenamefont {{Hu}}\ and\ \citenamefont
  {{Sawicki}}(2007)}]{Hu2007}%
  \BibitemOpen
  \bibfield  {author} {\bibinfo {author} {\bibfnamefont {W.}~\bibnamefont
  {{Hu}}}\ and\ \bibinfo {author} {\bibfnamefont {I.}~\bibnamefont
  {{Sawicki}}},\ }\bibfield  {title} {\bibinfo {title} {{Models of f(R) cosmic
  acceleration that evade solar system tests}},\ }\href
  {https://doi.org/10.1103/PhysRevD.76.064004} {\bibfield  {journal} {\bibinfo
  {journal} {\prd}\ }\textbf {\bibinfo {volume} {76}},\ \bibinfo {eid} {064004}
  (\bibinfo {year} {2007})},\ \Eprint {https://arxiv.org/abs/0705.1158}
  {arXiv:0705.1158 [astro-ph]} \BibitemShut {NoStop}%
\bibitem [{\citenamefont {{Dakin}}\ \emph {et~al.}(2019)\citenamefont
  {{Dakin}}, \citenamefont {{Hannestad}}, \citenamefont {{Tram}}, \citenamefont
  {{Knabenhans}},\ and\ \citenamefont {{Stadel}}}]{Dakin2019}%
  \BibitemOpen
  \bibfield  {author} {\bibinfo {author} {\bibfnamefont {J.}~\bibnamefont
  {{Dakin}}}, \bibinfo {author} {\bibfnamefont {S.}~\bibnamefont
  {{Hannestad}}}, \bibinfo {author} {\bibfnamefont {T.}~\bibnamefont {{Tram}}},
  \bibinfo {author} {\bibfnamefont {M.}~\bibnamefont {{Knabenhans}}},\ and\
  \bibinfo {author} {\bibfnamefont {J.}~\bibnamefont {{Stadel}}},\ }\bibfield
  {title} {\bibinfo {title} {{Dark energy perturbations in N-body
  simulations}},\ }\href {https://doi.org/10.1088/1475-7516/2019/08/013}
  {\bibfield  {journal} {\bibinfo  {journal} {\jcap}\ }\textbf {\bibinfo
  {volume} {2019}},\ \bibinfo {eid} {013} (\bibinfo {year} {2019})},\ \Eprint
  {https://arxiv.org/abs/1904.05210} {arXiv:1904.05210 [astro-ph.CO]}
  \BibitemShut {NoStop}%
\bibitem [{\citenamefont {{Fang}}\ \emph {et~al.}(2008)\citenamefont {{Fang}},
  \citenamefont {{Hu}},\ and\ \citenamefont {{Lewis}}}]{Fang2008}%
  \BibitemOpen
  \bibfield  {author} {\bibinfo {author} {\bibfnamefont {W.}~\bibnamefont
  {{Fang}}}, \bibinfo {author} {\bibfnamefont {W.}~\bibnamefont {{Hu}}},\ and\
  \bibinfo {author} {\bibfnamefont {A.}~\bibnamefont {{Lewis}}},\ }\bibfield
  {title} {\bibinfo {title} {{Crossing the phantom divide with parametrized
  post-Friedmann dark energy}},\ }\href
  {https://doi.org/10.1103/PhysRevD.78.087303} {\bibfield  {journal} {\bibinfo
  {journal} {\prd}\ }\textbf {\bibinfo {volume} {78}},\ \bibinfo {eid} {087303}
  (\bibinfo {year} {2008})},\ \Eprint {https://arxiv.org/abs/0808.3125}
  {arXiv:0808.3125 [astro-ph]} \BibitemShut {NoStop}%
\bibitem [{\citenamefont {Shvartsman}(1969)}]{Shvartsman1969}%
  \BibitemOpen
  \bibfield  {author} {\bibinfo {author} {\bibfnamefont {V.~F.}\ \bibnamefont
  {Shvartsman}},\ }\bibfield  {title} {\bibinfo {title} {{Density of relict
  particles with zero rest mass in the universe}},\ }\href@noop {} {\bibfield
  {journal} {\bibinfo  {journal} {Pisma Zh. Eksp. Teor. Fiz.}\ }\textbf
  {\bibinfo {volume} {9}},\ \bibinfo {pages} {315} (\bibinfo {year}
  {1969})}\BibitemShut {NoStop}%
\bibitem [{\citenamefont {Steigman}\ \emph {et~al.}(1977)\citenamefont
  {Steigman}, \citenamefont {Schramm},\ and\ \citenamefont
  {Gunn}}]{Steigman1977}%
  \BibitemOpen
  \bibfield  {author} {\bibinfo {author} {\bibfnamefont {G.}~\bibnamefont
  {Steigman}}, \bibinfo {author} {\bibfnamefont {D.~N.}\ \bibnamefont
  {Schramm}},\ and\ \bibinfo {author} {\bibfnamefont {J.~E.}\ \bibnamefont
  {Gunn}},\ }\bibfield  {title} {\bibinfo {title} {{Cosmological Limits to the
  Number of Massive Leptons}},\ }\href
  {https://doi.org/10.1016/0370-2693(77)90176-9} {\bibfield  {journal}
  {\bibinfo  {journal} {Phys. Lett. B}\ }\textbf {\bibinfo {volume} {66}},\
  \bibinfo {pages} {202} (\bibinfo {year} {1977})}\BibitemShut {NoStop}%
\bibitem [{\citenamefont {{Lewis}}\ \emph {et~al.}(2000)\citenamefont
  {{Lewis}}, \citenamefont {{Challinor}},\ and\ \citenamefont
  {{Lasenby}}}]{Lewis2000}%
  \BibitemOpen
  \bibfield  {author} {\bibinfo {author} {\bibfnamefont {A.}~\bibnamefont
  {{Lewis}}}, \bibinfo {author} {\bibfnamefont {A.}~\bibnamefont
  {{Challinor}}},\ and\ \bibinfo {author} {\bibfnamefont {A.}~\bibnamefont
  {{Lasenby}}},\ }\bibfield  {title} {\bibinfo {title} {{Efficient Computation
  of Cosmic Microwave Background Anisotropies in Closed
  Friedmann-Robertson-Walker Models}},\ }\href {https://doi.org/10.1086/309179}
  {\bibfield  {journal} {\bibinfo  {journal} {\apj}\ }\textbf {\bibinfo
  {volume} {538}},\ \bibinfo {pages} {473} (\bibinfo {year} {2000})},\ \Eprint
  {https://arxiv.org/abs/astro-ph/9911177} {arXiv:astro-ph/9911177 [astro-ph]}
  \BibitemShut {NoStop}%
\bibitem [{\citenamefont {{Zel'dovich}}(1970)}]{Zel1970}%
  \BibitemOpen
  \bibfield  {author} {\bibinfo {author} {\bibfnamefont {Y.~B.}\ \bibnamefont
  {{Zel'dovich}}},\ }\bibfield  {title} {\bibinfo {title} {{Gravitational
  instability: An approximate theory for large density perturbations.}},\
  }\href@noop {} {\bibfield  {journal} {\bibinfo  {journal} {\aap}\ }\textbf
  {\bibinfo {volume} {5}},\ \bibinfo {pages} {84} (\bibinfo {year}
  {1970})}\BibitemShut {NoStop}%
\bibitem [{\citenamefont {{Crocce}}\ \emph {et~al.}(2006)\citenamefont
  {{Crocce}}, \citenamefont {{Pueblas}},\ and\ \citenamefont
  {{Scoccimarro}}}]{Crocce2006}%
  \BibitemOpen
  \bibfield  {author} {\bibinfo {author} {\bibfnamefont {M.}~\bibnamefont
  {{Crocce}}}, \bibinfo {author} {\bibfnamefont {S.}~\bibnamefont
  {{Pueblas}}},\ and\ \bibinfo {author} {\bibfnamefont {R.}~\bibnamefont
  {{Scoccimarro}}},\ }\bibfield  {title} {\bibinfo {title} {{Transients from
  initial conditions in cosmological simulations}},\ }\href
  {https://doi.org/10.1111/j.1365-2966.2006.11040.x} {\bibfield  {journal}
  {\bibinfo  {journal} {\mnras}\ }\textbf {\bibinfo {volume} {373}},\ \bibinfo
  {pages} {369} (\bibinfo {year} {2006})},\ \Eprint
  {https://arxiv.org/abs/astro-ph/0606505} {arXiv:astro-ph/0606505 [astro-ph]}
  \BibitemShut {NoStop}%
\bibitem [{\citenamefont {Damianou}\ and\ \citenamefont
  {Lawrence}(2013)}]{Damianou:2013}%
  \BibitemOpen
  \bibfield  {author} {\bibinfo {author} {\bibfnamefont {A.}~\bibnamefont
  {Damianou}}\ and\ \bibinfo {author} {\bibfnamefont {N.}~\bibnamefont
  {Lawrence}},\ }\bibfield  {title} {\bibinfo {title} {Deep gaussian
  processes},\ }in\ \href {http://proceedings.mlr.press/v31/damianou13a.html}
  {\emph {\bibinfo {booktitle} {Proceedings of the Sixteenth International
  Conference on Artificial Intelligence and Statistics}}},\ \bibinfo {series}
  {Proceedings of Machine Learning Research}, Vol.~\bibinfo {volume} {31},\
  \bibinfo {editor} {edited by\ \bibinfo {editor} {\bibfnamefont {C.~M.}\
  \bibnamefont {Carvalho}}\ and\ \bibinfo {editor} {\bibfnamefont
  {P.}~\bibnamefont {Ravikumar}}}\ (\bibinfo  {publisher} {PMLR},\ \bibinfo
  {address} {Scottsdale, Arizona, USA},\ \bibinfo {year} {2013})\ pp.\ \bibinfo
  {pages} {207--215}\BibitemShut {NoStop}%
\bibitem [{\citenamefont {{Perdikaris}}\ \emph {et~al.}(2017)\citenamefont
  {{Perdikaris}}, \citenamefont {{Raissi}}, \citenamefont {{Damianou}},
  \citenamefont {{Lawrence}},\ and\ \citenamefont
  {{Karniadakis}}}]{Perdikaris:2017}%
  \BibitemOpen
  \bibfield  {author} {\bibinfo {author} {\bibfnamefont {P.}~\bibnamefont
  {{Perdikaris}}}, \bibinfo {author} {\bibfnamefont {M.}~\bibnamefont
  {{Raissi}}}, \bibinfo {author} {\bibfnamefont {A.}~\bibnamefont
  {{Damianou}}}, \bibinfo {author} {\bibfnamefont {N.~D.}\ \bibnamefont
  {{Lawrence}}},\ and\ \bibinfo {author} {\bibfnamefont {G.~E.}\ \bibnamefont
  {{Karniadakis}}},\ }\bibfield  {title} {\bibinfo {title} {{Nonlinear
  information fusion algorithms for data-efficient multi-fidelity modelling}},\
  }\bibfield  {journal} {\bibinfo  {journal} {Proc. R. Soc. A.}\ }\textbf
  {\bibinfo {volume} {473}},\ \href
  {https://doi.org/http://doi.org/10.1098/rspa.2016.0751}
  {http://doi.org/10.1098/rspa.2016.0751} (\bibinfo {year} {2017})\BibitemShut
  {NoStop}%
\bibitem [{\citenamefont {Salvatier}\ \emph {et~al.}(2016)\citenamefont
  {Salvatier}, \citenamefont {Wiecki},\ and\ \citenamefont
  {Fonnesbeck}}]{Salvatier2016}%
  \BibitemOpen
  \bibfield  {author} {\bibinfo {author} {\bibfnamefont {J.}~\bibnamefont
  {Salvatier}}, \bibinfo {author} {\bibfnamefont {T.}~\bibnamefont {Wiecki}},\
  and\ \bibinfo {author} {\bibfnamefont {C.}~\bibnamefont {Fonnesbeck}},\
  }\bibfield  {title} {\bibinfo {title} {{Probabilistic programming in Python
  using PyMC3}},\ }\bibfield  {journal} {\bibinfo  {journal} {PeerJ Computer
  Science}\ }\textbf {\bibinfo {volume} {2:e55}},\ \href
  {https://doi.org/10.7717/peerj-cs.55} {10.7717/peerj-cs.55} (\bibinfo {year}
  {2016})\BibitemShut {NoStop}%
\bibitem [{\citenamefont {Virtanen}\ \emph {et~al.}(2020)\citenamefont
  {Virtanen}, \citenamefont {Gommers}, \citenamefont {Oliphant}, \citenamefont
  {Haberland}, \citenamefont {Reddy}, \citenamefont {Cournapeau}, \citenamefont
  {Burovski}, \citenamefont {Peterson}, \citenamefont {Weckesser},
  \citenamefont {Bright}, \citenamefont {{van der Walt}}, \citenamefont
  {Brett}, \citenamefont {Wilson}, \citenamefont {Millman}, \citenamefont
  {Mayorov}, \citenamefont {Nelson}, \citenamefont {Jones}, \citenamefont
  {Kern}, \citenamefont {Larson}, \citenamefont {Carey}, \citenamefont {Polat},
  \citenamefont {Feng}, \citenamefont {Moore}, \citenamefont {{VanderPlas}},
  \citenamefont {Laxalde}, \citenamefont {Perktold}, \citenamefont {Cimrman},
  \citenamefont {Henriksen}, \citenamefont {Quintero}, \citenamefont {Harris},
  \citenamefont {Archibald}, \citenamefont {Ribeiro}, \citenamefont
  {Pedregosa}, \citenamefont {{van Mulbregt}},\ and\ \citenamefont {{SciPy 1.0
  Contributors}}}]{Virtanen2020}%
  \BibitemOpen
  \bibfield  {author} {\bibinfo {author} {\bibfnamefont {P.}~\bibnamefont
  {Virtanen}}, \bibinfo {author} {\bibfnamefont {R.}~\bibnamefont {Gommers}},
  \bibinfo {author} {\bibfnamefont {T.~E.}\ \bibnamefont {Oliphant}}, \bibinfo
  {author} {\bibfnamefont {M.}~\bibnamefont {Haberland}}, \bibinfo {author}
  {\bibfnamefont {T.}~\bibnamefont {Reddy}}, \bibinfo {author} {\bibfnamefont
  {D.}~\bibnamefont {Cournapeau}}, \bibinfo {author} {\bibfnamefont
  {E.}~\bibnamefont {Burovski}}, \bibinfo {author} {\bibfnamefont
  {P.}~\bibnamefont {Peterson}}, \bibinfo {author} {\bibfnamefont
  {W.}~\bibnamefont {Weckesser}}, \bibinfo {author} {\bibfnamefont
  {J.}~\bibnamefont {Bright}}, \bibinfo {author} {\bibfnamefont {S.~J.}\
  \bibnamefont {{van der Walt}}}, \bibinfo {author} {\bibfnamefont
  {M.}~\bibnamefont {Brett}}, \bibinfo {author} {\bibfnamefont
  {J.}~\bibnamefont {Wilson}}, \bibinfo {author} {\bibfnamefont {K.~J.}\
  \bibnamefont {Millman}}, \bibinfo {author} {\bibfnamefont {N.}~\bibnamefont
  {Mayorov}}, \bibinfo {author} {\bibfnamefont {A.~R.~J.}\ \bibnamefont
  {Nelson}}, \bibinfo {author} {\bibfnamefont {E.}~\bibnamefont {Jones}},
  \bibinfo {author} {\bibfnamefont {R.}~\bibnamefont {Kern}}, \bibinfo {author}
  {\bibfnamefont {E.}~\bibnamefont {Larson}}, \bibinfo {author} {\bibfnamefont
  {C.~J.}\ \bibnamefont {Carey}}, \bibinfo {author} {\bibfnamefont
  {{\.I}.}~\bibnamefont {Polat}}, \bibinfo {author} {\bibfnamefont
  {Y.}~\bibnamefont {Feng}}, \bibinfo {author} {\bibfnamefont {E.~W.}\
  \bibnamefont {Moore}}, \bibinfo {author} {\bibfnamefont {J.}~\bibnamefont
  {{VanderPlas}}}, \bibinfo {author} {\bibfnamefont {D.}~\bibnamefont
  {Laxalde}}, \bibinfo {author} {\bibfnamefont {J.}~\bibnamefont {Perktold}},
  \bibinfo {author} {\bibfnamefont {R.}~\bibnamefont {Cimrman}}, \bibinfo
  {author} {\bibfnamefont {I.}~\bibnamefont {Henriksen}}, \bibinfo {author}
  {\bibfnamefont {E.~A.}\ \bibnamefont {Quintero}}, \bibinfo {author}
  {\bibfnamefont {C.~R.}\ \bibnamefont {Harris}}, \bibinfo {author}
  {\bibfnamefont {A.~M.}\ \bibnamefont {Archibald}}, \bibinfo {author}
  {\bibfnamefont {A.~H.}\ \bibnamefont {Ribeiro}}, \bibinfo {author}
  {\bibfnamefont {F.}~\bibnamefont {Pedregosa}}, \bibinfo {author}
  {\bibfnamefont {P.}~\bibnamefont {{van Mulbregt}}},\ and\ \bibinfo {author}
  {\bibnamefont {{SciPy 1.0 Contributors}}},\ }\bibfield  {title} {\bibinfo
  {title} {{{SciPy} 1.0: Fundamental Algorithms for Scientific Computing in
  Python}},\ }\href {https://doi.org/10.1038/s41592-019-0686-2} {\bibfield
  {journal} {\bibinfo  {journal} {Nature Methods}\ }\textbf {\bibinfo {volume}
  {17}},\ \bibinfo {pages} {261} (\bibinfo {year} {2020})}\BibitemShut
  {NoStop}%
\bibitem [{\citenamefont {{Davis}}\ \emph {et~al.}(1985)\citenamefont
  {{Davis}}, \citenamefont {{Efstathiou}}, \citenamefont {{Frenk}},\ and\
  \citenamefont {{White}}}]{Davis1985}%
  \BibitemOpen
  \bibfield  {author} {\bibinfo {author} {\bibfnamefont {M.}~\bibnamefont
  {{Davis}}}, \bibinfo {author} {\bibfnamefont {G.}~\bibnamefont
  {{Efstathiou}}}, \bibinfo {author} {\bibfnamefont {C.~S.}\ \bibnamefont
  {{Frenk}}},\ and\ \bibinfo {author} {\bibfnamefont {S.~D.~M.}\ \bibnamefont
  {{White}}},\ }\bibfield  {title} {\bibinfo {title} {{The evolution of
  large-scale structure in a universe dominated by cold dark matter}},\ }\href
  {https://doi.org/10.1086/163168} {\bibfield  {journal} {\bibinfo  {journal}
  {\apj}\ }\textbf {\bibinfo {volume} {292}},\ \bibinfo {pages} {371} (\bibinfo
  {year} {1985})}\BibitemShut {NoStop}%
\bibitem [{\citenamefont {{Shen}}\ \emph {et~al.}(2024)\citenamefont {{Shen}},
  \citenamefont {{Kokron}}, \citenamefont {{DeRose}}, \citenamefont {{Tinker}},
  \citenamefont {{Wechsler}}, \citenamefont {{Banerjee}},\ and\ \citenamefont
  {{the Aemulus Collaboration}}}]{Shen2024}%
  \BibitemOpen
  \bibfield  {author} {\bibinfo {author} {\bibfnamefont {D.}~\bibnamefont
  {{Shen}}}, \bibinfo {author} {\bibfnamefont {N.}~\bibnamefont {{Kokron}}},
  \bibinfo {author} {\bibfnamefont {J.}~\bibnamefont {{DeRose}}}, \bibinfo
  {author} {\bibfnamefont {J.}~\bibnamefont {{Tinker}}}, \bibinfo {author}
  {\bibfnamefont {R.~H.}\ \bibnamefont {{Wechsler}}}, \bibinfo {author}
  {\bibfnamefont {A.}~\bibnamefont {{Banerjee}}},\ and\ \bibinfo {author}
  {\bibnamefont {{the Aemulus Collaboration}}},\ }\bibfield  {title} {\bibinfo
  {title} {{Aemulus $\nu$: Precision halo mass functions in w$\nu$CDM
  cosmologies}},\ }\href {https://doi.org/10.48550/arXiv.2410.00913} {\bibfield
   {journal} {\bibinfo  {journal} {arXiv e-prints}\ ,\ \bibinfo {eid}
  {arXiv:2410.00913}} (\bibinfo {year} {2024})},\ \Eprint
  {https://arxiv.org/abs/2410.00913} {arXiv:2410.00913 [astro-ph.CO]}
  \BibitemShut {NoStop}%
\bibitem [{\citenamefont {{Salucci}}(2019)}]{Salucci2019}%
  \BibitemOpen
  \bibfield  {author} {\bibinfo {author} {\bibfnamefont {P.}~\bibnamefont
  {{Salucci}}},\ }\bibfield  {title} {\bibinfo {title} {{The distribution of
  dark matter in galaxies}},\ }\href
  {https://doi.org/10.1007/s00159-018-0113-1} {\bibfield  {journal} {\bibinfo
  {journal} {\aapr}\ }\textbf {\bibinfo {volume} {27}},\ \bibinfo {eid} {2}
  (\bibinfo {year} {2019})},\ \Eprint {https://arxiv.org/abs/1811.08843}
  {arXiv:1811.08843 [astro-ph.GA]} \BibitemShut {NoStop}%
\bibitem [{\citenamefont {{Marinacci}}\ \emph {et~al.}(2018)\citenamefont
  {{Marinacci}}, \citenamefont {{Vogelsberger}}, \citenamefont {{Pakmor}},
  \citenamefont {{Torrey}}, \citenamefont {{Springel}}, \citenamefont
  {{Hernquist}}, \citenamefont {{Nelson}}, \citenamefont {{Weinberger}},
  \citenamefont {{Pillepich}}, \citenamefont {{Naiman}},\ and\ \citenamefont
  {{Genel}}}]{Marinacci2018}%
  \BibitemOpen
  \bibfield  {author} {\bibinfo {author} {\bibfnamefont {F.}~\bibnamefont
  {{Marinacci}}}, \bibinfo {author} {\bibfnamefont {M.}~\bibnamefont
  {{Vogelsberger}}}, \bibinfo {author} {\bibfnamefont {R.}~\bibnamefont
  {{Pakmor}}}, \bibinfo {author} {\bibfnamefont {P.}~\bibnamefont {{Torrey}}},
  \bibinfo {author} {\bibfnamefont {V.}~\bibnamefont {{Springel}}}, \bibinfo
  {author} {\bibfnamefont {L.}~\bibnamefont {{Hernquist}}}, \bibinfo {author}
  {\bibfnamefont {D.}~\bibnamefont {{Nelson}}}, \bibinfo {author}
  {\bibfnamefont {R.}~\bibnamefont {{Weinberger}}}, \bibinfo {author}
  {\bibfnamefont {A.}~\bibnamefont {{Pillepich}}}, \bibinfo {author}
  {\bibfnamefont {J.}~\bibnamefont {{Naiman}}},\ and\ \bibinfo {author}
  {\bibfnamefont {S.}~\bibnamefont {{Genel}}},\ }\bibfield  {title} {\bibinfo
  {title} {{First results from the IllustrisTNG simulations: radio haloes and
  magnetic fields}},\ }\href {https://doi.org/10.1093/mnras/sty2206} {\bibfield
   {journal} {\bibinfo  {journal} {\mnras}\ }\textbf {\bibinfo {volume}
  {480}},\ \bibinfo {pages} {5113} (\bibinfo {year} {2018})},\ \Eprint
  {https://arxiv.org/abs/1707.03396} {arXiv:1707.03396 [astro-ph.CO]}
  \BibitemShut {NoStop}%
\bibitem [{\citenamefont {{Schaye}}\ \emph {et~al.}(2015)\citenamefont
  {{Schaye}}, \citenamefont {{Crain}}, \citenamefont {{Bower}}, \citenamefont
  {{Furlong}}, \citenamefont {{Schaller}}, \citenamefont {{Theuns}},
  \citenamefont {{Dalla Vecchia}}, \citenamefont {{Frenk}}, \citenamefont
  {{McCarthy}}, \citenamefont {{Helly}}, \citenamefont {{Jenkins}},
  \citenamefont {{Rosas-Guevara}}, \citenamefont {{White}}, \citenamefont
  {{Baes}}, \citenamefont {{Booth}}, \citenamefont {{Camps}}, \citenamefont
  {{Navarro}}, \citenamefont {{Qu}}, \citenamefont {{Rahmati}}, \citenamefont
  {{Sawala}}, \citenamefont {{Thomas}},\ and\ \citenamefont
  {{Trayford}}}]{Schaye2015}%
  \BibitemOpen
  \bibfield  {author} {\bibinfo {author} {\bibfnamefont {J.}~\bibnamefont
  {{Schaye}}}, \bibinfo {author} {\bibfnamefont {R.~A.}\ \bibnamefont
  {{Crain}}}, \bibinfo {author} {\bibfnamefont {R.~G.}\ \bibnamefont
  {{Bower}}}, \bibinfo {author} {\bibfnamefont {M.}~\bibnamefont {{Furlong}}},
  \bibinfo {author} {\bibfnamefont {M.}~\bibnamefont {{Schaller}}}, \bibinfo
  {author} {\bibfnamefont {T.}~\bibnamefont {{Theuns}}}, \bibinfo {author}
  {\bibfnamefont {C.}~\bibnamefont {{Dalla Vecchia}}}, \bibinfo {author}
  {\bibfnamefont {C.~S.}\ \bibnamefont {{Frenk}}}, \bibinfo {author}
  {\bibfnamefont {I.~G.}\ \bibnamefont {{McCarthy}}}, \bibinfo {author}
  {\bibfnamefont {J.~C.}\ \bibnamefont {{Helly}}}, \bibinfo {author}
  {\bibfnamefont {A.}~\bibnamefont {{Jenkins}}}, \bibinfo {author}
  {\bibfnamefont {Y.~M.}\ \bibnamefont {{Rosas-Guevara}}}, \bibinfo {author}
  {\bibfnamefont {S.~D.~M.}\ \bibnamefont {{White}}}, \bibinfo {author}
  {\bibfnamefont {M.}~\bibnamefont {{Baes}}}, \bibinfo {author} {\bibfnamefont
  {C.~M.}\ \bibnamefont {{Booth}}}, \bibinfo {author} {\bibfnamefont
  {P.}~\bibnamefont {{Camps}}}, \bibinfo {author} {\bibfnamefont {J.~F.}\
  \bibnamefont {{Navarro}}}, \bibinfo {author} {\bibfnamefont {Y.}~\bibnamefont
  {{Qu}}}, \bibinfo {author} {\bibfnamefont {A.}~\bibnamefont {{Rahmati}}},
  \bibinfo {author} {\bibfnamefont {T.}~\bibnamefont {{Sawala}}}, \bibinfo
  {author} {\bibfnamefont {P.~A.}\ \bibnamefont {{Thomas}}},\ and\ \bibinfo
  {author} {\bibfnamefont {J.}~\bibnamefont {{Trayford}}},\ }\bibfield  {title}
  {\bibinfo {title} {{The EAGLE project: simulating the evolution and assembly
  of galaxies and their environments}},\ }\href
  {https://doi.org/10.1093/mnras/stu2058} {\bibfield  {journal} {\bibinfo
  {journal} {\mnras}\ }\textbf {\bibinfo {volume} {446}},\ \bibinfo {pages}
  {521} (\bibinfo {year} {2015})},\ \Eprint {https://arxiv.org/abs/1407.7040}
  {arXiv:1407.7040 [astro-ph.GA]} \BibitemShut {NoStop}%
\bibitem [{\citenamefont {{Guo}}\ \emph {et~al.}(2022)\citenamefont {{Guo}},
  \citenamefont {{Manzoni}}, \citenamefont {{Amendt}}, \citenamefont
  {{Conti}},\ and\ \citenamefont {{Hesthaven}}}]{Guo2022}%
  \BibitemOpen
  \bibfield  {author} {\bibinfo {author} {\bibfnamefont {M.}~\bibnamefont
  {{Guo}}}, \bibinfo {author} {\bibfnamefont {A.}~\bibnamefont {{Manzoni}}},
  \bibinfo {author} {\bibfnamefont {M.}~\bibnamefont {{Amendt}}}, \bibinfo
  {author} {\bibfnamefont {P.}~\bibnamefont {{Conti}}},\ and\ \bibinfo {author}
  {\bibfnamefont {J.~S.}\ \bibnamefont {{Hesthaven}}},\ }\bibfield  {title}
  {\bibinfo {title} {{Multi-fidelity regression using artificial neural
  networks: Efficient approximation of parameter-dependent output
  quantities}},\ }\href {https://doi.org/10.1016/j.cma.2021.114378} {\bibfield
  {journal} {\bibinfo  {journal} {Computer Methods in Applied Mechanics and
  Engineering}\ }\textbf {\bibinfo {volume} {389}},\ \bibinfo {pages} {114378}
  (\bibinfo {year} {2022})},\ \Eprint {https://arxiv.org/abs/2102.13403}
  {arXiv:2102.13403 [math.NA]} \BibitemShut {NoStop}%
\bibitem [{\citenamefont {{Diao}}\ and\ \citenamefont
  {{Mao}}(2023)}]{Diao2023}%
  \BibitemOpen
  \bibfield  {author} {\bibinfo {author} {\bibfnamefont {K.}~\bibnamefont
  {{Diao}}}\ and\ \bibinfo {author} {\bibfnamefont {Y.}~\bibnamefont {{Mao}}},\
  }\bibfield  {title} {\bibinfo {title} {{Multi-fidelity Emulator for
  Cosmological Large Scale 21 cm Lightcone Images: a Few-shot Transfer Learning
  Approach with GAN}},\ }\href {https://doi.org/10.48550/arXiv.2307.04976}
  {\bibfield  {journal} {\bibinfo  {journal} {arXiv e-prints}\ ,\ \bibinfo
  {eid} {arXiv:2307.04976}} (\bibinfo {year} {2023})},\ \Eprint
  {https://arxiv.org/abs/2307.04976} {arXiv:2307.04976 [astro-ph.CO]}
  \BibitemShut {NoStop}%
\bibitem [{\citenamefont {{Yang}}\ \emph {et~al.}(2025)\citenamefont {{Yang}},
  \citenamefont {{Bird}},\ and\ \citenamefont {{Ho}}}]{GokuEmu2025}%
  \BibitemOpen
  \bibfield  {author} {\bibinfo {author} {\bibfnamefont {Y.}~\bibnamefont
  {{Yang}}}, \bibinfo {author} {\bibfnamefont {S.}~\bibnamefont {{Bird}}},\
  and\ \bibinfo {author} {\bibfnamefont {M.-F.}\ \bibnamefont {{Ho}}},\ }\href
  {https://github.com/astro-YYH/GokuEmu} {\bibinfo {title} {{GokuEmu and the
  associated data}}} (\bibinfo {year} {2025}),\ \bibinfo {note} {{GitHub
  repository}}\BibitemShut {NoStop}%
\bibitem [{\citenamefont {{van Daalen}}\ \emph {et~al.}(2011)\citenamefont
  {{van Daalen}}, \citenamefont {{Schaye}}, \citenamefont {{Booth}},\ and\
  \citenamefont {{Dalla Vecchia}}}]{vanDaalen2011}%
  \BibitemOpen
  \bibfield  {author} {\bibinfo {author} {\bibfnamefont {M.~P.}\ \bibnamefont
  {{van Daalen}}}, \bibinfo {author} {\bibfnamefont {J.}~\bibnamefont
  {{Schaye}}}, \bibinfo {author} {\bibfnamefont {C.~M.}\ \bibnamefont
  {{Booth}}},\ and\ \bibinfo {author} {\bibfnamefont {C.}~\bibnamefont {{Dalla
  Vecchia}}},\ }\bibfield  {title} {\bibinfo {title} {{The effects of galaxy
  formation on the matter power spectrum: a challenge for precision
  cosmology}},\ }\href {https://doi.org/10.1111/j.1365-2966.2011.18981.x}
  {\bibfield  {journal} {\bibinfo  {journal} {\mnras}\ }\textbf {\bibinfo
  {volume} {415}},\ \bibinfo {pages} {3649} (\bibinfo {year} {2011})},\ \Eprint
  {https://arxiv.org/abs/1104.1174} {arXiv:1104.1174 [astro-ph.CO]}
  \BibitemShut {NoStop}%
\bibitem [{\citenamefont {{Mead}}\ \emph {et~al.}(2021)\citenamefont {{Mead}},
  \citenamefont {{Brieden}}, \citenamefont {{Tr{\"o}ster}},\ and\ \citenamefont
  {{Heymans}}}]{Mead2021}%
  \BibitemOpen
  \bibfield  {author} {\bibinfo {author} {\bibfnamefont {A.~J.}\ \bibnamefont
  {{Mead}}}, \bibinfo {author} {\bibfnamefont {S.}~\bibnamefont {{Brieden}}},
  \bibinfo {author} {\bibfnamefont {T.}~\bibnamefont {{Tr{\"o}ster}}},\ and\
  \bibinfo {author} {\bibfnamefont {C.}~\bibnamefont {{Heymans}}},\ }\bibfield
  {title} {\bibinfo {title} {{HMCODE-2020: improved modelling of non-linear
  cosmological power spectra with baryonic feedback}},\ }\href
  {https://doi.org/10.1093/mnras/stab082} {\bibfield  {journal} {\bibinfo
  {journal} {\mnras}\ }\textbf {\bibinfo {volume} {502}},\ \bibinfo {pages}
  {1401} (\bibinfo {year} {2021})},\ \Eprint {https://arxiv.org/abs/2009.01858}
  {arXiv:2009.01858 [astro-ph.CO]} \BibitemShut {NoStop}%
\end{thebibliography}%

\appendix

\section{\label{app:pplusf_and_largerbox}P+F vs. Larger-Volume Simulation}

\begin{figure}
    \includegraphics[width=\linewidth]{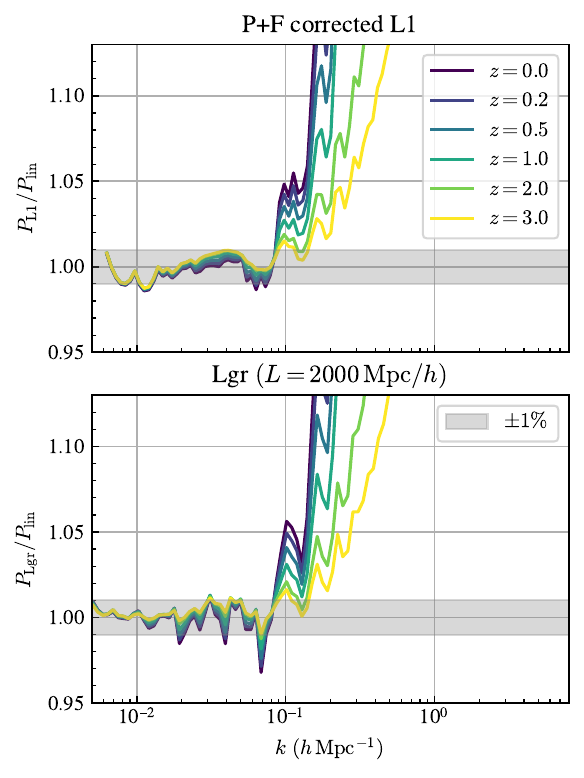}
    \caption{\label{fig:pplusf_vs_largerbox}\textit{Upper panel:} The ratio of the P+F-corrected L1 matter power spectrum of \texttt{Goku-N-0195} to the corresponding linear spectrum at various redshifts ($z=0, 0.2, 0.5, 1, 2, 3$). Each color represents a different redshift. The gray-shaded region highlights the range of relative differences within $1\%$. \textit{Lower panel:} The ratio of the matter power spectrum of the larger-volume simulation (Lgr) to the linear theory prediction for the same cosmology and redshifts.}
\end{figure}

In Fig.~\ref{fig:pplusf_vs_largerbox}, we compare the P+F-corrected matter power spectrum from the L1 simulation of \texttt{Goku-N-0195} with the spectrum from the larger-volume simulation (Lgr). The comparison is presented as ratios to the linear theory prediction in the two panels. Both the P+F-corrected L1 and the Lgr spectra exhibit good consistency with linear theory at large scales. However, noticeable differences emerge between the two. At the largest scales, around $k\sim 0.01h\,$Mpc$^{-1}$, Lgr aligns more closely with the linear theory. In contrast, at intermediate scales ($k\sim 0.07h\,$Mpc$^{-1}$), particularly at lower redshifts, Lgr shows larger deviations. In addition,the P+F spectrum is overall smoother compared to the Lgr spectrum.

\begin{figure}
    \includegraphics[width=\linewidth]{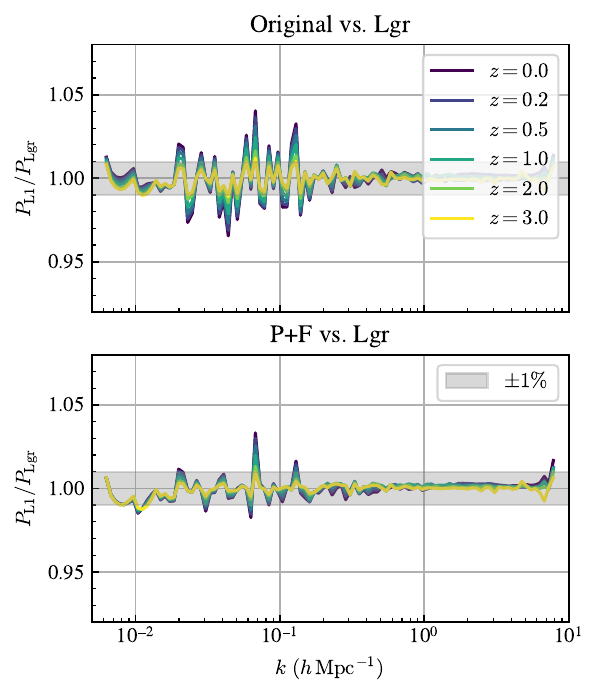}
    \caption{\label{fig:matter_pow_vs_Lgr}\textit{Upper panel:} The ratio of the original L1 matter power spectrum of \texttt{Goku-N-0195} to the corresponding larger-volume simulation at various redshifts ($z=0, 0.2, 0.5, 1, 2, 3$). Each color represents a different redshift. The gray-shaded region highlights the range of relative differences within $1\%$. \textit{Lower panel:} The ratio of the P+F-corrected L1 matter power spectrum to the larger-volume counterpart for the same cosmology and redshifts, showing improved consistency at large scales due to the P+F correction.}
\end{figure}

The upper panel of Fig.~\ref{fig:matter_pow_vs_Lgr} presents a convergence test between the original L1 and Lgr simulations of \texttt{Goku-N-0195}. The relative differences between are generally within $1\%$ at most scales, though they reach a few percent at some $k$-modes, particularly at lower redshifts. The lower panel plots the ratio of the P+F-corrected spectrum to the Lgr spectrum, showing significant improvements in consistency across all redshifts. This demonstrates the effectiveness of the P+F technique in reducing cosmic variance. The observed $3\%$ deviation at $k\sim 0.07h\,$Mpc$^{-1}$ and $z=0$ corresponds to the dip in the Lgr spectrum seen in Fig.~\ref{fig:pplusf_vs_largerbox}, and the P+F spectrum is likely the more accurate representation of the matter power spectrum at this scale.

The results suggest that the P+F correction is highly effective in reducing cosmic variance at large scales. The corrected spectrum not only exhibits improved smoothness but also shows better consistency with linear theory compared to the larger-volume simulation at specific scales.

\section{Error Function Estimation\label{app:error_function}}
The analytic formula used in Ref.~\cite{Ho2023} for the error function is (simplified such that there is no dimensionality dependence)
\begin{equation}
    \Phi_\text{old} (n_\text{L}, n_\text{H}) = \eta (\rho n_\text{L}^{-\beta_\text{L}} + n_\text{H}^{-\beta_\text{H}}).\label{eq:error_function_Ho}
\end{equation}
which provided a good fit to the error function in Ref.~\cite{Ho2023}. However, we found that the error function is not well described by $\Phi_\text{old}$ (Eq.~(\ref{eq:error_function_Ho})), as the properties of $\Phi_\text{old}$ are not inherently consistent with the error function in our case. For example, we expect the error function to go to zero as $n_\text{H}$ goes to infinity regardless of the value of $n_\text{L}$, i.e., there would be literally no interpolation error if an infinitely large number of HF simulations are used for prediction, but this is not the case for $\Phi_\text{old}$, which would asymptote to $\eta \rho n_\text{L}^{-\beta_L}$ as $n_\text{H}$ goes to infinity, implying that the error could be estimated to be a large value (when $n_\text{L}$ is small) even if a huge number of HF simulations are used. In addition, we should expect the error function to be finite when only one of the two variables goes to zero, since a single-fidelity emulator can still perform well as long as the training set is large enough. But $\Phi_\text{old}$ diverges as $n_\text{L}$ or $n_\text{H}$ goes to zero.

We therefore propose the new formula ($\Phi$ defined by Eq.~(\ref{eq:error_function})) to approximate the error function, whose properties are naturally consistent with the true error function. Specifically, $\Phi $ meets all the requirements that we expect from the error function:
\begin{itemize}
    \item $\Phi \rightarrow f(n_\text{H})$ as $n_\text{L}\rightarrow \infty$;
    \item $\Phi \rightarrow 0$ as $n_\text{H}\rightarrow \infty$;
    \item $\Phi \rightarrow \infty$ as $n_\text{L}\rightarrow 0$ and $n_\text{H}\rightarrow 0$;
    \item $\Phi \rightarrow g(n_\text{L})$ as $n_\text{H}\rightarrow 0$;
    \item $\Phi \rightarrow h(n_\text{H})$ as $n_\text{L}\rightarrow 0$.
\end{itemize}
The explicit forms of $f$, $g$, and $h$ can be derived from Eq.~(\ref{eq:error_function}). For example, $f(n_\text{H}) = \eta / (n_\text{H} + \alpha_2)^{\beta_2}$.

\begin{figure*}[t]
    \includegraphics[width=\textwidth]{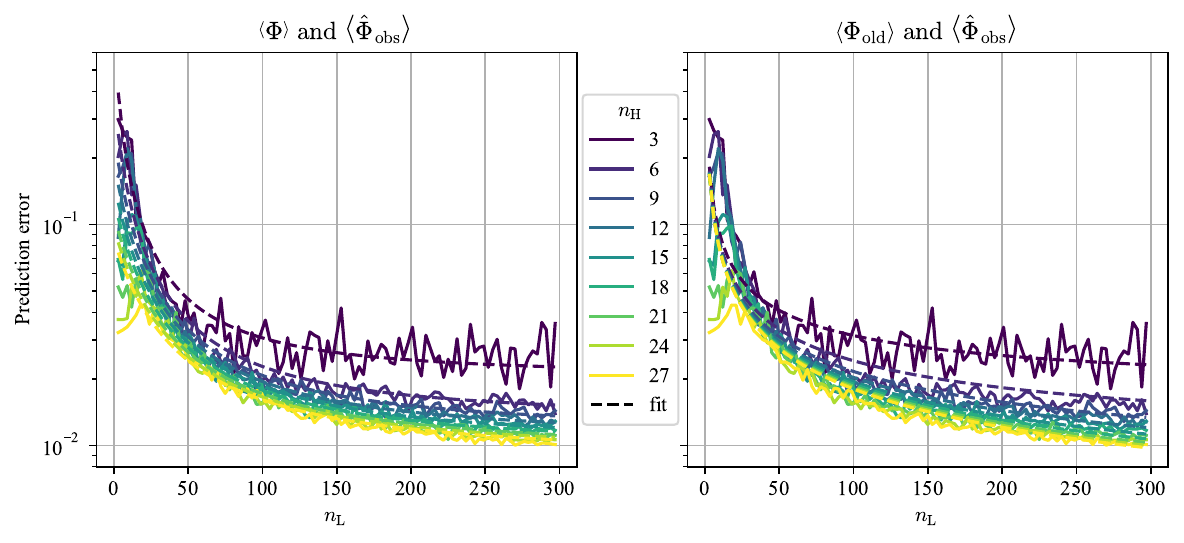}
    \caption{\label{fig:error_funcs}Expected (mean) prediction error as a function of the numbers of LF pairs ($n_\text{L}$) and HF ($n_\text{H}$) simulations. Solid curves represent the observed mean prediction error $\left<\hat\Phi_\text{obs}\right>$, and dashed curves are the estimated mean error functions $\left<\Phi\right>$ (\textit{left panel}) and $\left<\Phi_\text{old}\right>$ (\textit{right panel}).}
\end{figure*}

We compare the two formulae based on the data of \texttt{Goku-pre-N} (used to estimate the error function for \texttt{Goku-N}). Using MCMC analysis, we obtained the best-fit parameters for $\Phi_\text{old}$, with values listed in Table~\ref{tab:error_old_params}. As illustrated in Fig.~\ref{fig:error_funcs}, $\Phi$ more accurately reflects the error distribution than $\Phi_\text{old}$, particularly in high-error regions  (i.e., for small values of $n_\text{L}$ and/or $n_\text{H}$). For instance, for $n_\text{L} < 50$, the old formula fails to reflect the error's dependence on the number of HF points. Also, for intermediate sample sizes ($50< n_\text{L}<200$), $\left<\Phi_\text{old}\right>$ tends to overestimate the prediction error. Additionally, while differences between data and fit are small in high-sample regions, $\Phi_\text{old}$ may underestimate the error outside the data boundary (e.g., $n_\text{L}=564$ and $n_\text{H}=27$), as suggested by its derivative behavior.

\begin{table}
    \caption{\label{tab:error_old_params}Parameters of $\Phi_\text{old}$ (Eq.~(\ref{eq:error_function_Ho})) obtained from our MCMC analysis on \texttt{Goku-pre-N}, under the assumption $\log\hat{\Phi}_\text{old}(n_\text{L}, n_\text{H})\sim\mathcal{N}\left(\mu (n_\text{L}, n_\text{H}), \sigma^2\right)$ where $\mu (n_\text{L}, n_\text{H}) = \log \Phi_\text{old} (n_\text{L}, n_\text{H})$.}
    \begin{ruledtabular}
    \begin{tabular}{lcc}
    \textrm{Parameter} & \textrm{Prior} & \textrm{Median} \\
    \colrule
    $\eta$ & Normal($\mu = 1, \sigma=0.3$) & $0.0410^{+0.0025}_{-0.0023}$ \\
    $\rho$ & LogNormal($\mu=0$, $\sigma=1$) & $8.278^{+0.700}_{-0.663}$ \\
    $\beta_\text{L}$ & Normal($\mu=1$, $\sigma=0.2$) & $0.671^{+0.012}_{-0.012}$ \\
    $\beta_\text{H}$ & Normal($\mu=1$, $\sigma=0.2$) & $0.908^{+0.056}_{-0.052}$ \\
    $\sigma$ & HalfNormal($\sigma=0.1$) & $0.102^{+0.0014}_{-0.0014}$ \\
    \end{tabular}
    \end{ruledtabular} 
\end{table}

\section{Cosmologies of the HF Simulations\label{app:HF_cosmologies}}
The cosmological parameters of the HF simulations are summarized in Table~\ref{tab:HF_cosmo}.

\begin{table*}
    \caption{\label{tab:HF_cosmo}%
    The values of the cosmological parameters of the HF simulations. \texttt{W} and \texttt{N} are short for \texttt{Goku-W} and \texttt{Goku-N} respectively.}
    \begin{ruledtabular}
    \begin{tabular}{lcccccccccc}
    \textrm{Cosmology}&
    $\Omega_\text{m}$ & $\Omega_\text{b}$ & $h$ & $A_\text{s}/10^{-9}$ & $n_\text{s}$ & $w_0$ & $w_a$ & $\sum m_\nu/\text{eV}$ & $N_\text{eff}$ & $\alpha_\text{s}/10^{-2}$\\
    \colrule
    \texttt{W-0024} & 0.249 & 0.0425 & 0.6281 & $2.991$ & 0.915 & -0.648 & -0.688 & 0.494 & 4.140 & -0.02 \\
    \texttt{W-0025} & 0.311 & 0.0461 & 0.6677 & $1.055$ & 1.077 & -0.851 & -1.924 & 0.387 & 2.248 & 1.94 \\
    \texttt{W-0026} & 0.362 & 0.0504 & 0.7134 & $1.841$ & 0.801 & 0.0836 & -0.646 & 0.0796 & 3.057 & -3.31 \\
    \texttt{W-0054} & 0.276 & 0.0493 & 0.6605 & $1.588$ & 1.074 & -1.218 & -1.444 & 0.0500 & 2.683 & 2.62 \\
    \texttt{W-0055} & 0.293 & 0.0424 & 0.7147 & $1.724$ & 0.917 & 0.152 & -2.554 & 0.308 & 3.992 & -4.86 \\
    \texttt{W-0056} & 0.379 & 0.0530 & 0.6197 & $2.673$ & 0.869 & -0.363 & 0.003 & 0.526 & 3.266 & -0.00667 \\
    \texttt{W-0072} & 0.231 & 0.0472 & 0.7253 & $2.116$ & 0.834 & -1.130 & -0.142 & 0.221 & 4.091 & 2.23 \\
    \texttt{W-0073} & 0.303 & 0.0448 & 0.6020 & $2.588$ & 1.066 & -0.764 & -1.626 & 0.171 & 3.149 & -4.85 \\
    \texttt{W-0074} & 0.370 & 0.0502 & 0.7061 & $1.564$ & 0.911 & 0.156 & -2.069 & 0.468 & 2.217 & 1.18 \\
    \texttt{W-0207} & 0.265 & 0.0454 & 0.6940 & $2.956$ & 0.885 & -1.088 & -0.520 & 0.130 & 3.054 & -4.07 \\
    \texttt{W-0208} & 0.300 & 0.0533 & 0.6191 & $1.561$ & 1.004 & 0.121 & -1.070 & 0.320 & 2.370 & 2.99 \\
    \texttt{W-0209} & 0.383 & 0.0440 & 0.7121 & $1.716$ & 0.937 & -0.512 & -2.806 & 0.576 & 4.281 & -0.66 \\
    \texttt{W-0240} & 0.279 & 0.0421 & 0.6944 & $1.340$ & 1.063 & -0.392 & -2.844 & 0.284 & 3.750 & 3.93 \\
    \texttt{W-0241} & 0.306 & 0.0547 & 0.7076 & $2.207$ & 0.854 & -0.231 & -0.258 & 0.600 & 3.437 & 1.05 \\
    \texttt{W-0242} & 0.391 & 0.0477 & 0.6308 & $2.335$ & 0.938 & -1.256 & -1.640 & 0.0348 & 2.965 & -3.22 \\
    \texttt{W-0300} & 0.265 & 0.0499 & 0.7266 & $1.769$ & 0.805 & -0.0672 & -2.676 & 0.272 & 3.600 & 4.14 \\
    \texttt{W-0301} & 0.303 & 0.0527 & 0.6093 & $1.447$ & 1.030 & -0.875 & -1.229 & 0.555 & 2.275 & -1.07 \\
    \texttt{W-0302} & 0.379 & 0.0410 & 0.7002 & $2.383$ & 0.996 & -0.739 & -0.478 & 0.0116 & 3.919 & -2.81 \\
    \texttt{W-0522} & 0.222 & 0.0449 & 0.6163 & $2.087$ & 1.019 & -0.693 & -0.128 & 0.412 & 3.339 & -3.87 \\
    \texttt{W-0523} & 0.318 & 0.0481 & 0.7443 & $2.367$ & 0.987 & 0.185 & -1.845 & 0.384 & 4.489 & 3.10 \\
    \texttt{W-0524} & 0.377 & 0.0532 & 0.6754 & $1.377$ & 0.829 & -0.896 & -0.823 & 0.0908 & 2.591 & 1.37 \\
    \texttt{N-0144} & 0.285 & 0.0478 & 0.6582 & $2.289$ & 0.954 & -0.759 & -0.129 & 0.146 & 3.006 & 2.19 \\
    \texttt{N-0145} & 0.292 & 0.0495 & 0.7241 & $2.101$ & 0.987 & -0.982 & -0.922 & 0.0711 & 2.371 & -0.633 \\
    \texttt{N-0146} & 0.327 & 0.0452 & 0.6764 & $1.959$ & 0.974 & -1.268 & 0.185 & 0.0952 & 3.535 & -2.03 \\
    \texttt{N-0168} & 0.278 & 0.0465 & 0.6479 & $1.969$ & 0.963 & -0.874 & -0.305 & 0.140 & 2.783 & -2.75 \\
    \texttt{N-0169} & 0.296 & 0.0503 & 0.6819 & $1.939$ & 0.989 & -1.086 & -0.948 & 0.0744 & 3.532 & 1.74 \\
    \texttt{N-0170} & 0.349 & 0.0473 & 0.7176 & $2.440$ & 0.980 & -1.155 & 0.480 & 0.105 & 2.758 & 0.931 \\
    \texttt{N-0195} & 0.262 & 0.0505 & 0.6635 & $2.245$ & 0.975 & -0.973 & 0.0598 & 0.122 & 3.240 & 2.52 \\
    \texttt{N-0196} & 0.297 & 0.0457 & 0.6917 & $1.908$ & 0.955 & -0.851 & -0.855 & 0.103 & 3.185 & -2.51 \\
    \texttt{N-0197} & 0.336 & 0.0481 & 0.7364 & $2.113$ & 0.992 & -1.226 & -0.0093 & 0.0700 & 2.706 & -0.122 \\
    \texttt{N-0204} & 0.276 & 0.0502 & 0.6511 & $2.241$ & 0.956 & -1.092 & -0.400 & 0.0960 & 2.986 & -2.71 \\
    \texttt{N-0205} & 0.306 & 0.0456 & 0.6880 & $2.128$ & 0.991 & -1.158 & -0.632 & 0.0671 & 3.324 & 2.47 \\
    \texttt{N-0206} & 0.342 & 0.0483 & 0.7296 & $1.864$ & 0.973 & -0.879 & 0.281 & 0.141 & 2.711 & -0.420 \\
    \texttt{N-0336} & 0.261 & 0.0467 & 0.6901 & $1.949$ & 0.952 & -0.878 & -0.794 & 0.109 & 2.827 & 1.14 \\
    \texttt{N-0337} & 0.309 & 0.0479 & 0.6472 & $2.038$ & 0.991 & -1.238 & -0.334 & 0.0653 & 3.470 & -0.952 \\
    \texttt{N-0338} & 0.340 & 0.0492 & 0.7158 & $2.383$ & 0.979 & -1.003 & 0.172 & 0.145 & 2.517 & -1.26 \\

    \end{tabular}
    \end{ruledtabular}
\end{table*}

\end{document}